\documentclass[final,3p,times]{elsarticle}
\usepackage{amssymb}
\usepackage{amsmath}
\usepackage[fleqn]{mathtools}
\usepackage{esint}
\usepackage{latexsym}
\usepackage{newtxtext}
\usepackage{newtxmath}
\usepackage{color}
\DeclareMathAlphabet\rsfscr{U}{rsfso}{m}{n}
\biboptions{sort&compress}
\journal{Journal of Computational Physics}

\begin{document}

\begin{frontmatter}



\title{A robust high-resolution algorithm for quadrature-based moment methods applied to high-speed polydisperse multiphase flows}


% \author{} %% Author name

% \affiliation{organization={},%Department and Organization

\author[4]{Jacob W. Posey\corref{cor1}}
\cortext[cor1]{Corresponding author:}
\ead{jwposey@purdue.edu}
\author[2,3]{Rodney O. Fox}
\author[4,5]{Ryan W. Houim}

\affiliation[4]{organization={School of Mechanical Engineering, Purdue University},
            addressline={585 Purdue Mall},
            city={West Lafayette},
            postcode={47907},
            state={IN},
            country={USA}}

\affiliation[2]{organization={Department of Chemical and Biological Engineering, Iowa State University},
            addressline={618 Bissell Road},
            city={Ames},
            postcode={50011},
            state={IA},
            country={USA}}
\affiliation[3]{organization={Center for Multiphase Flow Research and Education, Iowa State University},
            addressline={Address},
            city={Ames},
            postcode={50011},
            state={IA},
            country={USA}}
\affiliation[5]{organization={School of Aeronautics and Astronautics, Purdue University},
            addressline={701 West Stadium Avenue},
            city={West Lafayette},
            postcode={47907},
            state={IN},
            country={USA}}

%% Abstract
\begin{abstract}
A high-resolution Eulerian method for simulating high-speed polydisperse granular multiphase flows has been developed.  The governing equations include a compressible gas that is coupled to mass-based moment equations for a polydisperse granular flow derived from the generalized population balance equation.  The model includes effects from particle collisions, drag, convective heat transfer, particle-fluid-particle pressure, and finite-size particle force terms. The mass moment integrals are closed using the generalized quadrature method of moments to allow for continuous size distributions. The governing equations are solved by using high-resolution reconstruction schemes and results from decoupled Riemann problems for the gas and particles as each quadrature node.  Success of the technique is demonstrated through a variety of numerical experiments including polydisperse multiphase Riemann shock-tube problems, shock--particle-curtain interactions, dust layer dispersal, dust layer dispersal by shock waves, and dispersal of spherical particle shells by high-pressure gas.  
\end{abstract}


% \begin{highlights}
% \end{highlights}

%% Keywords

\end{frontmatter}

\section{Introduction}

Granular multiphase flows are common in nature and are important to many phenomena. Particulates make up the tephra in volcanic plumes and pyroclastic flows \cite{Cimarelli2022}, water droplets in atmospheric clouds \cite{Berry1974}, and protoplanetary disks that form planets around young stars \cite{Youdin2005,Lambrechts2012}. Multiphase flows are also important to many industrial and defense applications. Reactive metallic particles are used as additives in propellants \cite{Ingenito2004,Risha2009} and other energetic formulations \cite{Trzcinski2015,Frost2018,Posey2021,Houim2022}. Dust explosions still wreak havoc in the processing and coal-mining industries.  Almost all of these multiphase flows are polydisperse and can involve a wide range of particle sizes.  Developing models and numerical methods for simulating polydisperse multiphase flows is not only important for scientific understanding, but also improving industrial safety and developing next-generation propulsion systems.

Many multiphase flow models approximate the particle size distribution as monodisperse. Given that most multiphase flows are polydisperse, the monodisperse simplification is accurate only in specific and carefully prepared laboratory applications. Effects from polydisperse size distributions can have a significant impact on the flow, which will be completely missed by monodisperse models.  Lightning is generated by polydispersity-driven triboelectric charging of small ash particles in volcanic plumes \cite{Cimarelli2014,MendezHarper2016,Cimarelli2022}.  Pyroclastic flows form size-segregated layers where a dilute flow of small particles flows on top of a dense layer of larger particles \cite{Neri2003}. Atmospheric cloud formation and evolution is driven by a broadening and increasingly bimodal distribution of water droplet sizes \cite{Berry1974}. Electrification of clouds in a thunderstorm is driven by triboelectric charging between falling large graupel particles and rising small ice crystals \cite{Saunders1993}. The leading theory for planetesimal formation in protoplanetary disks relies on the polydisperse streaming instability \cite{Squire2018,Squire2022})  where drag forces create clusters of protoplanetary dust that increase the likelihood of agglomeration driven by gravitational forces \cite{Youdin2005,McNally2020,McNally2021}.

Lagrangian approaches for particles can routinely handle polydisperse flows by simply assigning each particle its own size. However, these approaches can become computationally prohibitive in many practical scenarios with a large number of particles. Eulerian representations of the particle phase are more practical for scenarios with a very large number of particles. However, most Eulerian multiphase flow models remove information of the particle size distribution by directly making the monodisperse assumption \cite{Gidaspow1994,Lun1984,BrilliantovBook}. Simplified approaches for modeling polydisperse multiphase flows include using the size distribution to determine the surface area used for drag and convective heat transfer \cite{feroukas2023simplified}; however, all particles are assumed to share the same velocity and temperature.  Thus, these approaches do not allow the size distribution to evolve.  
Binning approaches to represent polydisperse multiphase flows and allow the size distribution to evolve \cite{binning,binning2,Neri2003}, but require many bins to represent realistic size distributions.   Quadrature-Based Moment Methods (QBMM) \cite{McGraw1997,Marchisio2013,Fox2024} are used to model the particle size distribution (PSD) by transporting higher-order statistical quantities of the size distribution such as variance, skewness, and kurtosis.  As a result, QBMM can represent realistic size distributions accurately with far fewer transport equations than binning approaches. Although recent extensions of QBMM to high-speed compressible multiphase flows use highly dissipative first-order numerical methods \cite{Fox2024}, this approach can accurately handle size distributions but is dissipative to flow features such as vortices and contact surfaces.   

In this work we extend the first-order accurate numerical method for QBMM to high order.  The resulting high-order QBMM approach can robustly and efficiently compute polydisperse size distributions with minimal numerical dissipation of flow features.
The numerical method is tested with a variety of challenging problems including multiphase Riemann problems, shock--particle-curtain interactions, and dispersal of dust layers by shock waves and high-pressure gas.   

\section{Model Formulation}

Here we list the transport equations for a compressible gas coupled to a polydisperse mixture of particles.  Full details of the governing equations and their derivation can be found in \cite{Houim2016,Fox2019,Fox2024,Fox2025,Posey2025}.

\subsection{Gas-Phase Transport Equations}

The gas-phase conservation equations for mass, species mass, momentum, and total energy are
\begin{equation}
    \frac{\partial \alpha_{g}\rho_{g}}{\partial t} + \nabla \cdot \left( \alpha_{g}\rho_{g}\boldsymbol{u}_{g} \right) = 0,
\end{equation}
\begin{equation}
\frac{\partial \alpha_{g}\rho_{g}Y_{g,i}}{\partial t} + \nabla \cdot \left( \alpha_{g}\rho_{g}Y_{g,i}\boldsymbol{u}_{g} \right) = 0,
\;\;\;\; i=1,...,N_{g},
\end{equation}
\begin{equation}
    \frac{\partial \alpha_{g}\rho_{g}\boldsymbol{u}_{g}}{\partial t} + \nabla \cdot \left( \alpha_{g}\rho_{g}\boldsymbol{u}_{g}\boldsymbol{u}_{g} \right) = -\alpha_{g}\nabla p_{g} + \boldsymbol{F}_{fs,g}^{P} + \mathcal{D}_{1}^{P} - \mathcal{L}_{1}^{P},
\end{equation}
\begin{equation}
    \frac{\partial \alpha_{g}\rho_{g}E_{g}}{\partial t} + \nabla \cdot \left( \alpha_{g}\boldsymbol{u}_{g}\left( \rho_{g}E_{g} + p_{g} \right) \right) = -p_{g}\nabla \cdot \left( \alpha_{p}\boldsymbol{u}_{p} \right) + F_{fs,g}^{E}  + \mathcal{D}_{1}^{E} - \mathcal{L}_{1}^{E} + \mathcal{H}_1,
\end{equation}
where $\alpha_{g}$ and $\alpha_{p}$ are the gas- and particle-phase volume fractions ($\alpha_{p}+\alpha_{g}=1$), $\rho_{g}$ is the gas-phase density, $\boldsymbol{u}_{g}$ and $\boldsymbol{u}_{p}$ are the gas- and particle-phase velocity vectors, $Y_{g,i}$ is the mass fraction of gas-phase species $i$, $p_{g}$ is the gas-phase pressure, $E_{g}$ is the gas-phase total specific energy, $e_{g}$ is the gas-phase specific internal energy, and $N_{g}$ is the number of gas-phase species. 
The terms $\mathcal{D}_1^P$ and $\mathcal{D}_1^E$ are momentum and energy exchanges from drag forces. The terms $\mathcal{L}_1^P$ and $\mathcal{L}_1^E$ are momentum and energy exchanges from lift forces. Convective heat transfer between the gas and particles is $\mathcal{H}_1$.

The gas-phase pressure is computed with the ideal gas equation of state
\begin{equation}
    p_{g}=\rho_{g}T_{g}R_u\sum_{i=1}^{N_{g}}\frac{Y_{g,i}}{MW_{i}}, 
\end{equation}
where $R_{u}$ is the universal gas constant, $MW_{i}$ is the molecular weight of species $i$, and $T_{g}$ is the gas-phase temperature.
The total and internal energy are
\begin{equation}
    E_g = \frac{1}{2} \boldsymbol{u}_g \cdot \boldsymbol{u}_g + e_{g}, \ \ \ \ e_{g} = \sum_{i=1}^{N_{g}}Y_{g,i}\left( h_{f,i}^{0} + \int_{T_{0}}^{T_{g}} c_{p,i}\left(s\right)ds \right) - \frac{p_{g}}{\rho_{g}},
\end{equation}
where $h_{f,i}^{0}$ is the enthalpy of formation for species $i$, $T_{0}$ is the reference temperature, $c_{p,i}$ is the constant-pressure specific heat capacity of species $i$. The specific heat capacity for each species is computed using NASA-9 thermodynamic polynomials \cite{McBride2002}. Mixture-averaged approximations were used to calculate the transport coefficients needed for drag and convective heat transfer \cite{Ern1994,Kee2003,Houim2011a}.

\subsection{Particle-Phase Transport Equations}
The particle-phase transport equations are given by a generalized population balance equation (GPBE) that models the time evolution of the particle number density function (NDF)
\cite{Marchisio2013}
\begin{equation}
    \label{GPBE}
    \frac{\partial f\left( t,\boldsymbol{x},\boldsymbol{v},\boldsymbol{\xi} \right)}{\partial t} + \nabla_{\mathbb{R}}\cdot\left( \boldsymbol{v}f \right) + \nabla_{\mathbb{V}}\cdot\left( \boldsymbol{A}f \right) + \nabla_{\mathbb{P}}\cdot\left( \boldsymbol{G}f \right) = C + F,
\end{equation}
where $\boldsymbol{x} = \left( x,y,z \right)^{T}$ is the position vector in real-space $\mathbb{R}^{3}$, $\boldsymbol{v}=\left( v_{x},v_{y},v_{z} \right)^{T}$ is the velocity vector in velocity-space $\mathbb{V}^{3}$, $\boldsymbol{\xi}=\left( \xi_{1},\xi_{2},...,\xi_{h} \right)^{T}$ is the internal coordinates vector in $h$-dimensional phase-space $\mathbb{P}^{h}$, $\boldsymbol{A}$ is the acceleration vector acting on the particles due to continuous body forces, $\boldsymbol{G}$ is the change in the internal coordinates vector due to continuous phenomena, $C$ is the change in the NDF due to discrete collisions, and $F$ is the change in the NDF due to frictional-collisional forces at large particle volume fractions. 

In this work, we use a two-dimensional phase space $\mathbb{P}^{2}$, with internal coordinates  $\boldsymbol{\xi}=\left( m,e \right)^{T}$, where $m$ and $e$ are particle mass and internal energy, respectively.  
We  assume that the internal energy of the particle is correlated only with its mass in the NDF. The NDF then simplifies to $f(\boldsymbol{v},m)$.  Integrating over velocity produces a univariate distribution over particle mass 
\begin{equation}
     f\left(m \right) = 
     \int\limits_{\mathbb{V}^{3}}\hspace{-0.4em} f\left(\boldsymbol{v},m
     \right)d\boldsymbol{v}
     ,
\end{equation}
where $f\left(m \right)$ is the mass NDF and $f\left(\boldsymbol{v},m
\right)$ is the joint mass-velocity NDF. (Here $f(m)$ is understood to also be a function of time and position.)  The mean velocity, $\boldsymbol{u}_p(m)$ and granular temperature, $\Theta_p(m)$, as a function of particle mass are
 \begin{equation}
 \boldsymbol{u}_p(m) = \frac{\int \boldsymbol{v} f(\boldsymbol{v},m) d\boldsymbol{v}}{\int{ f(\boldsymbol{v},m) d\boldsymbol{v}}}, \ \ \ \ 
 \Theta_p(m) = \frac{\int (\boldsymbol{v}'\cdot \boldsymbol{v}') f(\boldsymbol{v},m) d\boldsymbol{v}}{\int{ f(\boldsymbol{v},m) d\boldsymbol{v}}},
 \end{equation}
 where $\boldsymbol{v}'$ is the fluctuating particle velocity, which is assumed to be Maxwellian \cite{Fox2024},

The mass ($\mathcal{M}_{s/q}$), momentum ($\boldsymbol{\mathcal{U}}_{s}$), pseudo-thermal energy ($\mathcal{T}_{s}$), and internal energy ($\mathcal{E}_{s}$) moments are
\begin{align}
    \label{eqn:Mom_defs}
    \mathcal{M}_{n/q} = \int m^{n/q} f(m) dm, \ \ \ \boldsymbol{\mathcal{U}}_s = \int m^s \boldsymbol{u}_p(m) f(m) dm, \ \ \ \\ \nonumber \mathcal{T}_s = \frac{3}{2}\int m^s \Theta_p(m) f(m) dm, \ \ \ \mathcal{E}_s = \int m^s e_p(m) f(m) dm,
\end{align}
respectively, where $q$ selects the type of mass moment used (values of 1, 2, and 3 correspond to mass, area, and size moments, respectively), $s$ and $n$ are integers greater than or equal to 0.

The governing equations for the particle phase are found by taking statistical moments of the generalized population balance equation \cite{Fox2019,Fox2024}
\begin{equation}
    \frac{\partial\mathcal{M}_{n/q}}{\partial t} + \nabla\cdot F\left(\mathcal{M}_{n/q}\right) = 0,
    \;\;\;\; n = 0,1,\cdots
\end{equation}
\begin{equation}
    \frac{\partial\boldsymbol{\mathcal{U}}_{s}}{\partial t} + \nabla\cdot\left(F\left(\boldsymbol{\mathcal{U}}_{s}\right) + \mathcal{P}_{s}\boldsymbol{I}\right) = -\frac{\mathcal{M}_{s}}{\rho_{p}}\nabla p_{g}-\boldsymbol{F}_{fs,p,s}^{P} 
    + \boldsymbol{\mathcal{C}}^{P}_{p,s} - \boldsymbol{\mathcal{D}}_s^P + \boldsymbol{\mathcal{L}}_s^P, \;\;\;\; s=0,1,\cdots
\end{equation}
\begin{equation}
    \frac{\partial\mathcal{T}_{s}}{\partial t} + \nabla\cdot F\left({\mathcal{T}}_{s}\right) = - \mathcal{P}_{s}^{kc} \nabla \cdot \frac{\boldsymbol{\mathcal{U}}_{s}}{\mathcal{M}_{s}} - F_{fs,p,s}^{PTE} 
    + \mathcal{C}^{PTE}_{p,s}  + \mathcal{D}_{s}^{PTE} + \mathcal{F}_{s}^{PTE}, \;\;\;\; s=0,1,\cdots
\end{equation}
\begin{equation}
    \frac{\partial\mathcal{E}_{s}}{\partial t} + \nabla\cdot F\left(\mathcal{E}_{s}\right) = - \mathcal{C}^{E}_{s} - \mathcal{F}^{PTE}_{s} - \mathcal{H}_s, \;\;\;\; s=0,1,\cdots
\end{equation} 
where $\rho_p$, $\mathcal{P}_{s}$, and $\mathcal{P}_{s}^{kc}$ are the particle material density, $s$-order particle-pressure moment, $s$-order kinetic-collisional particle-pressure moment. 
Collision source terms for momentum, pseudo-thermal energy, and total kinetic energy are $\boldsymbol{C}^{P}_{p,s}$ ${C}^{PTE}_{p,s}$, and ${C}^{E}_{p,s}$, respectively.
The fluxes in the moment equations are
\begin{align}\label{eqn:momentFluxes}
    F\left(\mathcal{M}_{n/q}\right) = \int m^{n/q} \boldsymbol{u}_p(m) f(m) dm, \ \ \
    F\left(\boldsymbol{\mathcal{U}}_{s}\right) = \int m^s \boldsymbol{u}_p(m)\boldsymbol{u}_p(m) f(m) dm, \ \ \ \\
    \nonumber
    F\left({\mathcal{T}}_{s}\right) = \frac{3}{2} \int m^s \Theta_p(m) \boldsymbol{u}_p(m) f(m) dm, \ \ \
    \;\;\;\;
    F\left(\mathcal{E}_{s}\right) = 
    \int m^s e_p(m) \boldsymbol{u}_p(m) f(m) dm.
\end{align}

The mean variables of particles are determined from the first-order moments
\begin{equation}
    \alpha_p = \frac{\mathcal{M}_{q/q}}{\rho_p}, \;\;\;\; \boldsymbol{u}_p = \frac{\boldsymbol{\mathcal{U}}_1}{\mathcal{M}_{q/q}}, \;\;\;\; \Theta_p = \frac{2}{3}\frac{\mathcal{T}_1}{\mathcal{M}_{q/q}}, \;\;\;\; e_{p} = \frac{\mathcal{E}_{1}}{\mathcal{M}_{q/q}},
    \label{eq:mean_quantities}
\end{equation}
where $\boldsymbol{u}_p$ is the mean bulk velocity of the particle phase, $\Theta_p$ is the mean granular temperature, and $e_{p}$ is the mean internal energy.  The total number density of the particles is $N_d = \mathcal{M}_0$. The particle internal energy is 
\begin{equation}
    e_{p}(m) = e_{\textrm{ref}} + c_{v,p} \left(T_p(m) - T_{\textrm{ref}}\right),
\end{equation}
where $T_{\textrm{ref}}$ is a reference temperature, $e_{\textrm{ref}}$ is the particle internal energy at $T_{\textrm{ref}}$, and $c_{v,p}$ is the constant-volume specific heat capacity of the particle material, which assumed to be constant for this work.

\subsubsection{Particle Pressure}
The particle-pressure moments are
\begin{equation}
    \mathcal{P}_{s} = \int m^s (\theta_k + \theta_c + \theta_f) f(m) dm, \ \ \     \mathcal{P}_{s}^{kc} = \int m^s (\theta_k + \theta_c) f(m) dm, 
\end{equation}
where $\theta_k$ is the kinetic temperature,
($\theta_{k} = \Theta_p$), $\theta_f$ is the ``frictional'' temperature, and $\theta_c$ is the collisional temperature.
The frictional temperature is \cite{Johnson1987} 
\begin{equation}
    \label{theta_f}
    \theta_{f} = \frac{1}{\alpha_{p}\rho_{p}}
    \begin{cases}
        0, & \textrm{if } \alpha_{p} < \alpha_{p,\textrm{crit}} \\
        Fr \, \alpha_p\frac{\left( \alpha_{p} - \alpha_{p,\textrm{crit}} \right)^{r_1}}{\left( \alpha_{p,\textrm{max}} - \alpha_{p} \right)^{r_2}}, & \textrm{if } \alpha_{p} \geq \alpha_{p,\textrm{crit}}
    \end{cases}
\end{equation}
where $\alpha_{p,\textrm{crit}}$ is the particle volume fraction where frictional-collisional pressure is activated, $\alpha_{p,\textrm{max}}$ is the packing limit, and $Fr$, $r_1$, and $r_2$ are constants.  The values for $\alpha_{p,\textrm{crit}}$, $\alpha_{p,\textrm{max}}$, and  $Fr$, $r_1$, and $r_2$ are 0.5, 0.65, 0.1, 2, and 5, respectively. 

The ``collisional'' temperature is
\begin{equation}
    \theta_{c} = \int 2\left( 1+e(m,M) \right)\beta(M)\alpha_{p}g(m,M) \chi(m,M)^{3}\mu(m,M)y(m,M)E(m,M) dM,
\end{equation}
where $M$ is the mass of the colliding particle, $e_c(m,M)$ is the coefficient of restitution between particles of masses $m$ and $M$ (see Section \ref{Sec:CoR}),
 \begin{equation}
     \mu(m,M) = \frac{2d_p^3(M)}{d_p^3(m) + d_p^3(M)}, \ \ \ \chi(m,M) = \frac{d_p(m) + d_p(M)}{2 d_p(M)}, \ \ \ y(m,M) = \begin{cases}
         \frac{1}{2}\mu(m,M), \ \text{ if } m \leq M \\
         \frac{1}{2}\mu(M,m), \ \text{ if } m > M
     \end{cases}.
 \end{equation}
 $\beta(M)$ is the volume fraction of particles of mass $M$ in the distribution
\begin{equation}
    \beta(M)=\frac{Mf(M)}{\mathcal{M}_1}, \;\;\;\; \int\beta(m)dm=1.
\end{equation}
$E(m,M)$ is the energy function
\begin{equation*}
    E(m,M) = \Theta_p(m) + \Theta_p(M) + \frac{1}{3}||\boldsymbol{u}_p(m) - \boldsymbol{u}_p(M)||^2,
\end{equation*}
and $g(m,M)$ is the polydisperse radial distribution function \cite{Santos1999}
\begin{equation}
    g(m,M) = \frac{1}{1-\alpha_{p}} + \left( g_{0} - \frac{1}{1-\alpha_{p}} \right) \frac{\langle d_{p}^{2} \rangle}{\langle d_{p}^{3} \rangle} \frac{d_p(m)}{\chi(m,M)}, \ \ \ \langle d_p^n \rangle = \int d_p^n(m) f(m) dm,
\end{equation}
where $g_{0}$ is the monodisperse radial distribution function \cite{sinclair1989gas}
\begin{equation}
    \label{g0}
    \frac{1}{g_{0}} = 1-\left( \frac{\alpha_{p}}{\alpha_{p,\textrm{max}}} \right)^{1/3}.
\end{equation}

\subsubsection{Coefficient of Restitution}
\label{Sec:CoR}
An impact velocity dependent coefficient of restitution model is used to approximate effects from particle deformation and viscoelasticity
\cite{Ramirez1999,BrilliantovBook}
\begin{equation}
    \label{CoR_Pade}
    e_c(m,M) = \left(1 - H(u_{\textrm{impact},cr})\right)\frac{1+a_{1}U_{\textrm{impact}}^{1/5}}{1+a_{2}U_{\textrm{impact}}^{1/5}+a_{3}U_{\textrm{impact}}^{2/5}+a_{4}U_{\textrm{impact}}^{3/5}+a_{5}U_{\textrm{impact}}^{4/5}},
\end{equation}
where $H$ is the Heaviside unit step function, $U_{\textrm{impact}}=u_{\textrm{impact}}/u^{*}$, $u_{\textrm{impact}}$ is the impact velocity, $u^{*}$ is the characteristic impact velocity, $u_{\textrm{impact},cr}$ is a critical velocity above which $e_c(m,M) = 0$ due to plastic deformation within the particle \cite{Hassani-Gangaraj2018},  $a_{1}\approx 2.583$, $a_{2}\approx 3.583$, $a_{3}\approx 2.983$, $a_{4}\approx 1.148$, and $a_{5}\approx 0.326$ \cite{Ramirez1999,BrilliantovBook}.  The impact velocity is determined from the difference in mean velocities between particles of masses $m$ and $M$, $u_{\textrm{impact}} = \left|\left| \boldsymbol{u}_{p}(m) - \boldsymbol{u}_p(M) \right|\right|$.  The potential dependence of $u_{\textrm{impact}}$ on the granular temperature is currently neglected. 

\begin{figure}[tbp]
    \centering
    \includegraphics[width=.35\textwidth]{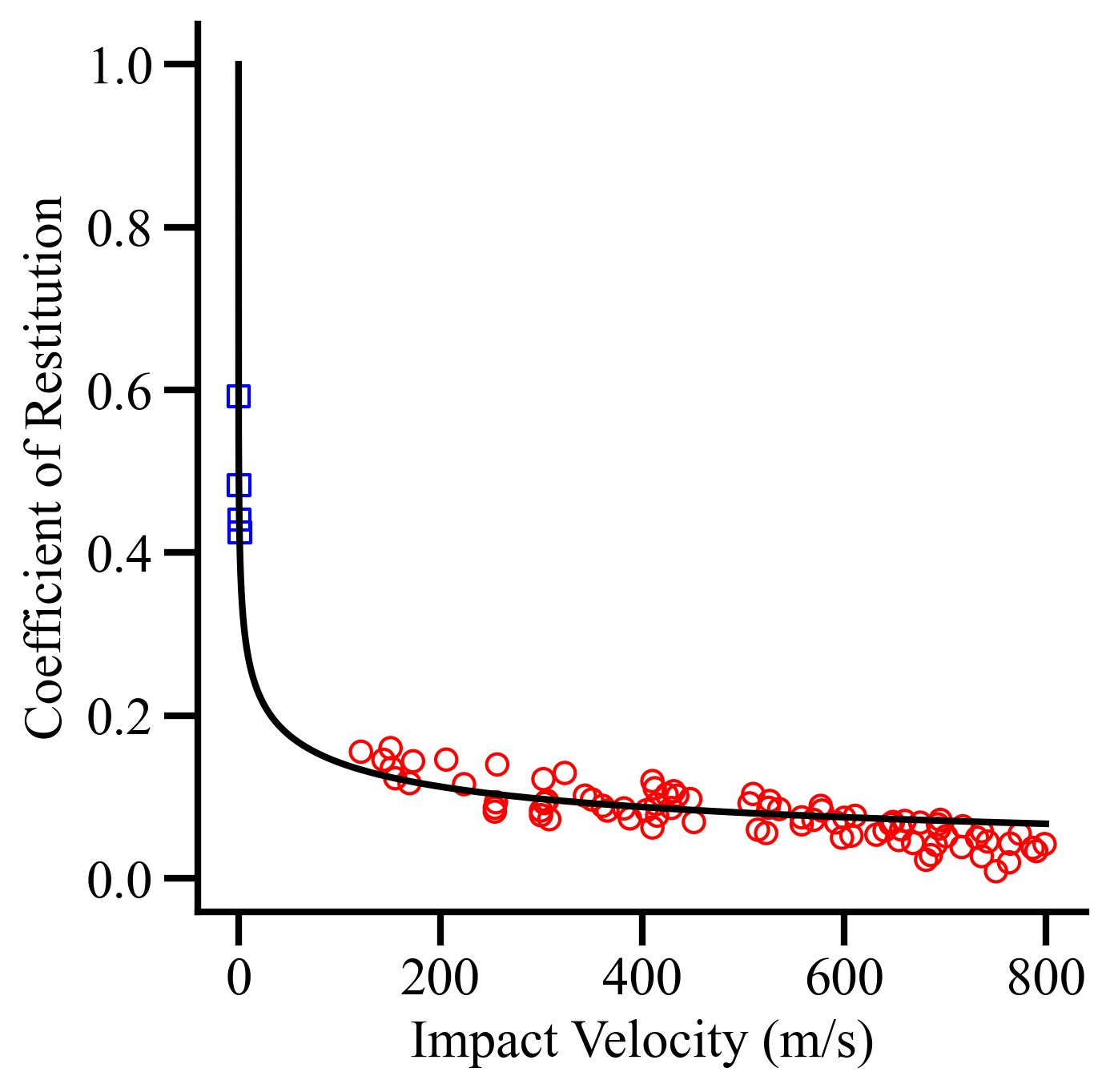}
\caption{Impact-velocity-dependent coefficient of restitution of aluminum particles (the black line is Eqn.~\eqref{CoR_Pade} and the blue \cite{Weir2005} and red \cite{Hassani-Gangaraj2018} symbols represent experimental data).} \label{CoR_data}
\end{figure}

A characteristic impact velocity of $u^{*} = 1.482$ m/s for colliding aluminum particles was found by fitting \eqref{CoR_Pade} to experimental measurements \cite{Weir2005,Hassani-Gangaraj2018}. A critical velocity of $u_{\textrm{impact},cr}=800$ m/s \cite{Hassani-Gangaraj2018} was used. Figure~\ref{CoR_data} shows the fitted coefficient of restitution as a function of impact velocity for aluminum particles. 

\subsection{Interphase and Interparticle Energy Exchange Terms}
\begin{figure}[tbp]
    \centering
    \includegraphics[width=.6\textwidth]{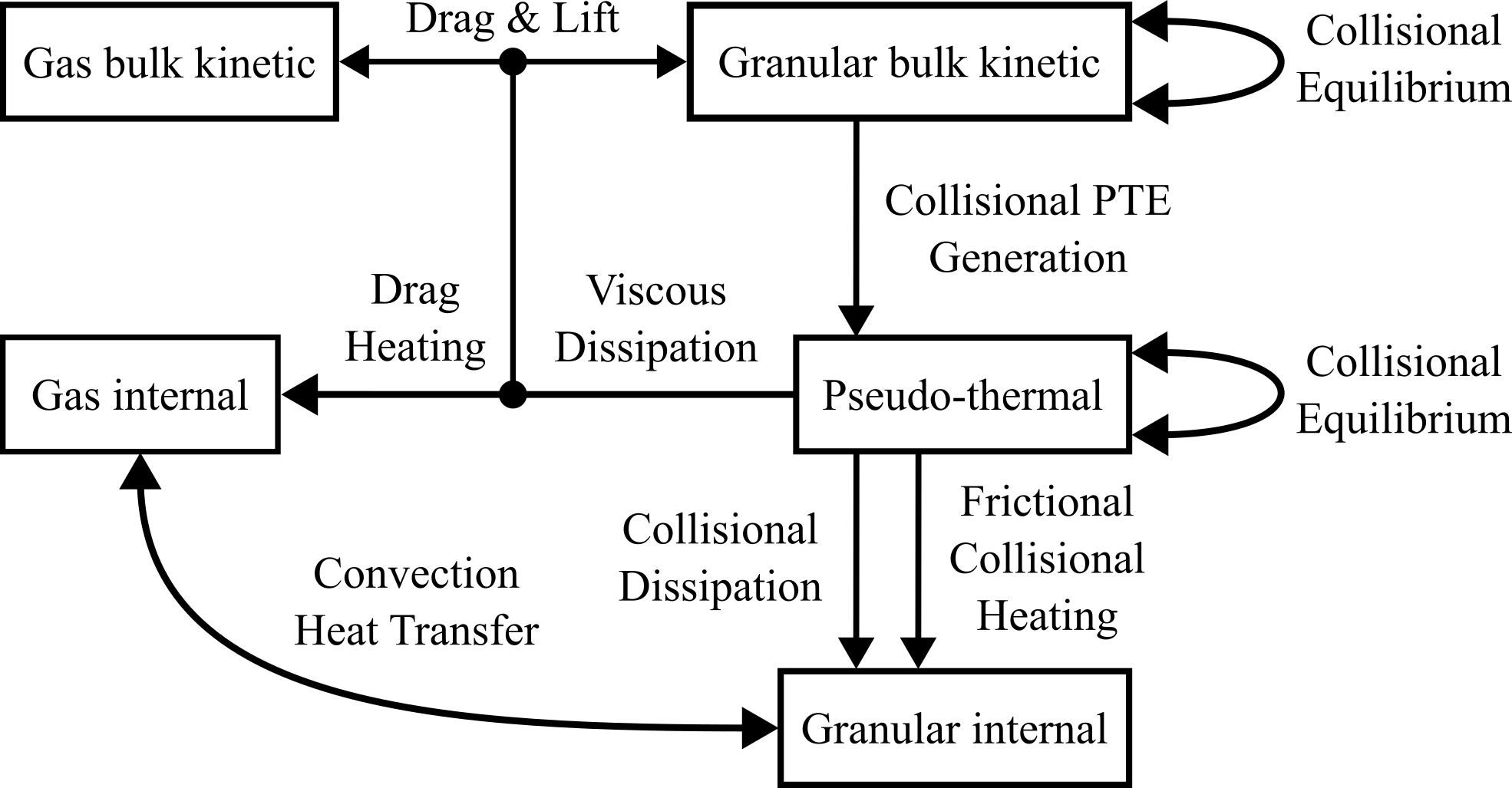}
    \caption{Diagram of energy transfer within the model, demonstrating conservation of energy.} \label{enTransfer}
\end{figure}

Figure~\ref{enTransfer} shows the energy transfer mechanisms between mixture of gas and polydisperse particles.  Lift forces, $\mathcal{L}$, reversibly transfer bulk kinetic energy between the phases.  Drag forces, $\mathcal{D}$, irreversibly transfer kinetic energy between and increase the internal energy of the gas.  Convection, $\mathcal{H}$, transfers heat between gas and particles.  Particle collisions, $\mathcal{C}$,  exchange bulk kinetic energy between particles of different masses as well as increase their internal energy. The model for each of these terms is discussed below.  A full derivation of these energy exchange terms can be found in \cite{Fox2024}.

\subsubsection{Drag and Lift}

The drag source terms are
\begin{equation}
  \begin{split}
    &\mathcal{D}_s^P = \int m^s\frac{(\boldsymbol{u}_p(m) - \boldsymbol{u}_g)}{\tau_p(m)} f(m)dm, \ \ \     \mathcal{D}_s^{{PTE}} = \int m^s\frac{3 m^s \Theta_p(m)}{\tau_p(m)} f(m)dm, \\
    &\mathcal{D}_s^E = \int m^s\frac{\left[\boldsymbol{u}_p(m) \cdot (\boldsymbol{u}_p(m) - \boldsymbol{u}_g) + 3\Theta_p(m)\right]}{\tau_p(m)} f(m)dm,   
\end{split}
\end{equation}
where $\tau_{p}(m)$ is the particle relaxation time due to drag
\begin{equation}
    \tau_{p}(m) = \frac{4\rho_{p}d_{p}^{2}(m)}{3\mu_{g}C_{D}(m)\textrm{Re}_{p}(m)}, \;\;\;\; \textrm{Re}_{p}(m) = \frac{\rho_{g}||\boldsymbol{u}_{p}(m)-\boldsymbol{u}_{g}||d_{p}(m)}{\mu_{g}},
\end{equation}
where $\mu_{g}$ is the gas phase viscosity and $\textrm{Re}_{p}(m)$ and are the particle Reynolds number and drag coefficient for particles of mass $m$. The Gidaspow-type drag correlation \cite{Gidaspow1994,Huilin2003} with a modified drag correlation in the dilute regime to account for high slip Mach number effects on the drag force \cite{Osnes2023} is blended with the Ergun equation to describe drag forces in fluidized-bed regimes \cite{Ergun}
\begin{equation}
    C_{D}(m) = (1-\varphi)C_{D,\textrm{Osnes}} + \varphi C_{D,\textrm{Ergun}}, \;\;\;\; C_{D,\textrm{Ergun}} = \frac{4}{3}\left( 150\frac{\alpha_{p}}{\alpha_{g}\textrm{Re}_{p}(m)} + 1.75 \right),
\end{equation}
where $\varphi$ is the blending function \cite{Huilin2003}
\begin{equation}
    \varphi = \frac{\arctan\left( 262.5\left( \alpha_p-0.2 \right) \right)}{\pi}+0.5.
\end{equation}
The details fo $C_{D,\textrm{Osnes}}$ can be found in \cite{Osnes2023}.

The $s$-order moment lift source terms are
\begin{equation}
   \begin{split}
   &\boldsymbol{\mathcal{L}}_s^P = \int \frac{m^s f(m)}{\alpha_p \rho_p}C_L \alpha_g \rho_g(\boldsymbol{u}_p(m) - \boldsymbol{u}_g) \times (\nabla \times \boldsymbol{u}_g) dm, \\
 &\boldsymbol{\mathcal{L}}_s^E = \int \frac{m^s f(m)}{\alpha_p \rho_p}C_L \alpha_g \rho_g \boldsymbol{u}_p(m)\cdot(\boldsymbol{u}_p(m) - \boldsymbol{u}_g) \times (\nabla \times \boldsymbol{u}_g) dm,
 \end{split}
 \label{eqn:lift}
\end{equation}
where $C_{L}=0.5$ is the lift coefficient.

\subsubsection{Convective Heat Transfer}
The $s$-order mass convection heat transfer source term is
\begin{equation}
    \mathcal{H}_{s} = \int m^s h(m)\left[T_g - T_p(m)\right] f(m) dm, \;\;\;\; h(m)=\frac{6 \lambda_g \textrm{Nu}_p(m)}{\rho_p d_p^2(m)},
\end{equation}
where $\lambda_{g}$ is the gas phase thermal conductivity and $\textrm{Nu}_{p}(m)$ is the Nusselt number \cite{Gunn1978}
\begin{equation}
    \textrm{Nu}_{p}(m) = \left( 7-10\alpha_{g}+5\alpha_{g}^{2} \right)\left( 1+0.7\textrm{Re}_p^{0.2}(m)\textrm{Pr}_{g}^{1/3} \right) + \left( 1.33-2.4\alpha_{g}+1.2\alpha_{g}^{2} \right)\textrm{Re}_{p}^{0.7}(m)\textrm{Pr}_{g}^{1/3},
\end{equation}
where $\textrm{Pr}_{g}$ is the Prandtl number.

\subsubsection{Particle Collisions and Friction}
A Bhatnagar-Gross-Krook (BGK) approximation is used to relax the velocities to a multi-variate Maxwellian distribution \cite{Andries2002}.  The collision source term for the momentum equations are
\begin{equation}
    \mathcal{C}_s^P = \iint m^s \kappa\left(m,M\right)\psi\left(m,M\right)\left(\boldsymbol{u}_p(M) - \boldsymbol{u}_p(m)\right) f(m) dM dm,
\end{equation}
\begin{equation}
    \mathcal{C}_s^E = \iint 3m^s\kappa(m,M)\psi(m,M) \left[ \psi(m,M)E(m,M) - \Theta_p(m) + \frac{1}{3}\boldsymbol{u}_p(m)\cdot\left(\boldsymbol{u}_p(M)-\boldsymbol{u}_p(m)\right) \right] f(m) dM dm,
\end{equation}
\begin{equation}
    \mathcal{C}_s^{PTE} = \iint 3m^s\kappa(m,M)\psi(m,M) \left[ \psi(m,M)E(m,M) - \Theta_p(m) \right] f(m) dM dm,
\end{equation}
where the collision rate between particles of sizes $m$ and $M$ is
\begin{equation}
   \kappa(m,M) = \sqrt{\frac{\pi}{2}}\left( d_p(m)+d_p(M) \right)^{2}f(M)g(m,M)\sqrt{E(m,M)}
\end{equation},
and the material parameter $\psi$ is
\begin{equation}
   \psi(m,M) = \frac{1}{4}\left( 1+e_c(m,M) \right)\mu(m,M).
\end{equation}

The $s$-order mass frictional collisions moment is \cite{Fox2024}
\begin{equation}
    \mathcal{F}_{s}^{PTE} = 
    -\int \frac{3 \Theta_p(m)}{2\tau_{fr}} m^s f(m) dm,
    \;\;\;\; \tau_{fr} = \frac{2c_{f}}{\max\left( \left|\nabla\cdot\boldsymbol{u}_p\right|,
    \tau_c
    \right)}\left( 1+\tanh\left( \frac{\alpha_{p}-\alpha_{p,\textrm{crit}}}{\Delta_{f}} \right) \right)^{-1},
\end{equation}
where $\tau_{fr}$ is the frictional collisions time scale and $\tau_c$ is the monodisperse collisional time scale
\begin{equation}
    \tau_c = \frac{d_p\sqrt{\pi}}{12\alpha_pg_0\sqrt{\Theta_p}}.
\end{equation}
The constants $c_f = 0.01$ and $\Delta_{f}=0.01$ control the strength of energy dissipation and width of the hyperbolic tangent function, respectively.

\subsubsection{Particle-Fluid-Pressure and Finite-Size Particle Forces}
The terms $\boldsymbol{F}_{fs,g}^{P}$ and $F_{fs,g}^{E}$ are forces and energy transfer due to finite-size particle effects \cite{Fox2019,Fox2025}
\begin{equation}
    \boldsymbol{F}_{fs,g}^{P} = \alpha_{p}\boldsymbol{F}_{pg} + \nabla \cdot \boldsymbol{P}_{pfp} - \nabla \cdot \left( \rho_{g}\alpha_{p}\left(\Theta_{p}\boldsymbol{I}+\boldsymbol{R}\right) \right),
\end{equation}
\begin{equation}
    F_{fs,g}^{E} = \alpha_{p}\boldsymbol{u}_{p}\cdot\boldsymbol{F}_{pg} + \boldsymbol{u}_{p} \cdot \left(\nabla \cdot \boldsymbol{P}_{pfp}\right) - \nabla \cdot \left( \rho_{g}\alpha_{p}\boldsymbol{u}_{p}\cdot\left(\Theta_{p}\boldsymbol{I}+\boldsymbol{R} \right)\right) -\nabla\cdot\boldsymbol{r} + \alpha_{p}D_{pg},
\end{equation}
where $\Theta_{p}$ is the particle phase granular temperature given in Eqn.~\eqref{eq:mean_quantities}, $\boldsymbol{R}$ is the slip pressure tensor \cite{Fox2019,Fox2025}
\begin{equation}
    \boldsymbol{R} = \frac{1}{5}\left( \boldsymbol{u}_{pg}\cdot\boldsymbol{u}_{pg} \right)\boldsymbol{I} + \frac{2}{5}\boldsymbol{u}_{pg}\boldsymbol{u}_{pg} .
\end{equation}
The particle-fluid-pressure (PFP), $\boldsymbol{P}_{pfp}$, is a stress due to unresolved action of particles altering the fluid, which, in turn, produces pressure-like forces on the neighboring particles \cite{Zhang2021,Fox2025}
\begin{equation}
    \boldsymbol{P}_{pfp} = \rho_{g}\alpha_{p}^{2}\boldsymbol{R} .
\end{equation}
The term $\boldsymbol{F}_{pg}$ consists of gas density gradient and mixture compression forces
\begin{equation}
    \boldsymbol{F}_{pg} = \left(\Theta_{p}\boldsymbol{I}+\alpha_{g}\boldsymbol{R}\right)\cdot\nabla\rho_{g} - \frac{2}{3}\rho_{g}\left( \nabla\cdot\boldsymbol{u}_{v} \right)\boldsymbol{u}_{pg}, \;\;\;\; \boldsymbol{u}_{v}=\alpha_{g}\boldsymbol{u}_{g}+\alpha_{p}\boldsymbol{u}_{p}
    , \;\;\;\; \boldsymbol{u}_{pg} = \boldsymbol{u}_{p}-\boldsymbol{u}_{g},
\end{equation}
where $\boldsymbol{u}_v$ is the volume-averaged mixture velocity, and $\boldsymbol{u}_{pg}$ is the slip velocity vector.  The terms  $\boldsymbol{r}$ and $D_{pg}$ are the finite-size particle energy flux and exchange terms are
\begin{equation}
    \boldsymbol{r} = 2\rho_{g}\alpha_{p}\Theta_{p}\boldsymbol{u}_{pg}, \;\;\;\; D_{pg} = 2\Theta_{p}\left( \boldsymbol{u}_{pg}\cdot\nabla\rho_{g} - \rho_{g}\nabla\cdot\boldsymbol{u}_{g} \right), 
\end{equation}
where
$\Theta_{p}$ is the mean granular temperature over for all particle masses, and $\boldsymbol{I}$ is the identity tensor.

The pressure non-disturbance condition requires that pressure and normal velocity be preserved by the numerical method at  contact waves and material interfaces \cite{Abgrall1996,Saurel1999}. Violating this produces pressure and velocity oscillations near contact surfaces \cite{Liou2008} and, in severe cases, code failure. 
For example, the gas-phase momentum equation with particles produces a pressure gradient term ($\nabla\alpha_{g}p_{g}$) and a seperate ``nozzling'' term ($p_{g}\nabla\alpha_{g}$).  It is difficult for these terms to perfectly balance near sharp granular interfaces when computed separately even if the pressure is uniform.  
This issue was solved in our previous work by combining both terms into one, $\alpha_{g}\nabla p_{g}$ \cite{Houim2016}, which trivially satisfies the non-disturbing condition at material interfaces. 
Similarly, it was found that the PFP and finite-size particle force terms needed to be rearranged to satisfy the non-disturbing condition.  The final forms of these source terms are
\begin{equation}
\begin{split}
    \boldsymbol{F}_{fs,g}^{P} = &-\frac{4}{15}\rho_g\alpha_p\alpha_g\boldsymbol{u}_{pg}\left(\nabla\cdot\boldsymbol{u}_g\right) + \rho_g\left[ \left( \alpha_p-\alpha_g \right)\boldsymbol{R}-\frac{2}{3}\alpha_p\boldsymbol{u}_{pg}\boldsymbol{u}_{pg}-\Theta_p\boldsymbol{I} \right]\cdot\nabla\alpha_p-\rho_g\alpha_p\nabla\Theta_p \\
    &- \rho_g\alpha_p\left( \frac{4}{15}\alpha_p+\frac{2}{5} \right)\boldsymbol{u}_{pg}\left( \nabla\cdot\boldsymbol{u}_p \right)+\frac{4}{5}\rho_g\alpha_p\alpha_g\boldsymbol{u}_{pg}\cdot\left(\boldsymbol{\varepsilon}_g-\boldsymbol{\varepsilon}_p\right),
\end{split}
\end{equation}
\begin{equation}
\begin{split}
    F_{fs,g}^{E} = &-\frac{4}{15}\rho_g\alpha_p\alpha_g\left(\boldsymbol{u}_p\cdot\boldsymbol{u}_{pg}\right)\left(\nabla\cdot\boldsymbol{u}_g\right)-\rho_g\alpha_p\left( \boldsymbol{u}_p+2\boldsymbol{u}_{pg} \right)\cdot\nabla\Theta_p+\frac{4}{5}\rho_g\alpha_p\alpha_g\boldsymbol{u}_{pg}\boldsymbol{u}_p:\left(\boldsymbol{\varepsilon}_g-\boldsymbol{\varepsilon}_p\right) \\
    &+\rho_g\left[ \left( \alpha_p-\alpha_g \right)\left(\boldsymbol{u}_p\cdot\boldsymbol{R}\right)-\frac{2}{3}\alpha_p\left(\boldsymbol{u}_p\cdot\boldsymbol{u}_{pg}\boldsymbol{u}_{pg}\right)-\Theta_p\left( \boldsymbol{u}_p+2\boldsymbol{u}_{pg} \right)\right]\cdot\nabla\alpha_p \\
    &- \left[\rho_g\alpha_p\left( \frac{4}{15}\alpha_p+\frac{2}{5} \right)\left(\boldsymbol{u}_p\cdot\boldsymbol{u}_{pg}\right)+3\rho_g\alpha_p\Theta_p\right]\left( \nabla\cdot\boldsymbol{u}_p \right) - \rho_g\alpha_p\boldsymbol{R}: \nabla\boldsymbol{u}_p,
\end{split}
\end{equation}
where $\boldsymbol{\varepsilon}_p=\frac{1}{2}\left[ \nabla\boldsymbol{u}_p+\left( \nabla\boldsymbol{u}_p \right)^T \right]$ and $\boldsymbol{\varepsilon}_g=\frac{1}{2}\left[ \nabla\boldsymbol{u}_g+\left( \nabla\boldsymbol{u}_g \right)^T \right]$ are the particle and gas phase rate-of-strain tensors.

The finite-particle-size source terms $\boldsymbol{F}_{fs,p,s}^{P}$ and $F_{fs,p,s}^{PTE}$ for the particle momentum and PTE moment equations are 
\begin{equation}
    \boldsymbol{F}_{fs,p,s}^{P} = \frac{\mathcal{M}_{s}}{\rho_{p}}\boldsymbol{F}_{pg} + \frac{\mathcal{M}_{s}}{\alpha_{p}\rho_{p}}\left( \nabla\cdot\boldsymbol{P}_{pfp} \right), \;\;\;\; F_{fs,p,s}^{PTE} = \frac{\mathcal{M}_{s}}{\rho_{p}}D_{pg},
\end{equation}
which may also be rearranged to satisfy the non-disturbing condition
\begin{equation}
\begin{split}
    \boldsymbol{F}_{fs,s,p}^{P} = &\frac{\mathcal{M}_s}{\rho_p}\left( \Theta_p\boldsymbol{I+\boldsymbol{R}} \right)\cdot\nabla\rho_g - \frac{\mathcal{M}_s}{\rho_p}\rho_g\left( \frac{2}{5}\alpha_p-\frac{2}{3}\alpha_g \right)\boldsymbol{u}_{pg}\left( \nabla\cdot\boldsymbol{u}_g \right) + 2\frac{\mathcal{M}_s}{\rho_p}\rho_g\left( \boldsymbol{R}-\frac{1}{3}\boldsymbol{u}_{pg}\boldsymbol{u}_{pg} \right)\cdot\nabla\alpha_p \\
    &- \frac{4}{15}\frac{\mathcal{M}_s}{\rho_p}\rho_g\alpha_p\boldsymbol{u}_{pg}\left( \nabla\cdot\boldsymbol{u}_p \right) + \frac{4}{5}\frac{\mathcal{M}_s}{\rho_p}\rho_g\alpha_p\boldsymbol{u}_{pg}\cdot\left( \boldsymbol{\varepsilon}_p-\boldsymbol{\varepsilon}_g \right),
\end{split}
\end{equation}
\begin{equation}
    \boldsymbol{F}_{fs,s,p}^{PTE} = 2\frac{\mathcal{M}_s}{\rho_p}\Theta_p\boldsymbol{u}_{pg}\cdot\nabla\rho_g - 2\frac{\mathcal{M}_s}{\rho_p}\Theta_p\rho_g\left( \nabla\cdot\boldsymbol{u}_g \right).
\end{equation}

\section{Quadrature Method of Moments}
In principle, an infinite series of moment equations are needed to exactly resolve the mass distribution function.  This is, of course, impossible and, as a result, the moment equations must be truncated.  Additionally, the mass-moment integrals need to be closed to solve the governing equations.  The quadrature method of moments (QMOM) is used to both truncate the governing equations and approximate the mass-moment integrals.  The integration of a generic function $\psi(m)$ is performed using an $N$-node Gaussian quadrature
\begin{equation}
    \int m^s \psi(m)  f(m) dm \approx \sum_{\alpha=1}^{N} m_{\alpha}^s \psi(m_{\alpha}) f_{\alpha}, 
\end{equation}
where $\psi(m)$ is a generic function of particle mass $m$ and $f_{\alpha}$ and $m_{\alpha}$ are the weights and abscissae for the Gaussian quadrature, respectively. Similarly, a double integral needed for collision terms is approximated as
\begin{equation}
    \iint m^s \psi(m,M) f(m) f(M) dM dm \approx \sum_{\alpha=1}^N \sum_{\gamma=1}^N m_{\alpha}^s \psi(m_{\alpha},m_{\gamma}) f_{\alpha} f_{\gamma}.
\end{equation}

The truncated governing equations are
\begin{equation}
    \frac{\partial\mathcal{M}_{n/q}}{\partial t} + \nabla\cdot\ F\left( \mathcal{M}_{n/q} \right) = 0, \;\;\;\;
    n = 0,1,\cdots,N_{\text{mass}}-1
    \label{eqn:partCont}
\end{equation}
\begin{equation}
    \frac{\partial\boldsymbol{\mathcal{U}}_{s}}{\partial t} + \nabla\cdot\left(F\left(\boldsymbol{\mathcal{U}}_{s}\right) + \mathcal{P}_{s}\boldsymbol{I}\right) = -\frac{\mathcal{M}_{s}}{\rho_{p}}\nabla p_{g}-\boldsymbol{F}_{fs,p,s}^{P}
     + \boldsymbol{\mathcal{C}}^{P}_{p,s} - \boldsymbol{\mathcal{D}}_s^P + \boldsymbol{\mathcal{L}}_s^P, \;\;\;\; s=0,1,\cdots,N-1,
\end{equation}
\begin{equation}
    \frac{\partial\mathcal{T}_{s}}{\partial t} + \nabla\cdot F\left({\mathcal{T}}_{s}\right) = - \mathcal{P}_{s}^{kc} \nabla \cdot \frac{\boldsymbol{\mathcal{U}}_{s}}{\mathcal{M}_{s}} - F_{fs,p,s}^{PTE} 
    + \mathcal{C}^{PTE}_{p,s}  + \mathcal{D}_{s}^{PTE}  + \mathcal{F}_{s}^{PTE}, \;\;\;\; s=0,1,\cdots,N-1,
\end{equation}
\begin{equation}
    \frac{\partial\mathcal{E}_{s}}{\partial t} + \nabla\cdot F\left(\mathcal{E}_{s}\right) = - \mathcal{C}^{E}_{s} - \mathcal{F}^{PTE}_{s} - \mathcal{H}_s, \;\;\;\; s=0,1,\cdots,N-1,
\end{equation} 
where $N_{\text{mass}}$ is the number of mass moments needed to determine the quadrature weights and abscissae and $N$ is the number of quadrature nodes.

\subsection{Mass Moment Inversion}
The weights and abscissae need to be determined to perform the quadrature.  It is advantageous to select these such that quadrature can exactly reproduce the mass moments
\begin{equation}
    \mathcal{M}_{n/q} = \sum_{\alpha=1}^N m_{\alpha}^{n/q}f_\alpha, \ \ \ n = 0, 1, \cdots, N_{\text{mass}}-1.
    \label{eqn:momInvDef}
\end{equation}
This leads to a system of nonlinear equations for the unknown $f_{\alpha}$ and $m_{\alpha}$.    

It is difficult to solve these nonlinear equations directly with Newton's method.  Instead, it is more robust to use the fundamental theorem of Gaussian quadrature where the optimal abscissae are the roots of an $N^{th}$-order orthogonal polynomial, $P_{N}$.  These orthogonal polynomials  
 have the recursion relation
\begin{equation}
    P_{\alpha+1}\left(m^{1/q}\right) = \left(m^{1/q}-a_{\alpha}\right)P_{\alpha}\left(m^{1/q}\right) - b_{\alpha}P_{\alpha-1}\left(m^{1/q}\right),
    \label{eqn:recursion}
\end{equation}
where $P_{-1}(m^{1/q}) = 0$, and $P_{0}(m^{1/q}) = 1$.  The coefficients $a_{\alpha}$ and $b_{\alpha}$ are determined by using orthogonality
\begin{equation}
    a_{\alpha} = \frac{\int m^{1/q} f(m) P_{\alpha}(m^{1/q})P_{\alpha}(m^{1/q}) dm}{\int f(m) P_{\alpha}(m^{1/q})P_{\alpha}(m^{1/q}) dm}, \ \ \     b_{\alpha} = \frac{\int f(m) P_{\alpha}(m^{1/q})P_{\alpha}(m^{1/q}) dm}{\int f(m) P_{\alpha-1}(m^{1/q})P_{\alpha-1}(m^{1/q}) dm}.
\end{equation}

The roots can be recast as an eigenvalue problem
\begin{equation*}
    m^{1/q} \begin{bmatrix}
        P_0(m^{1/q}) \\ P_1(m^{1/q}) \\ P_2(m^{1/q}) \\ \vdots \\ P_{N-2}(m^{1/q}) \\ P_{N-1}(m^{1/q})
    \end{bmatrix}
    = 
    \begin{bmatrix}
        a_0 & 1                           \\
        b_1 & a_1 & 1                     \\
        & b_2 & a_2 & 1                   \\
        & & \ddots & \ddots & \ddots      \\
        & & & \ddots & a_{N-2} & 1                \\
        & & & &  b_{N-1} & a_{N-1}
    \end{bmatrix} \begin{bmatrix}
        P_0(m^{1/q}) \\ P_1(m^{1/q}) \\ P_2(m^{1/q}) \\ \vdots \\ P_{N-2}(m^{1/q}) \\ P_{N-1}(m^{1/q})
    \end{bmatrix}+ 
        \begin{bmatrix}
        0 \\ 0 \\ 0 \\ \vdots \\ 0 \\ P_{N}(m^{1/q})
    \end{bmatrix} .
\end{equation*}
The system is transformed to give a tridiagonal symmetric Jacobi matrix that preserves the eigenvalues
\begin{equation}
    \label{JacobiMatrix}
    \boldsymbol{J} = 
    \begin{bmatrix}
    a_{0} & \sqrt{b_{1}} &  &  &  & \\
    \sqrt{b_{1}} & a_{1} & \sqrt{b_{2}} & &  & \\
     & \sqrt{b_{2}} & a_{2} & \ddots &  & \\
     &  & \ddots & \ddots & \ddots &  \\
     &  &  & \ddots & a_{N-2} & \sqrt{b_{N-1}} \\
     &  &  &  & \sqrt{b_{N-1}} & a_{N-1}
    \end{bmatrix}.
\end{equation}
Then $m_{\alpha}^{1/q} = m_1^{1/q}, m_2^{1/q}, \cdots, m_{N}^{1/q}$, are eigenvalues of $\boldsymbol{J}$.  Finally, the quadrature abscissae are $m_{\alpha} = (m_{\alpha}^{1/q})^q$ and the weights are
\begin{equation}
    f_{\alpha} = \mathcal{M}_{0/q} (\varphi_{\alpha1})^2, \ \ \alpha = 1, 2, \cdots, N
\end{equation}
where $\varphi_{\alpha1}$ is the first component of the $\alpha^{th}$ eigenvector with unit magnitude \cite{Wilf1962}. An iterative eigensystem solver \cite{Golub1996} was found to be robust for this application.

The coefficients $a_i$ and $b_i$ in the Jacobi matrix are determined recursively with the modified Chebyshev algorithm \cite{Wheeler1974,gautschi2004orthogonal,Marchisio2013}.  Let's define
\begin{equation}
    Z_{\alpha,n} = \int P_{\alpha}(m^{1/q})m^{n/q}f(m)dm.
\end{equation}    
 Using the polynomial recursion in Eq.~\eqref{eqn:recursion} gives
\begin{equation}
     Z_{\alpha,n} = Z_{\alpha-1,n+1} - a_{\alpha-1}Z_{\alpha-1,n} - b_{\alpha-1}Z_{\alpha-2,n}, \;\;\;\; Z_{-1,n} = 0, \;\;\;\; Z_{0,n} = \mathcal{M}_{n/q}, \ \ \ Z_{\alpha > n,n} = 0.
 \end{equation}
Then $a_{\alpha}$ and $b_{\alpha}$ are determined using orthogonality of the polynomials \cite{Marchisio2013,gautschi2004orthogonal}
\begin{equation}
    a_{\alpha} = \frac{Z_{\alpha,\alpha+1}}{Z_{\alpha,\alpha}}-\frac{Z_{\alpha-1,\alpha}}{Z_{\alpha-1,\alpha-1}},\;\;\;\; b_{\alpha} = \frac{Z_{\alpha,\alpha}}{Z_{\alpha-1,\alpha-1}},\;\;\;\; \alpha=1,...,N-1
\end{equation}
with $a_0 = \frac{\mathcal{M}_{1/q}}{\mathcal{M}_{0/q}}$ and $b_0 = 0$.

\subsubsection{Beta-GQMOM Algorithm} \label{Beta-GQMOM}
Using the Chebyshev algorithm discussed above to fill the Jacobi matrix requires $N_{\text{mass}}=2N$ transported mass moments necessary to determine the $N$ abscissae and weights. 
The Generalized Quadrature Method of Moments (GQMOM) reduces $N_{\text{mass}}$ to $2N-1$ by fitting a distribution function to an assumed functional form \cite{Fox2023} and utilizing that information to compute $a_{N-1}$. Here, we only describe how to use GQMOM.  More information on GQMOM and it derivation can be found in \cite{Fox2023}.  

The beta-GQMOM algorithm \cite{Fox2023} is used to reduce the number of transported mass moments from $N_{\text{mass}} = 2N-1$ in Eqn.~\eqref{eqn:partCont}.  
The beta distribution is fit to the transported moments
\begin{equation}
    f( m) \propto \left(\frac{m}{m_\textrm{max}}\right)^{\beta}\left[ 1-\left(\frac{m}{m_\textrm{max}}\right) \right]^{\alpha},
    \label{eqn:beta}
\end{equation}
where $\alpha$ and $\beta$ are shape parameters. The maximum particle mass, $m_{\textrm{max}}$, is an input to the beta-GQMOM algorithm.  The maximum particle mass can often be determined from the maximum particle diameter in size distribution during initialization. 
The first step in using beta-GQMOM is to use the modified Chebyshev algorithm to compute $a_0, a_1, a_2, \cdots,  a_{N-2}$, and $b_0, b_1, b_2, \cdots, b_{N-1}$ with the $N_{\text{mass}}$ transported moments.  Then compute
\begin{equation}
    \zeta_{0} = 0, \;\;\;\; \zeta_{1} = a_{0}, \;\;\;\; \zeta_{2i+1} = a_{i} - \zeta_{2i}, \;\;\;\; \zeta_{2i} = \frac{b_{i}}{\zeta_{2i-1}} \text{ for } \ \ \  \zeta_0, \zeta_1, \zeta_2, \cdots, \zeta_{2N-2}.
\end{equation}
Next compute the canonical moments
\begin{equation}
    p_{0} = 0, \;\;\;\; p_{i} = \frac{\zeta_{i}}{1-p_{i-1}}, \text{ for } \ \  p_0, p_1, p_2, \cdots, p_{2N-2}.
\end{equation}
The shape parameters for the beta distribution are
\begin{equation}
    \alpha = \frac{1-p_{1}-2p_{2}+p_{1}p_{2}}{p_{2}}, \;\;\;\; \beta = \frac{p_{1}-p_{2}-p_{1}p_{2}}{p_{2}}.
\end{equation}
Then the additional canonical moments can be computed using shape parameters
\begin{equation}
    \begin{split}
    &p_{2i-1} = 
    \begin{cases}
        p_{2N_{G}-1}\frac{p_{2i-1}^{J}}{p_{2N_{G}-1}^{J}}, & \textrm{if } p_{2N_{G}-1} \leq p_{2N_{G}-1}^{J} \textrm{ or } p_{2N_{G}-1}^{J} \geq p_{2i-1}^{J} \\
        \frac{p_{2N_{G}-1}\left( 1-p_{2i-1}^{J} \right)+p_{2i-1}^{J}-p_{2N_{G}-1}^{J}}{1-p_{2N_{G}-1}^{J}}, & \textrm{otherwise}
    \end{cases} \\
    &p_{2i} = 
    \begin{cases}
        p_{2N_{G}}\frac{p_{2i}^{J}}{p_{2N_{G}}^{J}}, & \textrm{if } p_{2N_{G}} \leq p_{2N_{G}}^{J} \textrm{ or } p_{2N_{G}}^{J} \geq p_{2i}^{J} \\
        \frac{p_{2N_{G}}\left( 1-p_{2i}^{J} \right)+p_{2i}^{J}-p_{2N_{G}}^{J}}{1-p_{2N_{G}}^{J}}, & \textrm{otherwise}
    \end{cases}
    \end{split}
\end{equation}
where $N_{G} = N-1$ and 
\begin{equation}
    p_{2i-1}^{J} = \frac{\beta + i}{2i+\alpha+\beta}, \;\;\;\; p_{2i}^{J} = \frac{i}{2i+1+\alpha+\beta}.
\end{equation}
Then we may recover the continued fractions
\begin{equation}
    \zeta_{i} = p_{i}\left(1-p_{i-1}\right),
\end{equation}
and finally find the remaining recurrence coefficients required to fill the Jacobi matrix in Eqn.~\eqref{JacobiMatrix}
\begin{equation}
    a_{i} = \zeta_{2i} + \zeta_{2i+1}, \;\;\;\; b_{i} = \zeta_{2i-1}\zeta_{2i}.
\end{equation}

If $N$-node quadrature is used to approximate all integrals with $N_{mass} = 2N-1$ transported mass moments (the minimal number), then the beta-GQMOM algorithm is only needed to compute $a_{N-1}$.  However, beta-GQMOM can be used to compute extra abscissae and weights to improve accuracy of the quadrature approximations witout increasing the number of transported mass moments.  This makes GQMOM useful in resolving nonlinear source terms such as nucleation \cite{Fox2023}. In such a scenario, the beta-GQMOM algorithm would be used to compute $a_{N-1}, a_{N}, a_{N+1}, \cdots$ and $b_{N}, b_{N+1}, \cdots$. In this work we only use GQMOM to compute $a_{N-1}$. Exploring use of GQMOM to increase the number of quadrature for source terms with fixed $N_{\text{mass}}$ is a topic of future work. 

\subsection{Binning}
A binning-like approach \cite{binning} can be used if the particle size distribution is truly bidisperse, tridisperse, etc.\  or if it is desired to fix the particle sizes.  The abscissae are assumed and fixed, which turns Eqn.~\eqref{eqn:momInvDef} into solving a linear system of equations for $f_{\alpha}$.  Binning substantially reduces accuracy of the distribution function because the abscissae are not in their optimal locations.  However, a tradeoff is reduced computational expanse due to the reduction of transported mass moments to $N_{\text{mass}} = N$ and transforming moment inversion to the solution of a system of linear equations.  There is no advantage to using diameter ($q=3$) or area ($q=2$) moments for binning.  Instead, mass moments ($q=1$) are directly used.

\subsubsection{Computing nodal primitive variables} \label{CQMOM}

Conditional QMOM (CQMOM) is used to find the conditional  variables (analogous to primitive variables in gas dynamics) at each quadrature node \cite{Yuan2011}. 
The conditional variables, $\boldsymbol{u}_{p,\alpha}$, $\Theta_{p,\alpha}$ $e_{p,\alpha}$ are defined such that quadrature reproduces the mass integrals in Eqn.~\eqref{eqn:Mom_defs} exactly
\begin{align}
        \frac{\boldsymbol{\mathcal{U}}_s}{m_r^s} &
        = \frac{1}{m_r^s}\sum_{\alpha=1}^N m_\alpha^{s} \boldsymbol{u}_{p,\alpha} f_{\alpha},  \ \ \ s = 0, 1, \cdots, N-1 \\
        \frac{\mathcal{T}_s}{m_r^s} & 
        = \frac{1}{m_r^s}\frac{3}{2}\sum_{\alpha=1}^{N}  m_{\alpha}^s \Theta_{p,\alpha} f_{\alpha},  \ \ \ s = 0, 1, \cdots, N-1 \\
        \frac{\mathcal{E}_s}{m_r^s} &
        = \frac{1}{m_r^s}\sum_{\alpha=1}^N  m_\alpha^s e_{p,\alpha}f_\alpha  ,  \ \ \ s = 0, 1, \cdots, N-1.
\end{align}
A reference mass, $m_r$ is used to scale each of the equations from becoming ill-conditioned \cite{Fox2024}  
\begin{equation}
    m_r = \sqrt{\left(m_1\right)^2 + \left( m_2\right)^2 + \left( m_3\right)^2 + \cdots + \left( m_N\right)^2 }.
\end{equation}
The granular temperatures are checked to ensure $\Theta_{p,\alpha}\geq 0$ to ensure the nodal values of $\Theta_{p,\alpha}$ are realizable.  The simulation is halted if any of the $\Theta_{p,\alpha} < 0$.

\subsection{Selecting the number of mass moments}
The required number of transported mass moments that needs to be transported, $N_{mass}$, depends on the moment-inversion algorithm employed and what type mass ($q=1$), area ($q=2$), or size ($q=3$) moments are used.  The GQMOM and  Binning approaches require at least $2N-1$ and $N$ mass moments, respectively.  The CQMOM algorithm for computing nodal variables requires the integer mass moment for every momentum, psuedo-thermal energy, or internal energy moment.  Thus, CQMOM requires at least $1+q(N-1)$ mass moments to ensure mass moment $n/q=N-1$ is directly transported. The final number of required transported mass moments is the maximum of these two constraints.  

Table~\ref{tab:Nmass} lists the required number of mass moments. Using size moments requires an additional $N-1$ moments for GQMOM and $N-2$ moments for QMOM. Thus,  size moments are more computationally intensive.  Our numerical experiments have found that using area moments to be more robust than mass moments for a wider variety of flows due to the large exponents which can create ill-conditioned Jacobi matrices during moment inversion.
\begin{table}
  \centering
  \caption{Required number of transported mass moments $N_{\text{mass}}$ for GQMOM and Binning.}
  \label{tab:Nmass}
  \begin{tabular}{ c c c c c}
    \hline
    $N$ & $q=1$ & $q=2$ & $q=3$ & Binning \\
    \hline
    3 & 5  & 5 & 7 & 3 \\
    4 & 7  & 7 & 10 & 4 \\
    5 & 9  & 9 & 13 & 5 \\
    6 & 11 & 11 & 16 & 6 \\
    \hline
  \end{tabular}
\end{table}

\section{Numerical Methods}

For the discussion of the numerical methods, we may recast the transport equations for both phases into one coupled system of partial differential equations (PDEs)
\begin{equation}
    \label{cons_system}
    \frac{\partial \boldsymbol{U}}{\partial t} + \nabla \cdot \boldsymbol{F} = \boldsymbol{h}\left( \boldsymbol{a},\nabla\boldsymbol{b},\nabla\cdot\boldsymbol{c} \right) + \boldsymbol{S},
\end{equation}
where $\boldsymbol{U}$ is the conserved variable vector, $\boldsymbol{F}$ is the flux vector.  The non-conservative spatial terms are represented by $\boldsymbol{h}$ with functional dependence on non-constant coefficients $\boldsymbol{a}$, gradients $\nabla\boldsymbol{b}$, and divergences $ \nabla\cdot\boldsymbol{c}$.  Inhomgeneous source terms due to  drag, collisions, convective heat transfer, etc.\ are represented by $\boldsymbol{S}$.

Second-order Strang operator splitting \cite{Strang1968} is used to integrate Eqn.~\eqref{cons_system}
\begin{equation}
    \label{Strang}
    \boldsymbol{U}^{t+2\Delta t} = \rsfscr{H}_{xyz}^{\Delta t}\left( \rsfscr{I}^{2\Delta t}\left(   \rsfscr{H}_{xyz}^{\Delta t}\left( \boldsymbol{U}^{t} \right) \right) \right),
\end{equation}
where $\rsfscr{H}_{xyz}^{\Delta t}$ indicates integration of the directionally unsplit hyperbolic terms for a time step of $\Delta t$, $\rsfscr{I}^{2\Delta t}$ indicates integration of the inhomogeneous source terms for a time step of $2\Delta t$.

The overall time-step size is determined following the Courant–Friedrichs–Lewy (CFL) condition \cite{Courant1928}
\begin{equation}
    \Delta t = CFL\left(\frac{S_{x,\textrm{max}}}{\Delta x} + \frac{S_{y,\textrm{max}}}{\Delta y} + \frac{S_{z,\textrm{max}}}{\Delta z}\right)^{-1},
\end{equation}
where $CFL$ is the CFL number.  The maximum wave speed in the $x$-direction is
\begin{equation}
S_{x,\textrm{max}} = 
        \max\left(|u_g| + c_g, |u_{p,1}| + c_{p,1}, |u_{p,2}| + c_{p,2}, \cdots, |u_{p,N}| + c_{p,N}\right), 
\end{equation}
where $c_{g}$ is the gas-phase speed of sound and $c_{p,\alpha}$ is the compaction wave speed (analogous to a sound speed for the particles) for quadrature node $\alpha$. The expressions for wave speeds in the $y$- and $z$-directions are similar. The gas-phase sound speed is
\begin{equation}
    c_g^2 = \gamma \frac{p_g}{\rho_g},
\end{equation}
where $\gamma$ is the ratio of specific heats.

\subsection{Compaction Wave Speed, $c_{p,\alpha}$}
A closed analytical relationship for compaction wave speed for the particles is impossible.  As a result, $c_{p,\alpha}$ must be approximated.  We make the assumption that the compaction wave speed may be computed at each quadrature node by treating it as a monodisperse granular mixture.
This assumption allows the use of closed relationships to compute the compaction wave speed at each quadrature node \cite{Houim2016}
\begin{equation}
    c_{p,\alpha}^{2} = \frac{1}{\rho_{p}} \left.\frac{\partial p_{p,\alpha}}{\partial \alpha_{p}}\right|_{\Theta_{p,\alpha}} + \frac{2}{3}\frac{\Theta_{p,\alpha}}{\rho_{p}^{2}\alpha_{p}^{2}}\left( \left.\frac{\partial p_{p,\alpha}}{\partial \Theta_{p,\alpha}}\right|_{\alpha_{p}} \right)^{2},
\end{equation}
where $p_{p,\alpha}$ is the particle pressure associated with the quadrature node $\alpha$
\begin{equation}
    \label{particlePress}
    p_{p,\alpha} = p_{p,k,\alpha} + p_{p,c,\alpha} + p_{p,f,\alpha} = f_{\alpha}m_{p,\alpha}\left( \theta_{k,\alpha}+\theta_{c,\alpha}+\theta_{f,\alpha} \right).
\end{equation}
It can be shown what this general expression for the compaction wave speed can be written as \cite{Houim2016}
\begin{equation}
    c_{p,\alpha}^{2} = c_{p,kc,\alpha}^{2} + c_{p,f,\alpha}^{2},
\end{equation}
where $c_{p,kc,\alpha}$ is the compaction wavespeed at quadrature node $\alpha$ due to the kinetic-collisional pressure ($p_{p,kc,\alpha}=p_{p,k,\alpha}+p_{p,c,\alpha}$) and $c_{p,f,\alpha}$ is the compaction wavespeed at quadrature node $\alpha$ due to the frictional pressure ($p_{p,f,\alpha}$). These wavespeeds are
\begin{equation}
    c_{p,kc,\alpha}^{2} = \frac{1}{\rho_{p}} \left.\frac{\partial p_{p,kc,\alpha}}{\partial \alpha_{p}}\right|_{\Theta_{p,\alpha}} + \frac{2}{3}\frac{\Theta_{p,\alpha}}{\rho_{p}^{2}\alpha_{p}^{2}}\left( \left.\frac{\partial p_{p,kc,\alpha}}{\partial \Theta_{p,\alpha}}\right|_{\alpha_{p}} \right)^{2}, \;\;\;\; c_{p,f,\alpha}^{2} =
    \begin{cases}
        0, & \textrm{if } \alpha_{p} < \alpha_{p,\textrm{crit}} \\
        \frac{1}{\rho_{p}}\frac{\partial p_{p,f,\alpha}}{\partial \alpha_{p}}, & \textrm{if } \alpha_{p} \geq \alpha_{p,\textrm{crit}}
    \end{cases}
\end{equation}
where the first derivative in $c_{p,kc,\alpha}$ is
\begin{equation}
    \left.\frac{\partial p_{p,kc,\alpha}}{\partial \alpha_{p}}\right|_{\Theta_{p,\alpha}} = \rho_{p}\Theta_{p,\alpha} + \alpha_{p}\rho_{p}\sum_{\gamma=1}^{N}2\left(1+e_{\alpha\gamma}\right)\beta_{\gamma}\left( 2g_{\alpha\gamma}+\alpha_{p}\frac{\partial g_{\alpha\gamma}}{\partial \alpha_{p}} \right)\chi_{\alpha\gamma}^{3}\mu_{\alpha\gamma}y_{\alpha\gamma}E_{\alpha\gamma},
\end{equation}
where the derivative of the polydisperse and monodisperse radial distribution functions are
\begin{equation}
    \frac{\partial g_{\alpha\gamma}}{\partial \alpha_{p}} = -\frac{1}{\left( 1-\alpha_{p} \right)^{2}} + \left( \frac{\partial g_{0}}{\partial \alpha_{p}} + \frac{1}{\left( 1-\alpha_{p} \right)^{2}} \right)\frac{\langle d_{p}^{2} \rangle}{\langle d_{p}^{3} \rangle}\frac{d_{p,\alpha}}{\chi_{\alpha\gamma}}, \;\;\;\; \frac{\partial g_{0}}{\partial \alpha_{p}} = \frac{g_{0}^{2}}{3\alpha_{p\textrm{max}}}\left( \frac{\alpha_{p,\textrm{max}}}{\alpha_{p}} \right)^{2/3}.
\end{equation}
The second derivative in $c_{p,\textrm{kinetic-collisional},\alpha}$ is
\begin{equation}
    \left.\frac{\partial p_{p,kc,\alpha}}{\partial \Theta_{p,\alpha}}\right|_{\alpha_{p}} = \alpha_{p}\rho_{p} + \alpha_{p}\rho_{p}\sum_{\gamma=1}^{N}2\left(1+e_{\alpha\gamma}\right)\beta_{\gamma}\alpha_{p}g_{\alpha\gamma}\chi_{\alpha\gamma}^{3}\mu_{\alpha\gamma}y_{\alpha\gamma}.
\end{equation}
The derivative of the frictional pressure is
\begin{equation}
    \label{dpfricpda}
    \frac{\partial p_{p,f,\alpha}}{\partial \alpha_{p}} = 
    Fr \, \frac{\left( \alpha_{p}-\alpha_{p,\textrm{crit}} \right)^{r_1}}{\left( \alpha_{p,\textrm{max}}-\alpha_{p} \right)^{r_2}} + r_1 Fr \, \alpha_p \frac{\left( \alpha_{p}-\alpha_{p,\textrm{crit}} \right)^{r_1-1}}{\left( \alpha_{p,\textrm{max}}-\alpha_{p} \right)^{r_2}} + r_2Fr \, \alpha_p\frac{\left( \alpha_{p}-\alpha_{p,\textrm{crit}} \right)^{r_1}}{\left( \alpha_{p,\textrm{max}}-\alpha_{p} \right)^{r_2+1}}.
\end{equation}

\subsection{Hyperbolic Processes, $\rsfscr{H}_{xyz}^{\Delta t}$}
The gas-phase hyperbolic terms are 
\begin{equation}
    \frac{\partial \alpha_{g}\rho_{g}}{\partial t} + \nabla \cdot \left( \alpha_{g}\rho_{g}\boldsymbol{u}_{g} \right) = 0, 
    \label{eqn:GasMassHypSplit}
\end{equation}
\begin{equation}
\frac{\partial \alpha_{g}\rho_{g}Y_{g,i}}{\partial t} + \nabla \cdot \left( \alpha_{g}\rho_{g}Y_{g,i}\boldsymbol{u}_{g} \right) = 0, \;\;\;\; i=1,...,N_{g}
\end{equation}
\begin{equation}
    \frac{\partial \alpha_{g}\rho_{g}\boldsymbol{u}_{g}}{\partial t} + \nabla \cdot \left( \alpha_{g}\rho_{g}\boldsymbol{u}_{g}\boldsymbol{u}_{g} \right) = -\alpha_{g}\nabla p_{g} + \boldsymbol{F}_{fs,g}^{P} - \boldsymbol{\mathcal{L}}_{1}^{P},
\end{equation}
\begin{equation}
    \frac{\partial \alpha_{g}\rho_{g}E_{g}}{\partial t} + \nabla \cdot \left( \alpha_{g}\boldsymbol{u}_{g}\left( \rho_{g}E_{g} + p_{g} \right) \right) = -p_{g}\nabla \cdot \left( \alpha_{p}\boldsymbol{u}_{p} \right) + F_{fs,g}^{E}  - \mathcal{L}_{1}^{E}. 
\end{equation}
The granular-phase hyperbolic terms are
\begin{equation}
    \frac{\partial\mathcal{M}_{s/q}}{\partial t} + \nabla\cdot F\left(\mathcal{M}_{s/q}\right) = 0, \;\;\;\; s = 0,1,\cdots, N_{\text{mass}}-1
    \label{eqn:partCont_Hyp}
\end{equation}
\begin{equation}
    \frac{\partial\boldsymbol{\mathcal{U}}_{s}}{\partial t} + \nabla\cdot\left(F\left(\boldsymbol{\mathcal{U}}_{s}\right) + \mathcal{P}_{s}\boldsymbol{I}\right) = -\frac{\mathcal{M}_{s}}{\rho_{p}}\nabla p_{g}-\boldsymbol{F}_{fs,p,s}^{P} + \boldsymbol{\mathcal{L}}_s^P, \;\;\;\; s=0,1,\cdots,N-1
\end{equation}
\begin{equation}
    \frac{\partial\mathcal{T}_{s}}{\partial t} + \nabla\cdot F\left({\mathcal{T}}_{s}\right) = - \mathcal{P}_{s}^{kc} \nabla \cdot \frac{\boldsymbol{\mathcal{U}}_{s}}{\mathcal{M}_{s}} - F_{fs,p,s}^{PTE}, \;\;\;\; s=0,1,\cdots,N-1
\end{equation}
\begin{equation}
    \frac{\partial\mathcal{E}_{s}}{\partial t} + \nabla\cdot F\left(\mathcal{E}_{s}\right) =  0, \;\;\;\; s=0,1,\cdots,N-1.
    \label{eqn:partEn_Hyp}
\end{equation}

A method-of-lines approach is used to time-march the hyperbolic terms using three-stage, third-order, Strong-Stability Preserving Runge-Kutta method \cite{Spiteri2002}.  The spatial derivatives and non-conservative terms are discretized using an extension of a high-order Godunov-based scheme developed for monodisperse multiphase flows \cite{Houim2016} discussed below. A generic $x$-derivative for computational cell $ijk$ is discretized as
\begin{equation}
    a \frac{\partial b}{\partial x} \approx a_{i,j,k}\frac{b_{i+1/2,j,k} - b_{i-1/2,j,k}}{\Delta x},
    \label{eqn:nonConsDer}
\end{equation}
where $a$ is a coefficient and $b$ is any term inside of a spatial derivative operator.  The coefficient terms are evaluated directly from the cell-centered values.  The terms on the cell edges $b_{i+1/2}$ and $b_{i-1/2}$ are evaluated from the results of separated gaseous and granular Riemann problems discussed below.

\subsection{Treatment of Very Small Particle Volume Fractions}
\label{sec:particleRemoval}
The moment-inversion algorithms become ill-conditioned when the volume fraction is very small.  At the end of every RK stage the particle volume fraction, $\alpha_p$, and number density $\mathcal{M}_0$ in every computational cell are checked for vanishingly small values.  If $\alpha_p < \alpha_{p,\min}$ or  $\mathcal{M}_{0} < \mathcal{M}_{0,\min}$ then the particles are removed from the cell.  In this work, $\alpha_{p,\min}$ and $\mathcal{M}_{0,\min}$ are $10^{-11}$ and $10^{5}$, respectively, unless stated otherwise.  The void left by the removed particles is filled by gas at the same pressure,  temperature, density, and velocity as the gas in the computatinal cell
\begin{equation}
    \boldsymbol{U}_{g,i}^{\text{new}}=\frac{1}{ 1-\alpha_{p,i}}\boldsymbol{U}_{g,i} ^{\text{old}} , \ \ \ \boldsymbol{U}_{p,i} = 0,
\end{equation}
where $\boldsymbol{U}_{g,i}$ and $\boldsymbol{U}_{i,i}$ are the vectors of gas-phase conserved variables, $(\alpha_g \rho_g, \alpha_g \rho_g Y_{g,i}, \alpha_g \rho_g \boldsymbol{u}_g , \alpha_g\rho_g E_g)^T$, and granular-phase conserved variables, $(\mathcal{M}_{s/q}, \mathcal{U}_s, \mathcal{T}_s, \mathcal{E}_s)^T$, in cell $i$, respectively. 

Computational cells are also checked if the packing limit is exceeded, $\alpha_p > \varepsilon\alpha_{p,\max}$, where $\varepsilon=99.99\%$ to avoid singular or complex values in Eqns.~\eqref{theta_f} and~\eqref{g0}.  If this condition is met, particles are removed to be slightly below the packing limit
\begin{equation}
    \boldsymbol{U}_{p,i}^{\text{new}}=\varepsilon \frac{ \alpha_{p,\max}}{\alpha_{p,i}}\boldsymbol{U}_{p,i}^{\text{old}}, \ \ \ \ \ \  \boldsymbol{U}_{g,i}^{\text{new}}=\left( 1-\varepsilon \frac{\alpha_{p,\max}}{1-\alpha_{p,i}} \right)\boldsymbol{U}_{g,i}^{\text{old}}.
\end{equation}

This procedure is similar to procedures used for monodisperse granular-gas flow \cite{Houim2016}.  It significantly increases robustness of the code.  Strategies such as forcing the particle volume fraction to a small value were tried, but were found to lack robustness and generality. Relatively little conservation error is introduced by this process because the volume fraction threshold is small and the particle removal is performed in a small number of cells. The particles are also removed if the mass moments become unrealizable due to numerical error during transport. Moment correction algorithms \cite{McGraw2012} will be explored in future work.

\subsection{Gas-Phase Riemann Problem}
The gas-phase Riemann problem is solved using the HLLC approximate Riemann solver \cite{Toro1994} with the Davis wave speed estimate \cite{Davis1988}.  HLLC has has been modified to return the state ($\boldsymbol{u}_{g,i+1/2}, p_{g,i+1/2}, \rho_{g,i+1/2}, E_{g,i+1/2}, Y_{g,i+1/2}$) rather than the flux terms so that the non-conservative terms can be assembled \cite{Houim2016}.  The edges of the gas-phase primitive variables ($p_g$, $T_g$, $Y_{i,g}$, and $\boldsymbol{u}_g$) are reconstructed using fifth-order MUSCL  \cite{Kim2005}. A low-Mach correction of the reconstructed velocity is used to reduce dissipation \cite{Thornber2008}. Details of the gas-phase flux can be found in \cite{Houim2016}. 

\subsection{Granular-Phase Riemann Problem}
A direct solution to the polydisperse granular Riemann problem for the moment equations cannot be computed analytically.  Instead, the polydisperse granular Riemann problem must be approximated.   Riemann solvers exist for monodisperse granular systems that includes number density ($n=0$), mean mass, mean momentum, mean PTE, and mean internal energy transport \cite{Houim2016}. Substituting the Gaussian quadrature into the granular hyperbolic terms for the number density ($n=0$) and $n=q/q=1$ for the mean quantities gives
\begin{equation}
    \sum_{\alpha=1}^{N}\left[\frac{\partial f_{\alpha}}{\partial t}+\nabla \cdot \boldsymbol{u}_{p,\alpha}f_{\alpha}\right]=0,
\end{equation}
\begin{equation}
    \sum_{\alpha=1}^{N}\left[\frac{\partial m_{\alpha}f_{\alpha}}{\partial t}+\nabla \cdot m_{\alpha} f_{\alpha}\boldsymbol{u}_{p,\alpha}\right]=0,
\end{equation}

\begin{equation}
    \sum_{\alpha=1}^N \left[\frac{\partial m_{\alpha}f_{\alpha}\boldsymbol{u}_{p,\alpha}}{\partial t} + \nabla \cdot m_{\alpha}f_{\alpha}\boldsymbol{u}_{p,\alpha}\boldsymbol{u}_{p,\alpha} + \nabla p_{p,\alpha} + \frac{m_{\alpha}f_{\alpha}}{\rho_p}\nabla p_g +\boldsymbol{F}_{fs,p,\alpha}^P - \boldsymbol{\mathcal{L}}_{\alpha}^P\right] = 0,
\end{equation}
\begin{equation}
    \sum_{\alpha=1}^N \left[\frac{3}{2}
    \left(\frac{\partial m_{\alpha}f_{\alpha}\Theta_{p,\alpha}}{\partial t} + \nabla \cdot m_{\alpha}f_{\alpha}\Theta_{p,\alpha}\boldsymbol{u}_{p,\alpha}\right) +p_{p,kc,\alpha} \nabla \cdot \boldsymbol{u}_{p,\alpha} + F_{fs,p,\alpha}^{PTE}\right] = 0,
\end{equation}
\begin{equation}
    \sum_{\alpha=1}^N \left[
    \frac{\partial m_{\alpha}f_{\alpha}e_{p,\alpha}}{\partial t} + \nabla \cdot m_{\alpha}f_{\alpha}e_{p,\alpha}\boldsymbol{u}_{p,\alpha} \right] = 0.
\end{equation}
We set each term in the sum to zero to force the entire sum to be zero.  For each quadrature node, this gives
\begin{equation}
    \frac{\partial f_{\alpha}}{\partial t}+\nabla \cdot \boldsymbol{u}_{p,\alpha}f_{\alpha}=0,
\end{equation}
\begin{equation}
    \frac{\partial m_{\alpha}f_{\alpha}}{\partial t}+\nabla \cdot m_{\alpha} f_{\alpha}\boldsymbol{u}_{p,\alpha}=0,
\end{equation}
\begin{equation}
    \frac{\partial m_{\alpha}f_{\alpha}\boldsymbol{u}_{p,\alpha}}{\partial t} + \nabla \cdot m_{\alpha}f_{\alpha}\boldsymbol{u}_{p,\alpha}\boldsymbol{u}_{p,\alpha} + \nabla p_{p,\alpha} = - \frac{m_{\alpha}f_{\alpha}}{\rho_p}\nabla p_g- \boldsymbol{F}_{fs,p,\alpha}^P + \boldsymbol{\mathcal{L}}_{\alpha}^P,
\end{equation}
\begin{equation}
     \frac{3}{2}
    \left(\frac{\partial m_{\alpha}f_{\alpha}\Theta_{p,\alpha}}{\partial t} + \nabla \cdot m_{\alpha}f_{\alpha}\Theta_{p,\alpha}\boldsymbol{u}_{p,\alpha}\right) = -p_{p,kc,\alpha} \nabla \cdot \boldsymbol{u}_{p,\alpha} - F_{fs,p,\alpha}^{PTE}, 
\end{equation}
\begin{equation}
    \frac{\partial m_{\alpha}f_{\alpha}e_{p,\alpha}}{\partial t} + \nabla \cdot m_{\alpha}f_{\alpha}e_{p,\alpha}\boldsymbol{u}_{p,\alpha}=0.
\end{equation}
Each set of equations for quadrature node $\alpha$ closely matches the monodisperse granular equations \cite{Houim2016}, which, in turn, allows the use of a modified monodisperse AUSM granular Riemann solver at each quadrature node. 

The strategy for computing the fluxes defined in Eqn.~\eqref{eqn:momentFluxes} is to solve a granular Riemann problem for each quadrature node and use their results to assemble the fluxes and source terms in Eqns.~\eqref{eqn:partCont_Hyp}-\eqref{eqn:partEn_Hyp}.
For example, the mass-velocity moment fluxes, defined by Eqn.~\eqref{eqn:momentFluxes}, at cell edge $i+1/2$ are
\begin{equation}
     F_{p,s,i+1/2}^P = \sum_{\alpha=1}^N 
     m_{\alpha,i+1/2}^s
     \boldsymbol{u}_{p,\alpha,i+1/2}\boldsymbol{u}_{p,\alpha,i+1/2}f_{\alpha,i+1/2}, \ \ \ \ s= 0, 1, 2, \cdots, N-1
\end{equation}
where $m_{\alpha,i+1/2}^s$,
      $f_{\alpha,i+1/2}$, and
     $\boldsymbol{u}_{p,\alpha,i+1/2}$, 
     are the abscissa, weight, and conditional velocity at cell edge $i+1/2$ that result from solving the granular Riemann problem for node $\alpha$.
The edge-value of the total particle volume fraction needed for the gas-phase fluxes can be found from the condition $\alpha_{p}\rho_{p}=\mathcal{M}_{1}$.  Thus
\begin{equation}
    \alpha_{p,i+1/2} = \sum_{\alpha = 1}^{N} \frac{f_{\alpha,i+1/2}m_{\alpha,i+1/2}}{\rho_{p}}.
\end{equation}
The average granular variables needed on the cell edges (e.g. $T_{p,i+1/2}$) are computed from the nodal values
\begin{equation}
    \phi_{p,i+1/2} = \frac{1}{\mathcal{M}_{1}}\sum_{\alpha=1}^{N} f_{\alpha,i+1/2}m_{\alpha,i+1/2}\phi_{p,\alpha,i+1/2}.
\end{equation}

\subsubsection{Nodal  Granular Riemann Solver}
\label{sec:granRiemann}
A modification of the AUSM flux scheme \cite{Liou1993,Liou1996,Liou2006} is used as the Riemann solver at each quadrature node due to its ability to handle scenarios with zero particle mass and that it can directly return the edge state needed to assemble the hyperbolic terms in the moment equations \cite{Houim2016}. 

The nodal mass flux rate is
\begin{equation}
    \Dot{m}_{\alpha,i+1/2} = \mathscr{F}_{\alpha} + c_{\alpha,1/2}M_{\alpha,1/2}
    \begin{cases}
        f_{\alpha}^{L}m_{p,\alpha}^{L}, & \textrm{if } M_{\alpha,1/2} > 0 \\
        f_{\alpha}^{R}m_{p,\alpha}^{R}, & \textrm{if } M_{\alpha,1/2} \leq 0
    \end{cases}
\end{equation}
where superscripts $L$ and $R$ refer to the left- and right-interpolated edge states (see Sec.~\ref{interp}.) and $c_{\alpha, 1/2}$ and $M_{\alpha,1/2}$ are the compaction wave speed and Mach number computed at the edge, defined below. $\mathscr{F}_{\alpha}$ is a dissipation term to help reduce oscillations in dense particle mixtures that approach the packing limit \cite{Houim2016} 
\begin{equation}
    \mathscr{F}_{\alpha} = \frac{\left( c_{\alpha,1/2}-\epsilon \right)\left( 1+\left| M_{\alpha,1/2} \right|\left( 1-\mathscr{G}/2 \right) \right)}{2}\frac{\max\left( \alpha_{p}^{L},\alpha_{p}^{R} \right)}{
    \alpha_{p,\textrm{max}}
    }\left( f_{\alpha}^{L}m_{p,\alpha}^{L} - f_{\alpha}^{R}m_{p,\alpha}^{R} \right),
\end{equation}
where $\epsilon=10^{-10}$ is a parameter to prevent division by zero, $\mathscr{G}$ is a dissipation variable to smoothly reduce the scheme to first-order as $\alpha_{p}$ approaches $\alpha_{p,\textrm{max}}$
\begin{equation}
    \mathscr{G} = \max\left( 2\left( 1-\mathscr{D}\zeta^{2} \right),0 \right), \;\;\;\; \zeta = 
    \begin{cases}
        \frac{\alpha_{p}^{M}-\alpha_{p,\textrm{crit}}}{\alpha_{p,\textrm{max}}-\alpha_{p,\textrm{crit}}}, & \textrm{if } \alpha_{p}^{M} > \alpha_{p,\textrm{crit}} \\
        0, & \textrm{if } \alpha_{p}^{M} < \alpha_{p,\textrm{crit}}
    \end{cases}
    \label{eqn:G-AUSM}
\end{equation}
where $\alpha_{p}^{M}$ is the maximum volume fraction used in the interpolation stencil and $\mathscr{D}$ is a dissipation-controlling parameter that controls how quickly $\mathscr{G}$ reduces to zero. $\mathscr{D}=1$ was found to work well for the numerical experiments discussed later.

The edge compaction wave speed is 
\begin{equation}
    \label{cM1/2}
    c_{\alpha,1/2} = \sqrt{\frac{f_{\alpha}^{L}m_{p,\alpha}^{L}\left( c_{p,\alpha}^{L} \right)^{2} + f_{\alpha}^{R}m_{p,\alpha}^{R}\left( c_{p,\alpha}^{R} \right)^{2}}{f_{\alpha}^{L}m_{p,\alpha}^{L} + f_{\alpha}^{R}m_{p,\alpha}^{R}}} + \epsilon.
\end{equation}
$c_{\alpha,1/2}$ and $M_{\alpha,1/2}$ are the compaction waves peed and Mach number of node $\alpha$. The edge-averaged Mach number is
\begin{equation}
M_{\alpha,1/2} = \mathscr{M}_{4}^{+}\left( M_{\alpha}^{L} \right) + \mathscr{M}_{4}^{-}\left( M_{\alpha}^{R} \right) -2\frac{K_{p}}{n_{a}}\max\left( 1-\sigma\overline{M}_{\alpha}^{2},0 \right)\frac{p_{p,\alpha}^{R}-p_{p,\alpha}^{L}}{\left( f_{\alpha}^{L}m_{p,\alpha}^{L}+f_{\alpha}^{R}m_{p,\alpha}^{R}+\epsilon \right)c_{\alpha,1/2}^{2}},
\end{equation}
where
\begin{equation}
    M_{\alpha}^{L} = \frac{u_{p,\alpha}^{L}}{c_{\alpha,1/2}}, \;\;\;\; M_{\alpha}^{R} = \frac{u_{p,\alpha}^{R}}{c_{\alpha,1/2}}, \;\;\;\; 
        \overline{M}_{\alpha}^{2} = \frac{\left( u_{p,\alpha}^{L} \right)^{2} + \left( u_{p,\alpha}^{R} \right)^{2}}{2c_{\alpha,1/2}^{2}},
\end{equation}
$K_{p}=0.25 + 0.75\left( 1-\mathscr{G}/2 \right)$, $n_{a}=1$, and $\sigma = 0.75\mathscr{G}/2$.

The particle velocity normal to the cell edge (independent of the dissipation term $\mathscr{F}_{\alpha}$) is defined from the interfacial compaction wavespeed and Mach number in Eqn.~\eqref{cM1/2}, 
\begin{equation}
    u_{\alpha,1/2} = c_{\alpha,1/2}M_{\alpha,1/2}.
\end{equation}

The Mach number splitting polynomials used to compute the interfacial Mach number are
\begin{equation}
    \mathscr{M}_{4}^{\pm}\left(M\right) = 
    \begin{cases}
        \mathscr{M}_{1}^{\pm}\left(M\right), & \textrm{if } \left| M \right| \geq 1 \\
        \mathscr{M}_{2}^{\pm}\left(M\right)\left( 1\mp 16\beta\mathscr{M}_{2}^{\mp}\left(M\right) \right), & \textrm{if } \left| M \right| < 1
    \end{cases} 
\end{equation}
where $\beta = 0.125$ and
\begin{equation}
    \mathscr{M}_{1}^{\pm}\left(M\right) = \frac{1}{2}\left( M \pm \left| M \right| \right), \;\;\;\; \mathscr{M}_{2}^{\pm}\left(M\right) = \pm\frac{1}{4}\left( M \pm 1 \right)^{2}.
\end{equation}

The interfacial pressure is
\begin{equation}
    \label{p1/2}
    \begin{split}
    p_{p,\alpha,i+1/2} = &-K_{u}f_{a}\left( c_{\alpha,1/2}-\epsilon \right)\mathscr{P}_{5}^{+}\left( M_{\alpha}^{L} \right)\mathscr{P}_{5}^{-}\left( M_{\alpha}^{R} \right)\left( f_{\alpha}^{R}m_{p,\alpha}^{R}u_{p,\alpha}^{R}-f_{\alpha}^{L}m_{p,\alpha}^{L}u_{p,\alpha}^{L}\right) \\ & +\mathscr{P}_{5}^{+}\left( M_{\alpha}^{L} \right)p_{p,\alpha}^{L} + \mathscr{P}_{5}^{-}\left( M_{\alpha}^{R} \right)p_{p,\alpha}^{R},
    \end{split}
\end{equation}
where $K_{u} = 0.75 + 0.25\left( 1-\mathscr{G}/2 \right)$.
The pressure splitting polynomials in Eqn.~\eqref{p1/2} are
\begin{equation}
    \mathscr{P}_{5}^{\pm}\left(M\right) = 
    \begin{cases}
        \mathscr{M}_{1}^{\pm}\left(M\right)/M, & \textrm{if } \left| M \right| \geq 1 \\
        \mathscr{M}_{2}^{\pm}\left(M\right)\left( \left( \pm 2-M \right)\mp 16\xi M\mathscr{M}_{2}^{\mp}\left(M\right) \right), & \textrm{if } \left| M \right| < 1
    \end{cases}
\end{equation}
where
\begin{equation}
    \xi = \frac{3}{16}\left( -4+5n_{a}^{2} \right),
\end{equation}
and $n_a = 1$.

The weight and abscissa for quadrature node $\alpha$ at edge $i+1/2$ are
\begin{equation}
    f_{\alpha,i+1/2} = 
    \begin{cases}
        f_{\alpha}^{L}, & \textrm{if } u_{\alpha,1/2} > 0 \\
        f_{\alpha}^{R}, & \textrm{if } u_{\alpha,1/2} \leq 0
    \end{cases} \;\;\;\;
    m_{p,\alpha,i+1/2} = 
    \begin{cases}
        m_{p,\alpha}^{L}, & \textrm{if } u_{\alpha,1/2} > 0 \\
        m_{p,\alpha}^{R}, & \textrm{if } u_{\alpha,1/2} \leq 0
    \end{cases}.
\end{equation}

The normal direction velocity component is found from the mass flux rate of the quadrature node
\begin{equation}
    u_{p,\alpha,i+1/2} = 
    \begin{cases}
        \Dot{m}_{\alpha,i+1/2}/f_{\alpha}^{L}m_{p,\alpha}^{L}, & \textrm{if } u_{\alpha,1/2} > 0 \\
        \Dot{m}_{\alpha,i+1/2}/f_{\alpha}^{R}m_{p,\alpha}^{R}, & \textrm{if } u_{\alpha,1/2} \leq 0
    \end{cases}.
\end{equation}
The tangential velocity components are simply upwinded
\begin{equation}
    v_{p,\alpha,i+1/2} = 
    \begin{cases}
        v_{p,\alpha}^{L}, & \textrm{if } u_{\alpha,1/2} > 0 \\
        v_{p,\alpha}^{R}, & \textrm{if } u_{\alpha,1/2} \leq 0
    \end{cases} \;\;\;\;
    w_{p,\alpha,i+1/2} = 
    \begin{cases}
        w_{p,\alpha}^{L}, & \textrm{if } u_{\alpha,1/2} > 0 \\
        w_{p,\alpha}^{R}, & \textrm{if } u_{\alpha,1/2} \leq 0
    \end{cases}.
\end{equation}
The granular and thermodynamic temperature for node $\alpha$ are
\begin{equation}
    \Theta_{p,\alpha,i+1/2} = 
    \begin{cases}
        \Theta_{p,\alpha}^{L}, & \textrm{if } u_{\alpha,1/2} > 0 \\
        \Theta_{p,\alpha}^{R}, & \textrm{if } u_{\alpha,1/2} \leq 0
    \end{cases} \;\;\;\;
    T_{p,\alpha,i+1/2} = 
    \begin{cases}
        T_{p,\alpha}^{L}, & \textrm{if } u_{\alpha,1/2} > 0 \\
        T_{p,\alpha}^{R}, & \textrm{if } u_{\alpha,1/2} \leq 0
    \end{cases}.
\end{equation}

\subsubsection{Interpolation of Granular Variables to cell Edges} \label{interp}

The nodal granular variables ($f_{\alpha}, m_{\alpha}, \boldsymbol{u}_{p,\alpha}, \Theta_{p,\alpha}$, and  $T_{p,\alpha}$) are reconstructed to the cell edges.  High-order reconstruction of the mass moments and abscissae are prone to moment corruption and other unrealizability issues \cite{Desjardins2008,Vikas2011}.  Thus, as is usual for higher-order QBMM implementations, the abscissae, $m_{\alpha}$, are reconstructed using first-order interpolation to aid in maintaining realizability
\begin{equation}
    m_{\alpha,i+1/2}^L = m_{\alpha,i}, \ \ \ m_{\alpha,i+1/2}^R = m_{\alpha,i+1}.
\end{equation}

\begin{figure}[tbp]
    \centering
    \includegraphics[width=.8\textwidth]{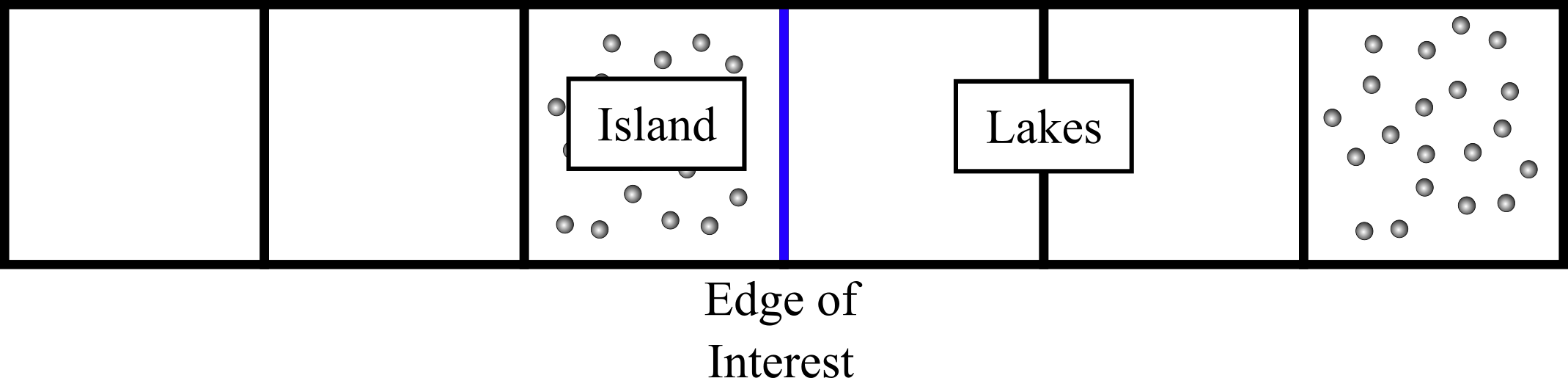}
    \caption{Primitive variable interpolation stencil with multiple vacuum interface, generating an ``island" and ``lakes" within the stencil.} \label{stencil_island}
\end{figure}

\begin{figure}[tbp]
    \centering
    \includegraphics[width=.8\textwidth]{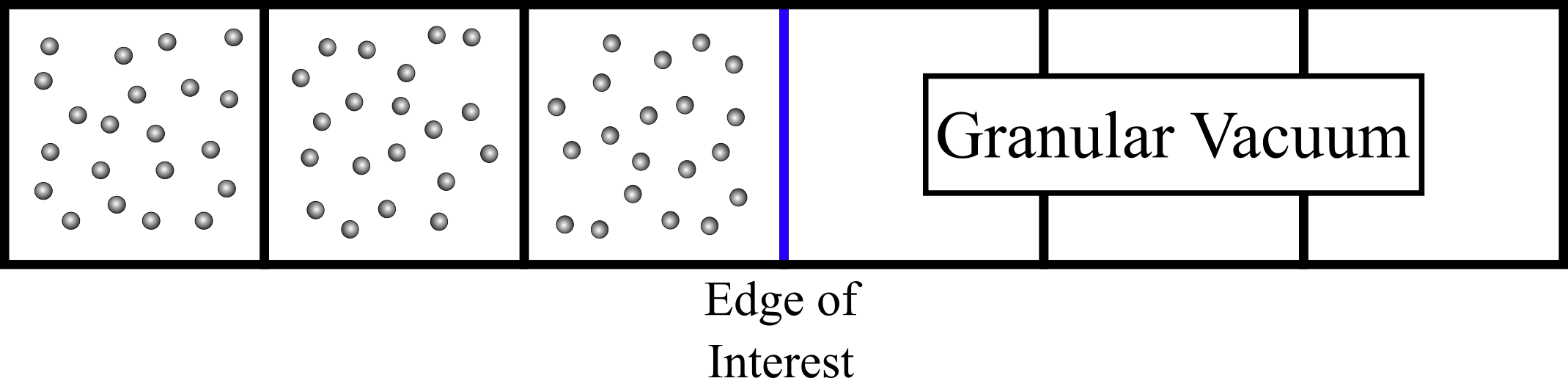}
    \caption{Primitive variable interpolation stencil with a vacuum interface.} \label{stencil_edge}
\end{figure}

The other nodal variables ($f_{\alpha},  \boldsymbol{u}_{p,\alpha}, \Theta_{p,\alpha}$, and  $T_{p,\alpha}$) are interpolated using fifth-order WENO \cite{Liu1994,Jiang1996} with a TVD slope limiter \cite{Houim2011a,Kim2005}.  Using fifth-order WENO at all locations for these variables was found to cause numerical instability in regions near particle ``islands'' or ``lakes'' or if there is a large variation in $m_{\alpha}$ along the interpolation stencil. As shown in Fig.~\ref{stencil_island}, particle ``islands'' are regions where an isolated cell with particles is surrounded by cells without any particles (``vacuums''). Particle ``lakes'' are isolated cells without particles surrounded by cells that contain particles.  In either of these situations, the edge reconstruction is dominated by the vacuum cells and is unreliable.  Numerical experimentation revealed that stencils containing particle cloud edges  encounter little difficulty.  (See Fig.~\ref{stencil_edge}.) 

The particle islands and lakes are found by 
counting the number edges that have particles on one side and no particles on the other side for all edges contained within the interpolation stencil.
\begin{equation}
    N_{\text{edge}} = \sum_{j=i-N_L}^{i+N_R-1}\Delta_{j+1/2}, \ \ \ \Delta_{j+1/2} = \begin{cases}
        1,& \text{ if } (\alpha_{p,j}-\alpha_{p,\min})(\alpha_{p,j+1}-\alpha_{p,\min})<0 \\
        0,& \text{ otherwise}
    \end{cases}
\end{equation}
where $N_L$ and $N_R$ are the number of cells in the interpolation stencil to the left and right of cell $i$, respectively, when interpolating to edge $i+1/2$. 
If $N_{\text{edge}} > 1$ for fifth-order WENO, then third-order WENO \cite{Liu1994,Jiang1996} is attempted.   First-order interpolation is used if  $N_{\text{edge}} > 1$ for third-order WENO.

\begin{figure}[tbp]
    \centering
    \includegraphics[width=.8\textwidth]{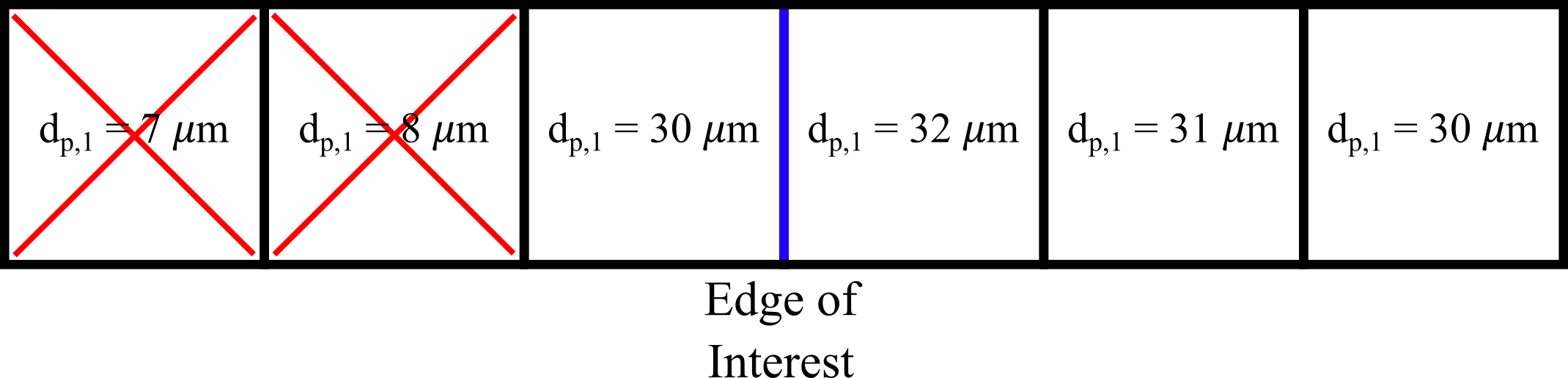}
    \caption{Primitive variable interpolation stencil with a large variation in the first abscissae, creating the opportunity for an invalid interpolation estimate.} \label{stencil_variation}
\end{figure}

Numerical experiments have also shown that high-order interpolation in stencils with  a large variation of particle diameter can lead to numerical instability (see Fig.~\ref{stencil_variation}).  Similar to the procedure for degrading the order of edge reconstruction near particle islands and lakes, we check for large variations in nodal particle diameter in the interpolation stencil.  First, we define
\begin{equation}
    \delta_{\alpha} = \frac{1}{N_{s}}\sum_{\substack{j=1-N_L,\\\alpha_{p,j}\neq 0}}^{i+N_R}\left| d_{p,\alpha,j}-d_{p,\alpha,i} \right|,
\end{equation}
where $\delta_{\alpha}$ is the stencil difference parameter for quadrature node $\alpha$, $N_{s}$ is the number of cells in the stencil without granular vacuums, $d_{p,\alpha,j}$ is the particle diameter corresponding to quadrature node $\alpha$ in cell $j$ of the stencil, and $d_{p,\alpha,i}$ is the particle diameter corresponding to quadrature node $\alpha$ at the edge of interest (cell $i$). This difference is computed for all quadrature nodes.   If the difference parameter for any of the quadrature nodes ($\delta_{1},\delta_{2},...,\delta_{N}$) are greater than a user-specified threshold (5\% was found to work well), the interpolation order for the particle phase variables is degraded. Similar to degradation near islands and wakes,  fifth-order WENO is first attempted.  Third-order WENO is attempted if the diameter change is too large in the fifth-order WENO stencil.  First-order interpolation is used if the diameter variation is too large in the third-order WENO stencil.

The final step of edge interpolation for nodal variables is to pass them through an extra Total Variation Diminishing (TVD) limiter that is active in highly packed states \cite{Houim2016,Kim2005} 
\begin{equation}
    Q_{i+1/2}^{L}=Q_{i}+0.5\left( Q_{i}-Q_{i-1} \right)\phi_{TVD}, \;\;\;\; \phi_{TVD}=\max\left( 0,\min\left( \mathscr{G},\mathscr{G}\frac{Q_{i+1}-Q_{i}}{Q_{i}-Q_{i-1}},2\frac{\hat{Q}_{i+1/2}^{L}-Q_{i}}{Q_{i}-Q_{i-1}} \right) \right),
    \label{eqn:GranTVD}
\end{equation}
where $Q_{i+1/2}^{L}$ is a generic left-interpolated variable, $\phi_{TVD}$ is the TVD slope limiter, and $\hat{Q}_{i+1/2}^{L}$ is the  left-interpolated variable after stencil degradation.  

Reducing the order of accuracy near islands, lakes, and large changes in particle diameter and the extra TVD limiter are only applied to the nodal particle variables.

\subsubsection{Rusanov Flux of Degraded Edges}
The Rusanov flux \cite{Rusanov1962} is used to protect the moments from ill-conditioned fluxes in regions where the cell edge interpolation is degraded down to first-order
\begin{equation}
    \label{Rusanov}
    \boldsymbol{F}_{p,i+1/2}=\frac{\boldsymbol{F}_p^L+\boldsymbol{F}_p^R}{2} - \frac{S}{2}\left( \boldsymbol{U}_p^R-\boldsymbol{U}_p^L \right),
\end{equation}
where $\boldsymbol{F}_p$ and $\boldsymbol{U}_p$ are the particle phase flux and conserved variables vectors, respectively, and $S$ is the Rusanov wavespeed
\begin{equation}
    S = \max\left( \left| u_{p,1}^L \right| + c_{p,1}^L , \left| u_{p,2}^L \right| + c_{p,2}^L, ..., \left| u_{p,N}^L \right| + c_{p,N}^L, \left| u_{p,1}^R \right| + c_{p,1}^R , \left| u_{p,2}^R \right| + c_{p,2}^R, ..., \left| u_{p,N}^R \right| + c_{p,N}^R \right).
\end{equation}
The cell-edge primitive variables are found from Eqn.~\ref{Rusanov} \cite{Shen2016}.
The volume fraction at edge $i+1/2$ is
\begin{equation}
    \alpha_{p,i+1/2} = \frac{\alpha_p^R+\alpha_p^L}{2} - \frac{1}{2S}\left( \alpha_p^Ru_p^R-\alpha_p^Lu_p^L \right).
\end{equation}
The velocity normal to the edge is
\begin{equation}
    u_{p,i+1/2} = \frac{\alpha_{p}^{R}\left( S-u_{p}^{R} \right)u_{p}^{R}+\alpha_{p}^{L}\left( S+u_{p}^{L} \right)u_{p}^{L} + p_{p}^{L}/\rho_{p}-p_{p}^{R}/\rho_{p}}{\alpha_{p}^{R}\left( S-u_{p}^{R} \right)+\alpha_{p}^{L}\left( S+u_{p}^{L} \right)}.
\end{equation}
The tangential velocities are
\begin{equation}
    v_{p,i+1/2} = \frac{\alpha_{p}^{R}\left( S-u_{p}^{R} \right)v_{p}^{R}+\alpha_{p}^{L}\left( S+u_{p}^{L} \right)v_{p}^{L}}{\alpha_{p}^{R}\left( S-u_{p}^{R} \right)+\alpha_{p}^{L}\left( S+u_{p}^{L} \right)}, \;\;\;\; w_{p,i+1/2} = \frac{\alpha_{p}^{R}\left( S-u_{p}^{R} \right)w_{p}^{R}+\alpha_{p}^{L}\left( S+u_{p}^{L} \right)w_{p}^{L}}{\alpha_{p}^{R}\left( S-u_{p}^{R} \right)+\alpha_{p}^{L}\left( S+u_{p}^{L} \right)}.
\end{equation}
The granular temperature is
\begin{equation}
    \Theta_{p,i+1/2} = \frac{2}{3}\left[\frac{\alpha_{p}^{R}\left( S-u_{p}^{R} \right)E_{p}^{R}+\alpha_{p}^{L}\left( S+u_{p}^{L} \right)E_{p}^{L} + p_{p}^{L}u_{p}^{L}/\rho_{p}-p_{p}^{R}u_{p}^{R}/\rho_{p}}{\alpha_{p}^{R}\left( S-u_{p}^{R} \right)+\alpha_{p}^{L}\left( S+u_{p}^{L} \right)} - \frac{1}{2}\boldsymbol{u}_{p,i+1/2}\cdot\boldsymbol{u}_{p,i+1/2}\right],
\end{equation}
where $E_p$ is the particle phase total energy
\begin{equation}
    E_p = \frac{1}{2}\boldsymbol{u}_{p}\cdot\boldsymbol{u}_{p} + \frac{3}{2}\Theta_p.
\end{equation}
The material temperature is
\begin{equation}
    T_{p,i+1/2} = \frac{\alpha_{p}^{R}\left( S-u_{p}^{R} \right)T_{p}^{R}+\alpha_{p}^{L}\left( S+u_{p}^{L} \right)T_{p}^{L}}{\alpha_{p}^{R}\left( S-u_{p}^{R} \right)+\alpha_{p}^{L}\left( S+u_{p}^{L} \right)}.
\end{equation}

\subsubsection{Hyperbolic Solution Algorithm}
{
The steps to advance to the next Runge-Kutta stage for the hyperbolic operator $\rsfscr{H}_{xyz}^{\Delta t}$ are
\begin{enumerate}
    \item Remove particles from cells that have either $\alpha_{p} < \alpha_{p,\text{min}}$ or exceed the packing limit as discussed in Section~\ref{sec:particleRemoval}.
    \item Compute gas-phase primitive variables, quadrature abscissae and weights using beta-GQMOM, and the nodal variables using CQMOM at each cell center as discussed in Sections~\ref{Beta-GQMOM} and~\ref{CQMOM}.
    \item Compute the gas and granular sound speeds at each cell and determine the time-step size based on the CFL criterion. 
    \item Interpolate the gas-phase primitive variables and granular-phase nodal variables, weights, and abscissae to the cell edges in the $x$-, $y$-, and $z$- directions.
    \begin{enumerate}
    \item Interpolate the gas-phase variables using fifth-order MUSCL or any other appropriate method.
    \item Interpolate the abscissae using first-order upwind. 
    \item Check the fifth-order WENO stencil of each cell edge for islands or lakes on the stencil. Mark applicable stencils for reduction of the interpolation order-of-accuracy for the weights and conditional variables discussed in Section~\ref{interp}. \label{step1}
    \item Interpolate the quadrature weights and nodal variables using the highest-order scheme permitted by the interpolation reduction check: fifth-order WENO, third-order WENO, or first-order upwind.
    \item Compute $\mathscr{G}$ (Eqn.~\eqref{eqn:G-AUSM}) at each cell edge.
    \item Perform the extra TVD limiter (Eqn.~\ref{eqn:GranTVD}) on the quadrature weights and conditional variables.
    \end{enumerate}
    \item Use HLLC to compute the gas-phase state at the cell edges from the interpolated gas-phase variables. 
    \item Use the modified AUSM solver (or Rusanov Riemann solver if the interpolation has been reduced to first-order) to compute the nodal variables, abscissae, and weights at the cell edges for each quadrature node discussed in Sec.~\ref{sec:granRiemann}. 
    \item Compute the lift force (Eqn.~\eqref{eqn:lift}) with second-order central differing.
    \item Use the results of the gas and granular Riemann problems and cell-centered values to compute the spatial derivatives and non-conservative terms generalized in Eqn.~\eqref{eqn:nonConsDer}.  
    \item Assemble the spatial derivative terms in the hyperbolic split PDEs (Eqns.~\eqref{eqn:GasMassHypSplit}-\eqref{eqn:partEn_Hyp}).
    \item Advance the Runge-Kutta Stage.
\end{enumerate}
}

\subsection{Source Terms, $\rsfscr{I}^{\Delta t}$}
The source terms for effects including drag, convective heat transfer, collisions, etc.\ lead to a coupled set of ordinary differential equations
\begin{align}
    &\frac{ d\alpha_g \rho_g}{dt} = 0, \\
    &\frac{ d\alpha_g \rho_g \boldsymbol{u}_g}{dt} = \mathcal{D}_1^P, \\
    &\frac{d \alpha_g \rho_g E_g}{dt} = \mathcal{D}_1^E + \mathcal{H}_1, \\
    &\frac{d \mathcal{M}_{n/q}}{dt}  = 0, \ \ \ &n = 0, 1, 2, \cdots, N_{\text{mass}}-1 \\
    &\frac{d \mathcal{U}_{s}}{dt}  = \mathcal{C}_{s}^P - \mathcal{D}_s^P, \ \ \ &s = 0, 1, 2, \cdots, N-1 \\
    &\frac{d \mathcal{T}_{s}}{dt}  = \mathcal{C}_{s}^{PTE} + \mathcal{D}_s^{PTE} + \mathcal{F}_s^{PTE}, \ \ \ &s = 0, 1, 2, \cdots, N-1 \\
    &\frac{d \mathcal{E}_{s}}{dt}  = - \mathcal{C}_{s}^{E} - \mathcal{F}_s^{PTE} - \mathcal{H}_s, \ \ \ &s = 0, 1, 2, \cdots, N-1
\end{align}
Note that the weights and abscissae are fixed during integration of the source terms.  Substituting the quadrature rule into the moment equations and setting each term in the sum equal to zero gives
\begin{align}
    \frac{ d\boldsymbol{u}_g}{dt} &= \frac{\mathcal{D}_1^P}{\alpha_g \rho_g}, \\
    \frac{d  E_g}{dt} &= \frac{\mathcal{D}_1^E}{\alpha_g \rho_g} + \frac{\mathcal{H}_1}{\alpha_g \rho_g}, \\
    \frac{d \boldsymbol{u}_{p,\alpha}}{dt} & = \frac{\mathcal{C}_{1}^P(m_{\alpha}) - \mathcal{D}_{1}^P(m_{\alpha})}{m_{\alpha} f_{\alpha}}, \ \ \ & \alpha =  1, 2, \cdots, N \\
    \frac{d \Theta_{p,\alpha}}{dt} & = \frac{2}{3} \frac{\mathcal{C}_{1}^{PTE}(m_{\alpha}) + \mathcal{D}_1^{PTE}(m_{\alpha}) + \mathcal{F}_1^{PTE}(m_{\alpha})}{m_{\alpha} f_{\alpha}}, \ \ \ &\alpha=  1, 2, \cdots, N \\
    \frac{d e_{p,\alpha}}{dt} & = - \frac{\mathcal{C}_{1}^{E}(m_{\alpha}) - \mathcal{F}_1^{PTE}(m_{\alpha}) - \mathcal{H}_1(m_{\alpha})}{m_{\alpha} f_{\alpha}}, \ \ \ &\alpha =  1, 2, \cdots, N \\
    \frac{d f_{\alpha}}{dt} &=0, \ \  \frac{d m_{\alpha}}{dt} = 0, \ \ \frac{d \alpha_p}{dt} = 0, \ \  \frac{d \rho_g}{dt} = 0. \ \ &\alpha =  1, 2, \cdots, N
\end{align}
The ordinary differential equation (ODE) system solved during the $\rsfscr{I}^{\Delta t}$ operator is further split into ODEs for the drag forces, convection heat transfer, particle--particle collisions, and frictional particle collisions
\begin{equation}
    \rsfscr{I}^{\Delta t}\left( \boldsymbol{U} \right) = \rsfscr{I}_{D}^{\Delta t}\left( \rsfscr{I}_{H}^{\Delta t}\left( \rsfscr{I}_{C}^{\Delta t}\left( \rsfscr{I}_{F}^{\Delta t} 
    \left(\rsfscr{I}_{F}^{\Delta t} 
    \left( \rsfscr{I}_{C}^{\Delta t}
    \left(\rsfscr{I}_{H}^{\Delta t}
    \left(\rsfscr{I}_{D}^{\Delta t}\left( \boldsymbol{U} \right)\right)\right)
    \right) \right) \right) \right)\right), 
\end{equation}
where $\rsfscr{I}_{D}^{\Delta t}$, $\rsfscr{I}_{H}^{\Delta t}$, $\rsfscr{I}_{C}^{\Delta t}$, and $\rsfscr{I}_{F}^{\Delta t}$ represent the integration of source terms over arising from drag, convective heat transfer, collisions, and friction, respectively.  Splitting the integration of the various source terms allows the use of analytical expressions, which is more robust. 

\subsubsection{Drag, $\rsfscr{I}_{D}^{\Delta t}$}

The coupled ODE system for the gas--particle drag forces is \cite{Fox2024}
\begin{equation}
    \label{drag_ugODE}
    \frac{d\boldsymbol{u}_{g}}{dt} = -\boldsymbol{u}_{g}\sum_{\alpha=1}^{N}\frac{1}{\tau_{p}(m_{\alpha})}\frac{f_{\alpha}m_{\alpha}}{\alpha_{g}\rho_{g}} + \sum_{\alpha=1}^{N}\frac{\boldsymbol{u}_{p,\alpha}}{\tau_{p}(m_{\alpha})}\frac{f_{\alpha}m_{\alpha}}{\alpha_{g}\rho_{g}}, \;\; \frac{d\boldsymbol{u}_{p,\alpha}}{dt} = -\frac{\boldsymbol{u}_{p,\alpha}}{\tau_{p}(m_{\alpha})} + \frac{\boldsymbol{u}_{g}}{\tau_{p}(m_{\alpha})}, \;\;
    \frac{d\Theta_{p,\alpha}}{dt} = -2\frac{\Theta_{p,\alpha}}{\tau_{p}(m_{\alpha})}.
\end{equation}
These equations form a coupled system of ODEs that can be rewritten as
\begin{equation}
    \label{drag_system}
    \frac{d\boldsymbol{u}}{dt} = \boldsymbol{A}\boldsymbol{u}, \;\;\;\; \frac{d\boldsymbol{v}}{dt} = \boldsymbol{A}\boldsymbol{v}, \;\;\;\; \frac{d\boldsymbol{w}}{dt} = \boldsymbol{A}\boldsymbol{w},
\end{equation}
where $\boldsymbol{u}=\left( u_{g}, u_{p,1},u_{p,2},...,u_{p,N} \right)^{T}$ is the vector of $x$-velocities, $\boldsymbol{v}=\left( v_{g}, v_{p,1},v_{p,2},...,v_{p,N} \right)^{T}$ is the vector of $y$-velocities, $\boldsymbol{w} = \left( w_{g}, w_{p,1},w_{p,2},...,w_{p,N} \right)^{T}$ is the vector $z$-velocities, and $\boldsymbol{A}$ is the coefficient matrix
\begin{equation}
    \boldsymbol{A} = 
    \begin{pmatrix}
    -\sum_{\alpha}^{N}\frac{f_{\alpha}m_{\alpha}}{\alpha_{g}\rho_{g}}\frac{1}{\tau_{p}(m_{\alpha})} & \frac{f_{1}m_{1}}{\alpha_{g}\rho_{g}}\frac{1}{\tau_{p}(m_{1})} & \hdots & \frac{f_{N}m_{N}}{\alpha_{g}\rho_{g}}\frac{1}{\tau_{p}(m_{N})} \\
    \frac{1}{\tau_{p}(m_{1})} &  -\frac{1}{\tau_{p}(m_{1})} & \hdots & 0 \\
    \vdots & \vdots & \ddots & \vdots \\
    \frac{1}{\tau_{p}(m_{N})} & 0 & \hdots & -\frac{1}{\tau_{p}(m_{N})} \\
    \end{pmatrix}.
\end{equation}
Eqns.~\eqref{drag_system} can be integrated  analytically from time $t$ to $t+\Delta t$ using a matrix exponential solution \cite{Zwillinger1989} if the drag time scales $\tau_{p}(m_{\alpha})$ are held constant.  The integrated velocities are
\begin{equation}
    \boldsymbol{u}^{t+\Delta t} = e^{\boldsymbol{A}\Delta t}\boldsymbol{u}^{t}, \;\;\;\; \boldsymbol{v}^{t+\Delta t} = e^{\boldsymbol{A}\Delta t}\boldsymbol{v}^{t}, \;\;\;\; \boldsymbol{w}^{t+\Delta t} = e^{\boldsymbol{A}\Delta t}\boldsymbol{w}^{t}.
\end{equation}
The matrix exponential of $\boldsymbol{A}\Delta t$ is found from a Padé approximant \cite{Moler2003}. Integrating the ODE for viscous damping of granular temperature analytically from time $t$ to $t+\Delta t$, 
\begin{equation}
    \Theta_{p,\alpha}^{t+\Delta t} = \Theta_{p,\alpha}^{t}e^{-2\Delta t/\tau_{p,\alpha}}.
\end{equation}
The total decrease in gas and particle bulk kinetic energy is added to the gas as sensible energy (or ``drag heating'')
\begin{equation}
    \alpha_g \rho_g e_g^{t+\Delta t} = \alpha_g \rho_g e_{g}^t + \left[(\alpha_g \rho_g ke_g^t + KE_p^t + PTE_p^t) - (\alpha_g \rho_g ke_g^{t+\Delta t} + KE_p^{t+\Delta t} +PTE_p^{t+\Delta t})\right],
\end{equation}
where $ke_g = \boldsymbol{u}_g \cdot \boldsymbol{u}_g/2$ is the gas-phase kinetic energy per unit mass and 
\begin{equation}
    KE_p = \frac{1}{2}\sum_{\alpha=1}^N m_{\alpha} \boldsymbol{u}_{p,\alpha} \cdot \boldsymbol{u}_{p,\alpha} f_{\alpha}, \;\;\;\; PTE_p = \frac{3}{2}\sum_{\alpha=1}^N m_{\alpha} \Theta_{p,\alpha}f_{\alpha}.
\end{equation}

\subsubsection{Convection Heat Transfer, $\rsfscr{I}_{H}^{\Delta t}$}

The coupled ODE system for convective is \cite{Fox2024}
\begin{equation}
    \label{conv_TgODE}
    \frac{dT_{g}}{dt} = -T_{g}\sum_{\alpha=1}^{N}\frac{h(m_{\alpha})}{c_{v,g}}\frac{f_{\alpha}m_{\alpha}}{\alpha_{g}\rho_{g}} + \sum_{\alpha=1}^{N}\frac{h(m_{\alpha})}{c_{v,g}}\frac{f_{\alpha}m_{\alpha}}{\alpha_{g}\rho_{g}}T_{p,\alpha}, \;\;\;\; \frac{dT_{p,\alpha}}{dt} = -\frac{h(m_{\alpha})}{c_{v,p}}T_{p,\alpha} + \frac{h(m_{\alpha})}{c_{v,p}}T_{g}.
\end{equation}
These equations form a linear coupled system 
\begin{equation}
    \label{conv_system}
    \frac{d\boldsymbol{T}}{dt} = \boldsymbol{A}\boldsymbol{T},
\end{equation}
where $\boldsymbol{T} = \left( T_{g},T_{p,1},...,T_{p,N} \right)^{T}$ is the vector of temperatures, and $\boldsymbol{A}$ is the coefficient matrix
\begin{equation}
    \boldsymbol{A} = 
    \begin{pmatrix}
    -\sum_{\alpha}^{N}\frac{f_{\alpha}m_{\alpha}}{\alpha_{g}\rho_{g}}\frac{h(m_{\alpha})}{c_{v,g}} & \frac{f_{1}m_{p,1}}{\alpha_{g}\rho_{g}}\frac{h(m_{1})}{c_{v,g}} & \hdots & \frac{f_{N}m_{N}}{\alpha_{g}\rho_{g}}\frac{h(m_{N})}{c_{v,g}} \\
    \frac{h(m_{1})}{c_{v,p}} &  -\frac{h(m_{1})}{c_{v,p}} & \hdots & 0 \\
    \vdots & \vdots & \ddots & \vdots \\
    \frac{h(m_{N})}{c_{v,p}} & 0 & \hdots & -\frac{h(m_{N})}{c_{v,p}} \\
    \end{pmatrix}.
\end{equation}
This ODE system can be integrated analytically from time $t$ to $t+\Delta t$ if the convective heat transfer coefficient and specific heats are fixed during
\begin{equation}
    \boldsymbol{T}^{t+\Delta t} = e^{\boldsymbol{A}\Delta t}\boldsymbol{T}^{t} .
\end{equation}
Again, $e^{\boldsymbol{A}\Delta t}$ is computed using a Padé approximant \cite{Moler2003}.

\subsubsection{Particle--Particle Collisions, $\rsfscr{I}_{C}^{\Delta t}$}
The coupled ODEs from particle collisions are \cite{Fox2024}
\begin{equation}
    \label{collision_u}
    \frac{d\boldsymbol{u}_{p,\alpha}}{dt} = -\boldsymbol{u}_{p,\alpha}\sum_{\gamma=1,\gamma\neq \alpha}^{N}\kappa_{\alpha\gamma}\psi_{\alpha\gamma} + \sum_{\gamma=1,\gamma\neq \alpha}^{N}\kappa_{\alpha\gamma}\psi_{\alpha\gamma}\boldsymbol{u}_{p,\gamma},
\end{equation}
where $\psi_{\alpha\gamma}$ is a factor accounting for inelastic collision of particles $\alpha$ and $\gamma$
\begin{equation}
    \psi_{\alpha\gamma} = \frac{1}{4}\left[ 1+e_c({m_{\alpha},m_{\gamma}}) \right] \mu(m_{\alpha},m_{\gamma}), \ \ \ \kappa_{\alpha\gamma}=\kappa(m_{\alpha},m_{\gamma}).
\end{equation}

Collisions convert some of the mean particle kinetic energy into random kinetic energy.  Thus, there is a source of particle granular temperatures from collisions
\begin{equation}
    \label{collision_theta}
    \begin{split}
        \frac{d\Theta_{p,\alpha}}{dt} = -\frac{1}{4}\kappa_{\alpha\alpha}\left( 1-e_c^{2}(m_{\alpha},m_{\alpha}) \right)\Theta_{p,\alpha} &+ \sum_{\gamma=1,\gamma\neq \alpha}^{N}\left( -2\kappa_{\alpha\gamma}\psi_{\alpha\gamma}\Theta_{p,\alpha} + 2\kappa_{\alpha\gamma}\psi_{\alpha\gamma}^{2}\left( \Theta_{p,\alpha}+\Theta_{p,\gamma} \right)\right) \\
        &+ \sum_{\gamma=1,\gamma\neq \alpha}^{N}\frac{2}{3}\kappa_{\alpha\gamma}\psi_{\alpha\gamma}^{2}\left( \boldsymbol{u}_{p,\alpha}-\boldsymbol{u}_{p,\gamma} \right)\cdot\left( \boldsymbol{u}_{p,\alpha}-\boldsymbol{u}_{p,\gamma} \right).
    \end{split}
\end{equation}
The first term in Eqn.~\eqref{collision_theta} accounts for collisional dissipation due to same-size interactions (also called ``like'' collisions) and is Haff's cooling law \cite{Haff1983}.

Eqns.~\eqref{collision_u} and~\eqref{collision_theta} can be recast in a matrix form
\begin{equation}
    \label{collisional_u_matrix}
    \frac{d\boldsymbol{u}}{dt} = \boldsymbol{A}\boldsymbol{u}, \;\;\;\; \frac{d\boldsymbol{v}}{dt} = \boldsymbol{A}\boldsymbol{v}, \;\;\;\; \frac{d\boldsymbol{w}}{dt} = \boldsymbol{A}\boldsymbol{w}, \;\;\;\; \frac{d\boldsymbol{\Theta}}{dt} = \boldsymbol{B}\boldsymbol{\Theta} + \boldsymbol{D},
\end{equation}
where $\boldsymbol{u} = \left( u_{p,1},u_{p,2},...,u_{p,N} \right)^{T}$ is the vector of nodal $x$-velocities, $\boldsymbol{v} = \left( v_{p,1},v_{p,2},...,v_{p,N} \right)^{T}$ is the vector of nodal $y$-velocities, $\boldsymbol{w} = \left( w_{p,1},w_{p,2},...,w_{p,N} \right)^{T}$ is the vector of nodal $z$-velocities, and $\boldsymbol{\Theta} = \left( \Theta_{p,1},\Theta_{p,2},...,\Theta_{p,N} \right)^{T}$ is the vector of nodal granular temperatures. The coefficient matrix $\boldsymbol{A}$ is
\begin{equation}
    \boldsymbol{A} = 
    \begin{pmatrix}
    -\sum_{\alpha\neq 1}^{N}\kappa_{1\alpha}\psi_{1\alpha} & \kappa_{12}\psi_{12} & \hdots & \kappa_{1N}\psi_{1N} \\
    \kappa_{21}\psi_{21} &  -\sum_{\alpha\neq 2}^{N}\kappa_{2\alpha}\psi_{2\alpha} & \hdots & \kappa_{2N}\psi_{2N} \\
    \vdots & \vdots & \ddots & \vdots \\
    \kappa_{N1}\psi_{N1} & \kappa_{N2}\psi_{N2} & \hdots & -\sum_{\alpha\neq N}^{N}\kappa_{N\alpha}\psi_{N\alpha} \\
    \end{pmatrix}.
\end{equation}

The coefficient matrix $\boldsymbol{B}$ in Eqn.~\eqref{collisional_u_matrix} is split into coefficients for like collisions and unlike collisions (collisions between particles of a different mass), $\boldsymbol{B} = \boldsymbol{B}_{\textrm{like}} + \boldsymbol{B}_{\textrm{unlike}}$, where the like collisions coefficient matrix is
\begin{equation}
    \boldsymbol{B}_{\textrm{like}} = 
    \begin{pmatrix}
    -\frac{1}{4}\kappa_{11}\left( 1-e_{11}^{2} \right) & 0 & \hdots & 0 \\
    0 &  -\frac{1}{4}\kappa_{22}\left( 1-e_{22}^{2} \right) & \hdots & 0 \\
    \vdots & \vdots & \ddots & \vdots \\
    0 & 0 & \hdots & -\frac{1}{4}\kappa_{NN}\left( 1-e_{NN}^{2} \right) \\
    \end{pmatrix},
\end{equation}
and the unlike collisions coefficient matrix is
\begin{equation}
    \boldsymbol{B}_{\textrm{unlike}} = 
    \begin{pmatrix}
    -\sum_{\alpha\neq 1}^{N}2\kappa_{1\alpha}\left( \psi_{1\alpha}-\psi_{1\alpha}^{2} \right) & 2\kappa_{12}\psi_{12}^{2} & \hdots & 2\kappa_{1N}\psi_{1N}^{2} \\
    2\kappa_{21}\psi_{21}^{2} &  -\sum_{\alpha\neq 2}^{N}2\kappa_{2\alpha}\left( \psi_{2\alpha}-\psi_{2\alpha}^{2} \right) & \hdots & 2\kappa_{2N}\psi_{2N}^{2} \\
    \vdots & \vdots & \ddots & \vdots \\
    2\kappa_{N1}\psi_{N1}^{2} & 2\kappa_{N2}\psi_{N2}^{2} & \hdots & -\sum_{\alpha\neq N}^{N}2\kappa_{N\alpha}\left( \psi_{N\alpha}-\psi_{N\alpha}^{2} \right) \\
    \end{pmatrix}.
\end{equation}
The inhomogeneous source term $\boldsymbol{D}$ in Eqn.~\eqref{collisional_u_matrix} due to differences in particle mean velocity generating PTE is
\begin{equation}
    \boldsymbol{D} = 
    \begin{pmatrix}
    \sum_{\alpha\neq 1}^{N}\frac{2}{3}\kappa_{1\alpha}\psi_{1\alpha}^{2}\left( \boldsymbol{u}_{p,1}-\boldsymbol{u}_{p,\alpha} \right)\cdot\left( \boldsymbol{u}_{p,1}-\boldsymbol{u}_{p,\alpha} \right) \\
    \sum_{\alpha\neq 2}^{N}\frac{2}{3}\kappa_{2\alpha}\psi_{2\alpha}^{2}\left( \boldsymbol{u}_{p,2}-\boldsymbol{u}_{p,\alpha} \right)\cdot\left( \boldsymbol{u}_{p,2}-\boldsymbol{u}_{p,\alpha} \right) \\
    \vdots \\
    \sum_{\alpha\neq N}^{N}\frac{2}{3}\kappa_{N\alpha}\psi_{N\alpha}^{2}\left( \boldsymbol{u}_{p,N}-\boldsymbol{u}_{p,\alpha} \right)\cdot\left( \boldsymbol{u}_{p,N}-\boldsymbol{u}_{p,\alpha} \right) \\
    \end{pmatrix}.
\end{equation}

The nodal velocity ODE system for collisions are inhomogeneous and are not analytically integrable. The collision ODEs are integrated using the Variable-Coefficient ODE (VODE) stiff solver \cite{Brown1989}. The analytic Jacobian is used to increase robustness. 

After the mean kinetic and pseudo-thermal energies are changed due to collisions, the lost energy is distributed to the internal energies of each node according to their mass
\begin{equation}
    \alpha_p\rho_p e_{p,\alpha}^{t+\Delta t} = \alpha_p\rho_p e_{p,\alpha}^{t} + \frac{m_\alpha f_\alpha}{\alpha_p\rho_p}\left[(KE_p^t + PTE_p^t) - (KE_p^{t+\Delta t} +PTE_p^{t+\Delta t})\right].
\end{equation}

\subsubsection{Particle Frictional Collisions, $\rsfscr{I}_{F}^{\Delta t}$}
The particle pressure is dominated by the frictional pressure when the particles become densely packed
 \cite{Fox2024,Gidaspow1994}. This generates a corresponding frictional collisional source term.  The ODE for the frictional attenuation of the nodal granular temperature is
\begin{equation}
    \frac{d\Theta_{p,\alpha}}{dt} = -\frac{1}{\tau_{fr,\alpha}}\Theta_{p,\alpha}.
\end{equation}
The analytical solution to this ODE is
\begin{equation}
    \Theta_{p,\alpha}^{t+\Delta t} = \Theta_{p,\alpha}^{t}e^{-\Delta t/\tau_{fr,\alpha}}.
\end{equation}
The reduction of nodal pseudo-thermal energy by friction increases the nodal internal energy
\begin{equation}
    \alpha_p\rho_p e_{p,\alpha}^{t+\Delta t} = \alpha_p\rho_p e_{p,\alpha}^{t} + \frac{m_\alpha f_\alpha}{\alpha_p\rho_p}\left[PTE_p^t -PTE_p^{t+\Delta t}\right].
\end{equation}

\section{Numerical Tests}

Here we demonstrate the capabilities and robustness of the model and high-order numerical methods with a series of polydisperse test problems in one- and two-dimensions. 
The De Brouckere (volume-averaged) mean diameter 
\begin{equation}
    d_{4,3} = \frac{\int d_{p}^{4}(m)f(m)dm}{\int d_{p}^{3}(m)f(m)dm}=\frac{\sum_{\alpha=1}^{N}f_{\alpha}d_{p}^{4}(m_\alpha)}{\sum_{\alpha=1}^{N}f_{\alpha}d_{p}^{3}(m_\alpha)},
\end{equation}
is used to characterize the mean particle size for the polydisperse mixtures.

\begin{figure}[tbp]
    \centering
    \includegraphics[width=.4\textwidth]{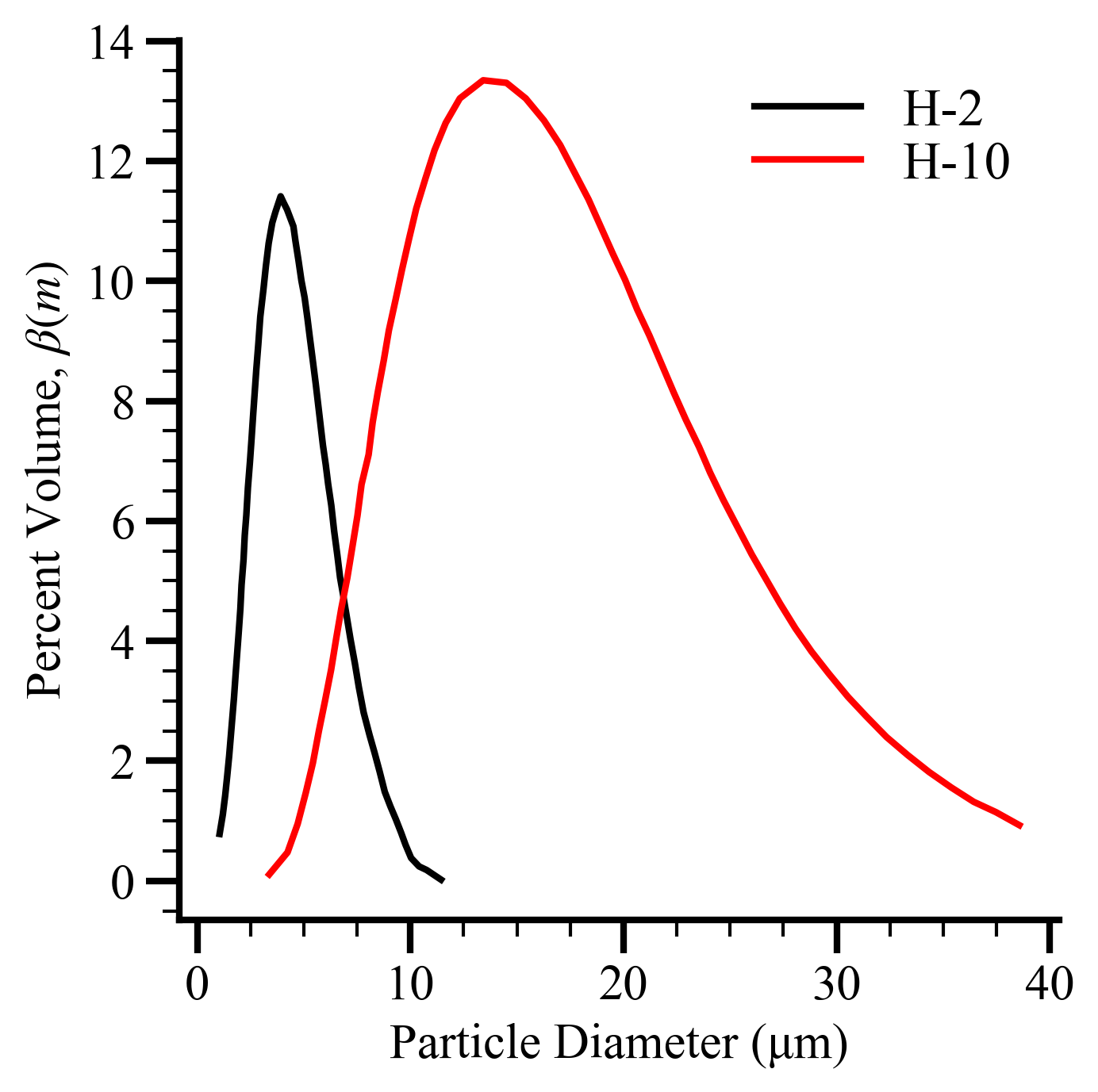}
    \includegraphics[width=.4\textwidth]{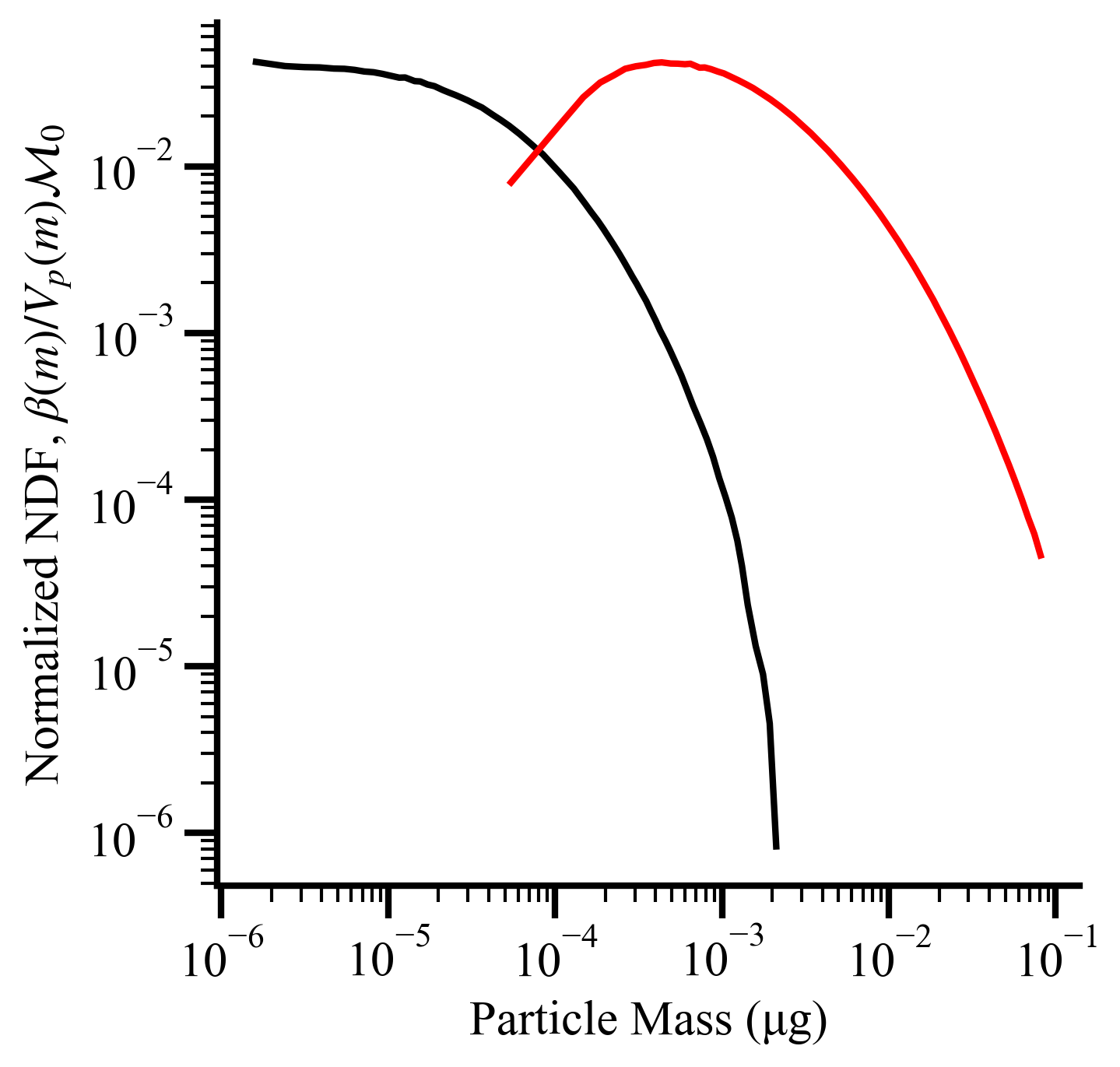}
    \caption{Particle size distributions of H-2 \cite{Blais2020} and H-10 \cite{Julien2015} aluminum given in (left) volume fraction form as presented   and (right) number density function form normalized by $\mathcal{M}_0$. $V_p(m)=\frac{\pi}{6}d_p^3(m)$ is the particle volume.} \label{psd}
\end{figure}

The H-2 \cite{Blais2020} and H-10 \cite{Julien2015} size distributions of aluminum particles, shown in Fig.~\ref{psd}, are used. The experimental PSDs provided on the left of Fig.~\ref{psd} are converted to NDFs shown on the right of Fig.~\ref{psd} to compute $\mathcal{M}_n$. Three quadrature nodes are used ($N=3$) with area moments ($q=2$) unless stated otherwise.
Beta-GQMOM is used for moment inversion with a maximum diameter of 100 $\mu$m corresponding to $m_{\max} = 1.414$ $\mu$g. For the H-2 distribution, the abscissae are 1.8, 4, and 7.4 $\mu$m in diameter and the volume fractions, $\beta(m)$, (from which the weights may be found) are 17.2\%, 63.2\%, and 19.6\%, respectively. For H-10, the abscissae are 7.13, 15.43 and 29.07 $\mu$m in diameter and the volume fractions, $\beta(m)$, are 25.2\%, 55.8\%, and 19.0\%, respectively. These distributions have mean diameters $d_{4,3}=6.225$ and $24.656$ $\mu$m, respectively. The particles have a density $\rho_{p}=2700$ kg/m$^3$ and specific heat $c_{v,p} = 1176$ J/kgK.  The CFL number is 0.5 unless stated otherwise. 

Adaptive mesh refinement (AMR) is employed in multidimensional problems using the AMReX library \cite{Zhang2019}. In this work, a cell is tagged for refinement if $\alpha_{p,i,j,k}>0$ or if
\begin{equation}
  \begin{gathered}
      \frac{1}{p_{g,i,j,k}}\max\left( \left| p_{g,i+1,j,k}-p_{g,i-1,j,k} \right|,\left| p_{g,i,j+1,k}-p_{g,i,j-1,k} \right|,\left| p_{g,i,j,k+1}-p_{g,i,j,k-1} \right| \right) > 0.5, \\
      \frac{1}{\rho_{\textrm{mix},i,j,k}}\max\left( \left| \rho_{\textrm{mix},i+1,j,k}-\rho_{\textrm{mix},i-1,j,k} \right|,\left| \rho_{\textrm{mix},i,j+1,k}-\rho_{\textrm{mix},i,j-1,k} \right|,\left| \rho_{\textrm{mix},i,j,k+1}-\rho_{\textrm{mix},i,j,k-1} \right| \right) > 0.1, 
  \end{gathered}
\end{equation}
where $\rho_{\textrm{mix}} = \alpha_{g}\rho_{g}+\alpha_{p}\rho_{p}$ is the mixture density.

\subsection{Advection of a Particle Curtain}
This test is to examine the ability of the algorithm to preserve the non-disturbance condition across sharp phase discontinuities. A particle curtain of H-10 particles with a volume fraction of $\alpha_{p}=40\%$ is advected in a bath of pure helium ($Y_{He}=1$) with uniform pressure, velocity, and temperature of 1 atm, 100 m/s, and 300 K, respectively. The particle volume fraction is $\alpha_p = 0.4$ if $0.4 \text{ (m)}< x < 0.6 \text{ (m)}$ and $\alpha_p = 0$ elsewhere. The domain is 2 m in length and is discretized with 400 cells. Inflow and outflow boundary conditions are used on the left and right sides, respectively.  All terms in the model including drag, convective heat transfer, and particle collisions were included in this test to make it more stringent. 

\begin{figure}[tbp]
    \centering
    \includegraphics[width=.4\textwidth]{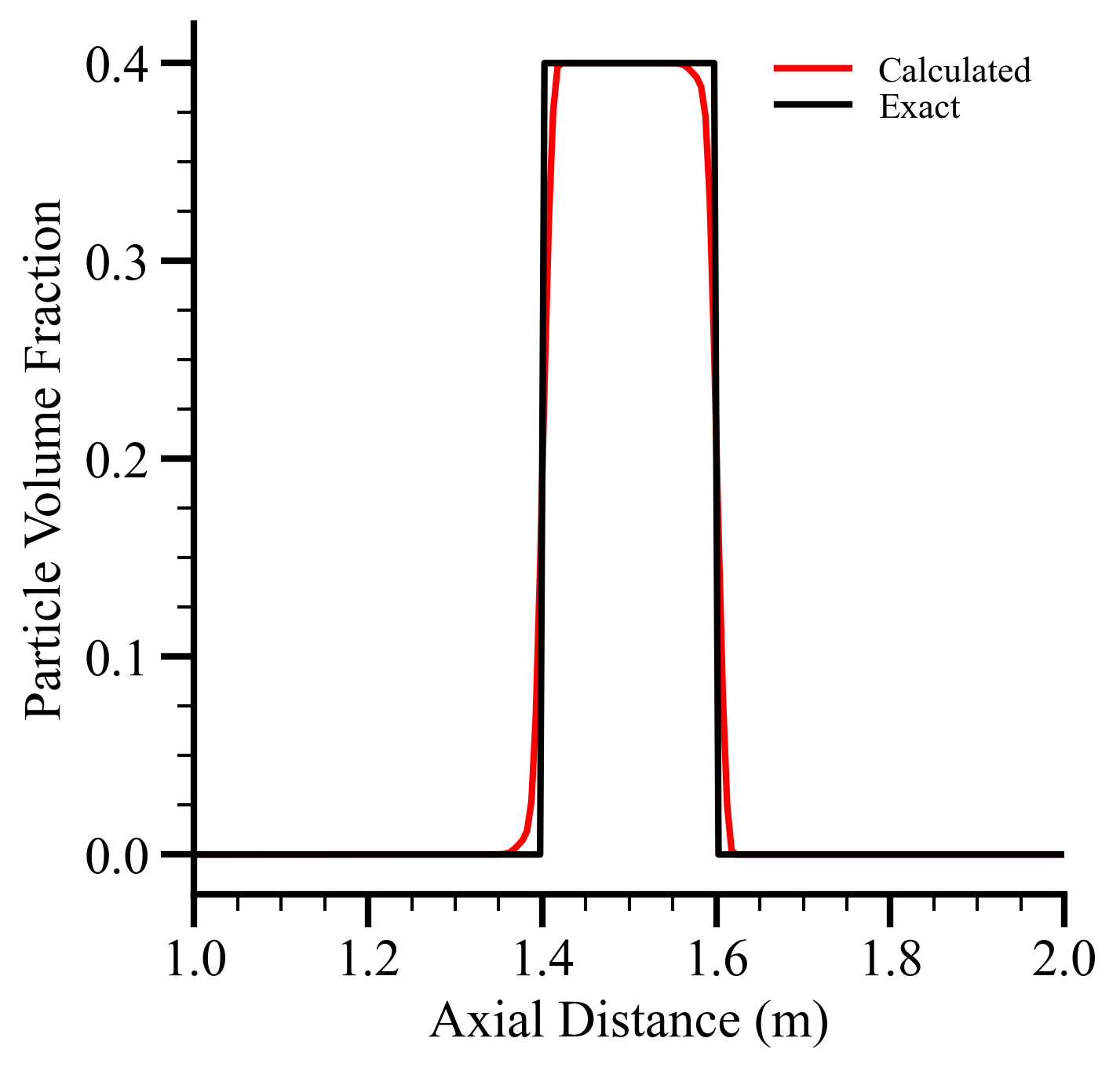}
    \includegraphics[width=.45662228\textwidth]{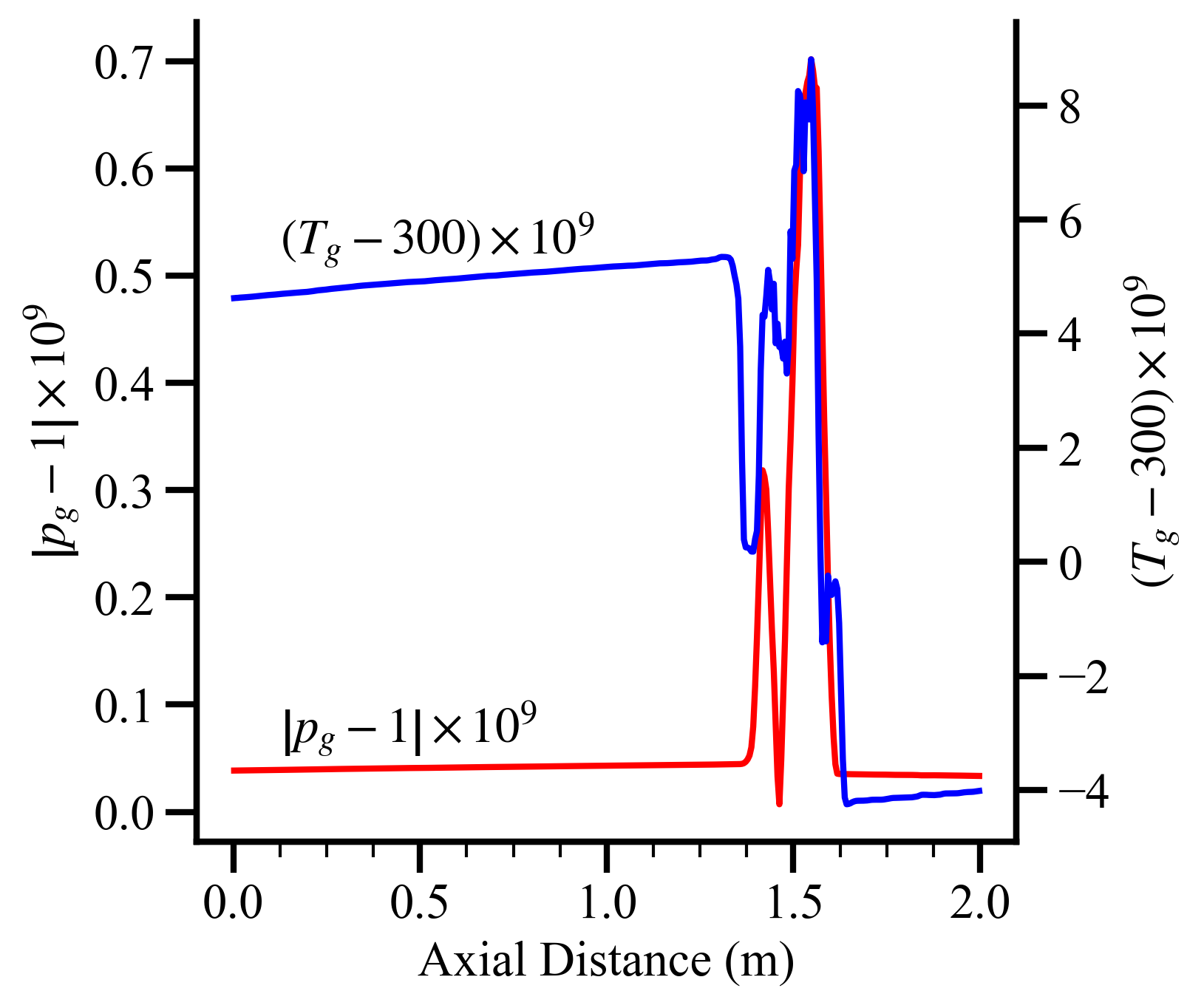}
    \caption{Results of the linear advection test at 0.01 s showing (left) calculated and exact particle volume fraction profile and (right) error in gas pressure and temperature over entire domain.} \label{advec}
\end{figure}

Ideally, the curtain should advect without diffusion of the sharp discontinuity on either side and without disturbing the constant pressure and temperature fields. The computed solution after the curtain has been advected  a distance of 1~m (shown in Fig.~\ref{advec}) shows minimally diffused edges due to the high-order spatial algorithm. The error in the pressure and temperature fields are $\mathcal{O}(10^{-9})$.  This is an acceptable amount of error acceptable given the inherent numerical tolerances in solvers related to the iterative eigensystem solver used in moment inversion, Pade approximants used in integrating drag and convective heat transfer, and convergence criterion for the stiff ODE solvers used to integrate particle collisions.

\subsection{Dilute Polydisperse Shock Tube}
We test the model in the dilute and nearly collisionless regime where drag forces dominate the particle-phase dynamics \cite{Miura1982,Houim2016,Saito2003}. A one-dimensional multiphase shock tube of length $L = 0.257798$ m. A diaphragm, placed at $x = 0.129$ m, separates high-pressure air without particles and low-pressure air with particles with the H-10 size distribution. The initial conditions on both sides of the diaphragm are
\begin{gather}
\begin{matrix}
     \text{Left Side} \ \ \ & \text{Right Side} \\
     p_g^L = 10 \text{ atm} \ \ \ & p_g^R = 1 \text{ atm} \\
     T_g^L = 300 \text{ K} \ \ \ & T_g^R = 300 \text{ K} \\
     u_g^L = 0 \text{ m/s} \ \ \ & u_g^R = 0 \text{ m/s}\\
     Y_{O2}^L = 0.233 \ \ \ & Y_{O2}^R = 0.233 \\
     Y_{N2}^L = 0.767 \ \ \ & Y_{N2}^R = 0.767 \\
      \alpha_p^L = 0 \ \ \ &\alpha_p^R = 4.825\times 10^{-4} \\
     T_p^L = 300 \text{ K} \ \ \ &T_p^R = 300 \text{ K} \\
     u_p^L = 0 \text{ m/s} \ \ \ &u_p^R = 0 \text{ m/s}   
\end{matrix}
\end{gather}
The total number of cells was 400, corresponding to a grid spacing of $\Delta x\approx 644.5$ $\mu$m. Symmetry conditions are used on the left and right boundary conditions.

\begin{figure}[tbp]
    \centering
    \includegraphics[width=.32\textwidth]{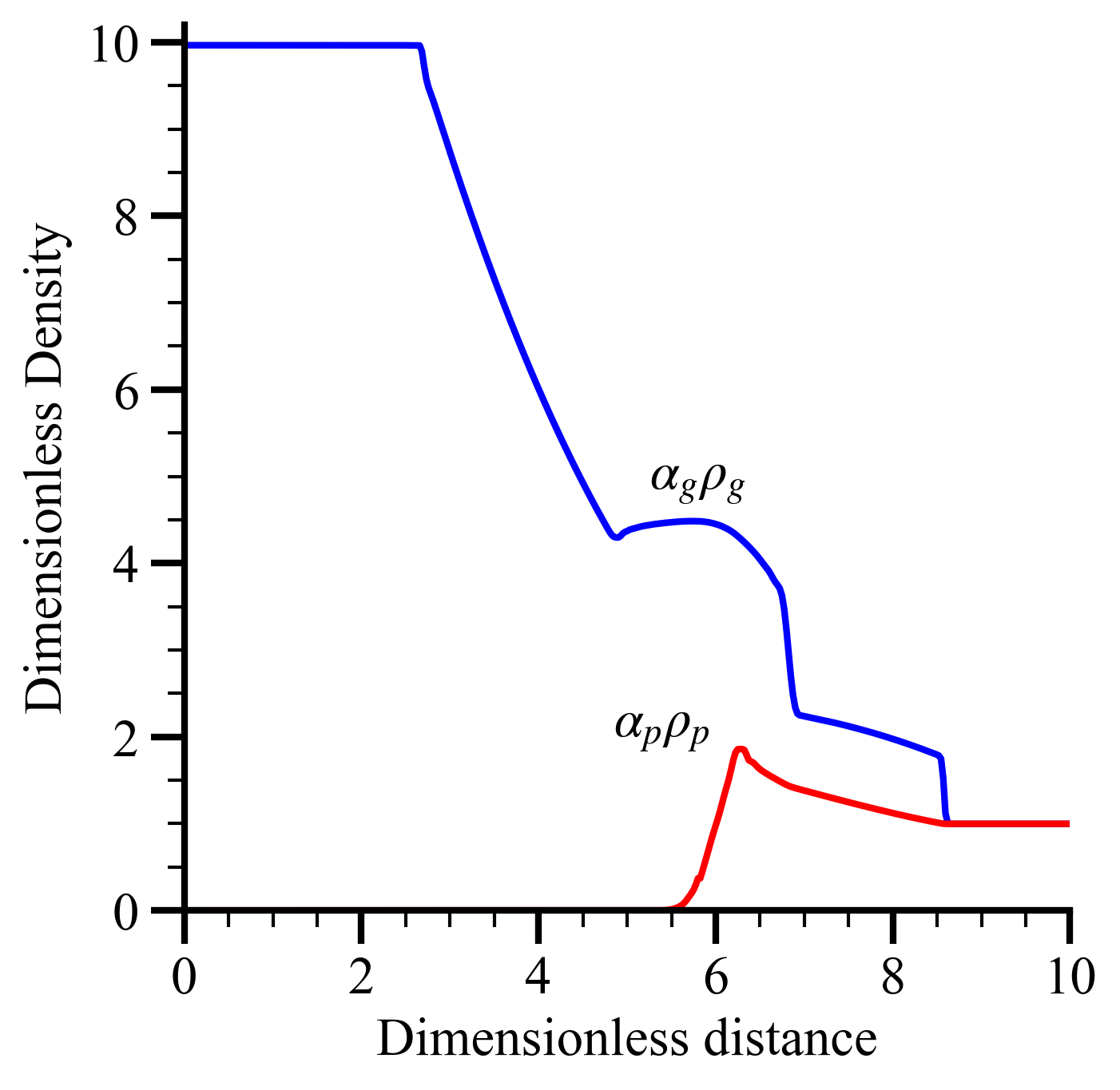}
    \includegraphics[width=.32\textwidth]{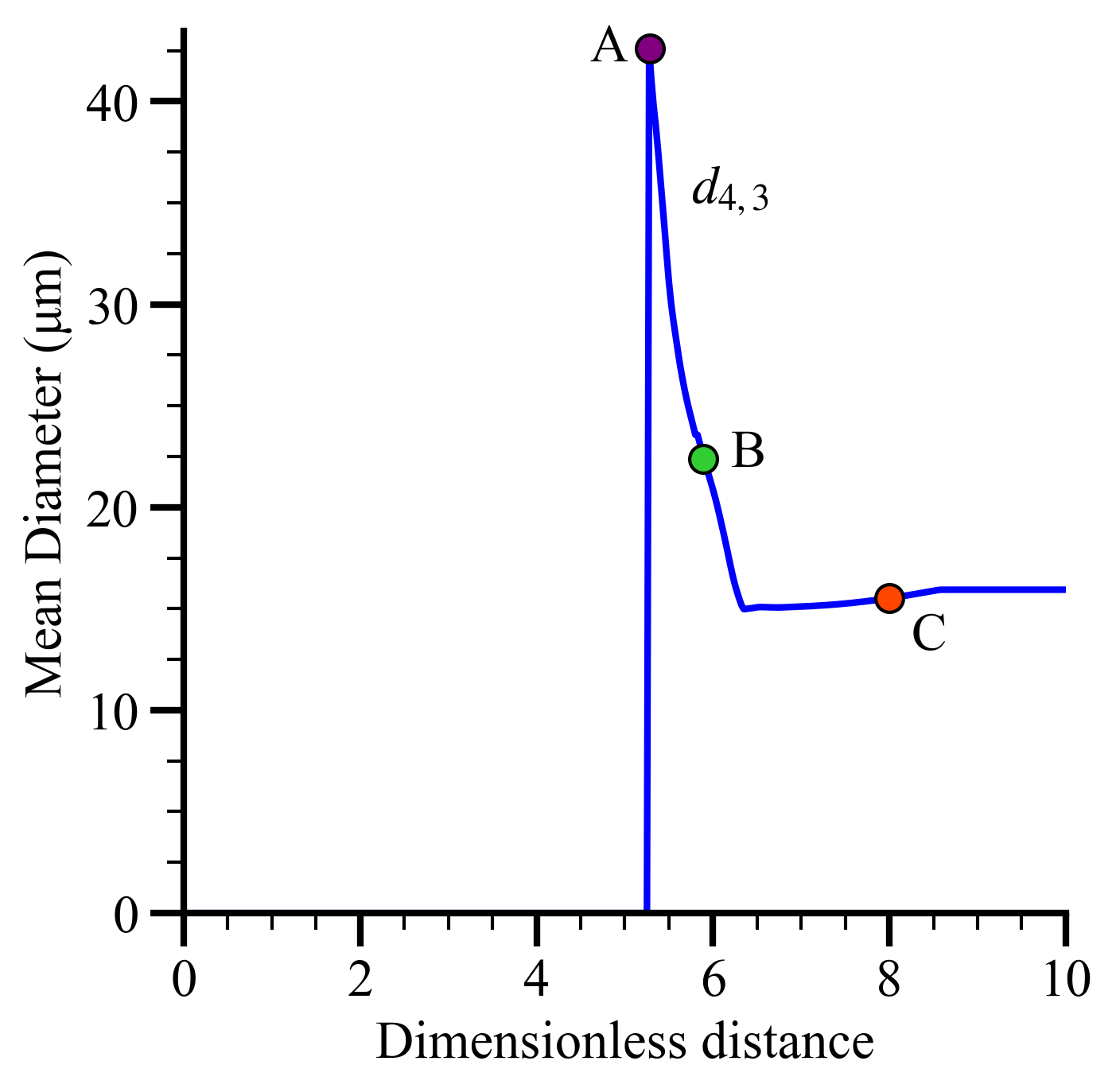}
    \includegraphics[width=.31042053\textwidth]{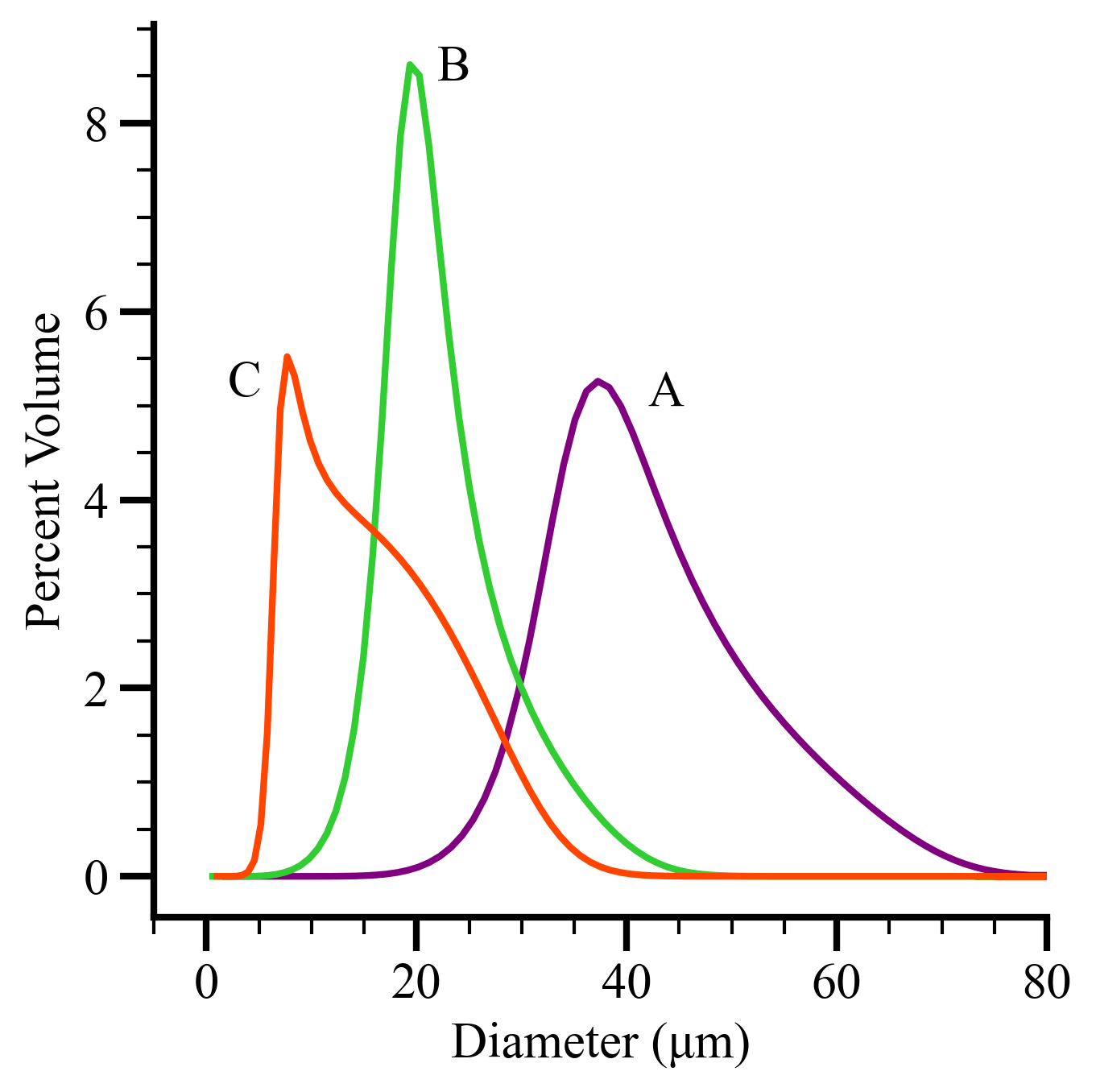}
    \caption{Computed results of the dilute two-phase shock tube problem at $t=184$ s showing (left) bulk densities for the gas ($\alpha_g \rho_g$) and particle phases ($\alpha_p \rho_p$), (center) volume-averaged mean diameter of the particles, and (right) particle size distributions ($\beta(m)=m f(m)/\mathcal{M}_1$) reconstructed by beta-GQMOM at the points labeled in the center $d_{4,3}$ plot.} \label{TubeResults}
\end{figure}

\begin{figure}[tbp]
    \centering
    \includegraphics[width=.45\textwidth]{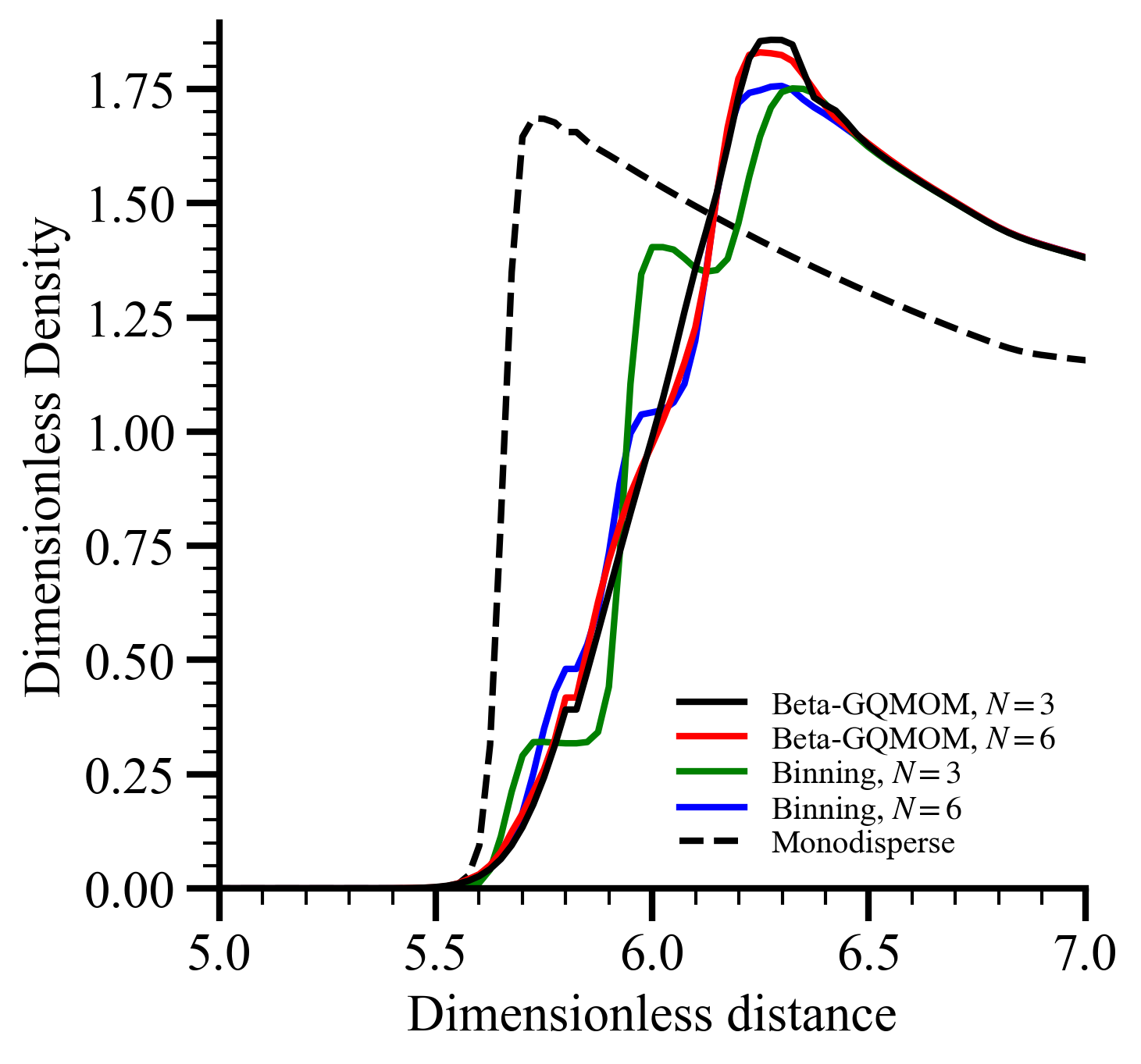}
    \includegraphics[width=.45\textwidth]{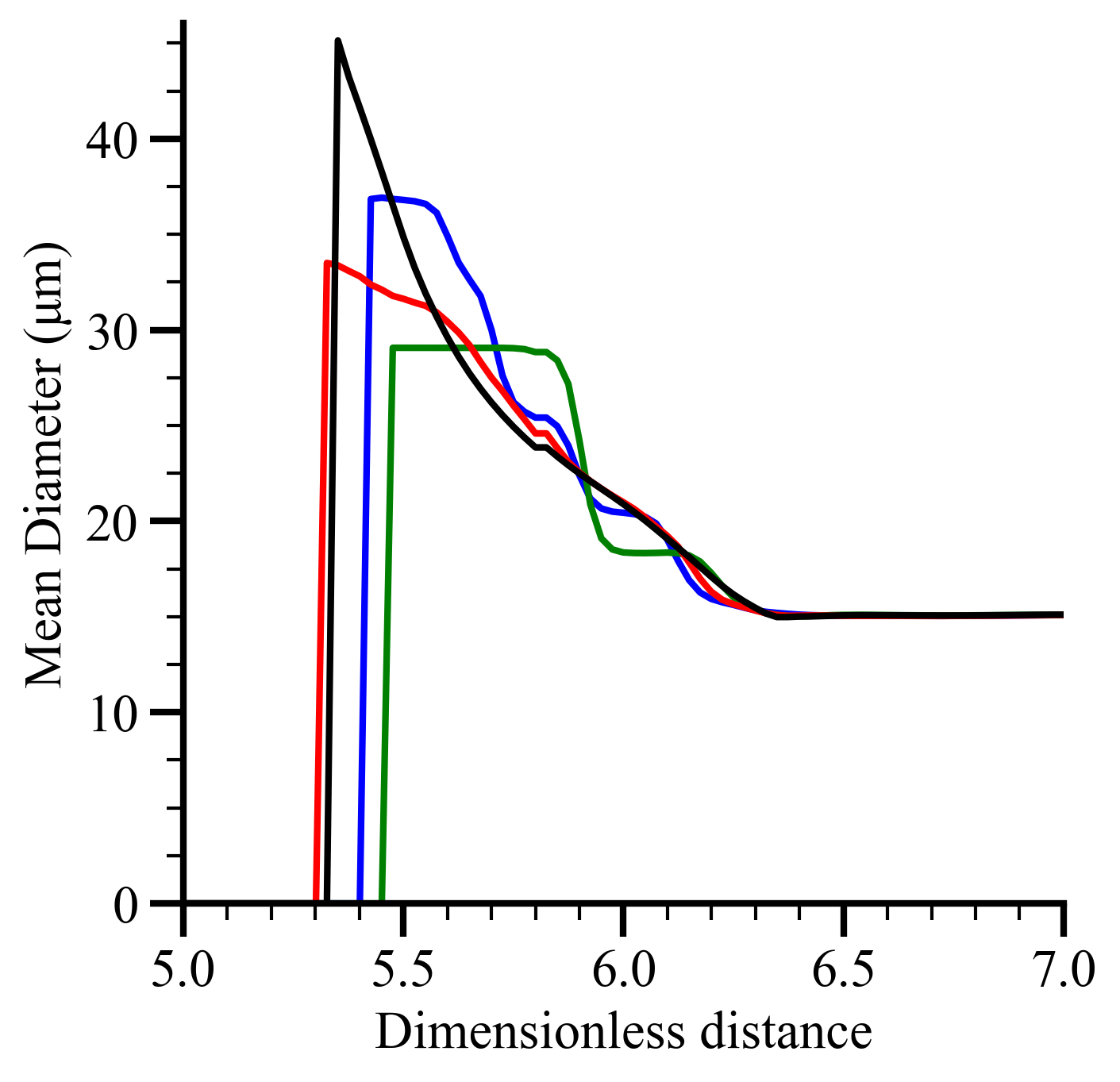}
    \caption{Computed results of the dilute two-phase shock tube problem at $t=184$ s for different moment inversion algorithms and $N$ showing (left) particle bulk density ($\alpha_p \rho_p$) and (right) volume-averaged mean diameter of the particles.} \label{TubeResults2}
\end{figure}

Figure~\ref{TubeResults} shows the computed of the polydisperse shock tube problem solution at $t=184$ $\mu$s.  The length scale is non-dimensionalized by  $X=10x/L$ such that the domain scales between 0 and 10 \cite{Fedorov2010,Houim2016}.

The burst of the diaphragm sends a shock wave into right side of the shock tube where the particle and low-pressure air mixture is located.  The high velocity gases behind the shock wave accelerate the particles due to drag.  A spatial gradient of the mean particle velocity is produced behind the shock from the drag forces having more time to accelerate the particles that are further away from the shock.  This particle velocity gradient causes the gradual increase in bulk density behind the shock.  

The drag time scales are much shorter for the smaller particles than they are for the larger particles in a polydisperse mixture.  As a result, smallest particles are the first to start moving behind the shock.  Thus, the increase of volume fraction immediately behind the shock is from the smaller particles in the mixture.  This causes the volume-averaged mean particle diameter, $d_{4,3}$, to decrease as well as the low-diameter spike in  particle size distribution close to the shock at point C.  

The sharp increase in $d_{4,3}$ and volume fraction peak at $X=6.5$ corresponds to the  location where all of the smallest particles have all moved downstream. Thus, particle mixtures to the left of $X=6.5$ do not contain small particles.  This causes mean diameter and particle size distribution shift towards larger and larger particles at points B and A.  The volume fraction decreases gradually to the left of $X=6.5$ to no particles at $X=5.5$ due to size segregation effects where larger particles require more distance behind the shock to accelerate and move upstream. 

The gas-phase bulk densities ($\alpha_g \rho_g$ and $\alpha_p \rho_p$) shown in Fig.~\ref{TubeResults} closely match the monodisperse results from \cite{Houim2016}.  An exception is that the monodisperse results show a much sharper edge of $\alpha_p \rho_p$. The monodisperse results \cite{Houim2016} do not have the effect where smaller particles leave the mixture first while larger particles stay behind.  Instead, all of the particles move together in a given location.  This causes the edge of the particle volume fraction to be very sharp if a monodisperse model is used.

The simulation is performed with the binning approach to moment inversion for $N=3$ and $6$ to demonstrate the beta-GQMOM algorithm's capability to represent a continuous PSD with a modest increase in the number of mass moment PDEs (see Table~\ref{tab:Nmass}). Results with $N=6$ beta-GQMOM and monodisperse (with diameter $24.656$ $\mu$m) are included to compare against extremes. The initial $N=6$ abscissae are 5.7, 9.35, 16.47, 23.79, 31.51, 37.47 $\mu$m in diameter and the corresponding volume fractions are 6.27\%, 34.87\%, 30.7\%, 21.81\%, 5.32\%, 1.03\%.
Zoomed in plots of $\alpha_p\rho_p$ and $d_{4,3}$ are shown in Fig.~\ref{TubeResults2}.
Beta-GQMOM with $N=3$ shows good comparison to $N=6$, the change in upstream mean diameter is attributed to adding larger diameter nodes with the increase in $N$ due to the H-10 distribution's kurtosis and positive skewness.
Given their discrete representation of the PSD, the binning results exhibit noticeable staircasing in bulk density and mean diameter, although binning with $N=6$ better approximates the bulk density predicted by beta-GQMOM.
The monodisperse bulk density is sharp and absent of size segregation effects as expected from previous work \cite{Houim2016}.

\subsection{Shock and Particle-Curtain Interaction}
The previous numerical experiments had very dilute mixtures of particles, and, as a result, collisions were not important. 
Here we perform a moderately dense numerical experiment of a shock wave interacting with a particle curtain of H-10 particles.  
This problem is motivated by the monodisperse shock and particle curtain interaction experiments \cite{Wagner2012,Daniel2022}.  
The shock wave has a strength of Mach 2 and propagates into air at 1 atm and 300 K.  The particle volume fraction in the curtain is ($\alpha_{p}=25\%$), which is a volume fraction regime where effects from collisions can be important.   The curtain is $L=4$ mm in width.  The gas is air at 1 atm and 300 K. The domain is $-75L\leq x \leq 75L$.  The curtain and shock are placed at $x=0$ and $x=2$ cm, respectively. The grid resolution is 40 cells across the initial curtain width. 

\begin{figure}[tbp]
    \centering
    \includegraphics[width=.6\textwidth]{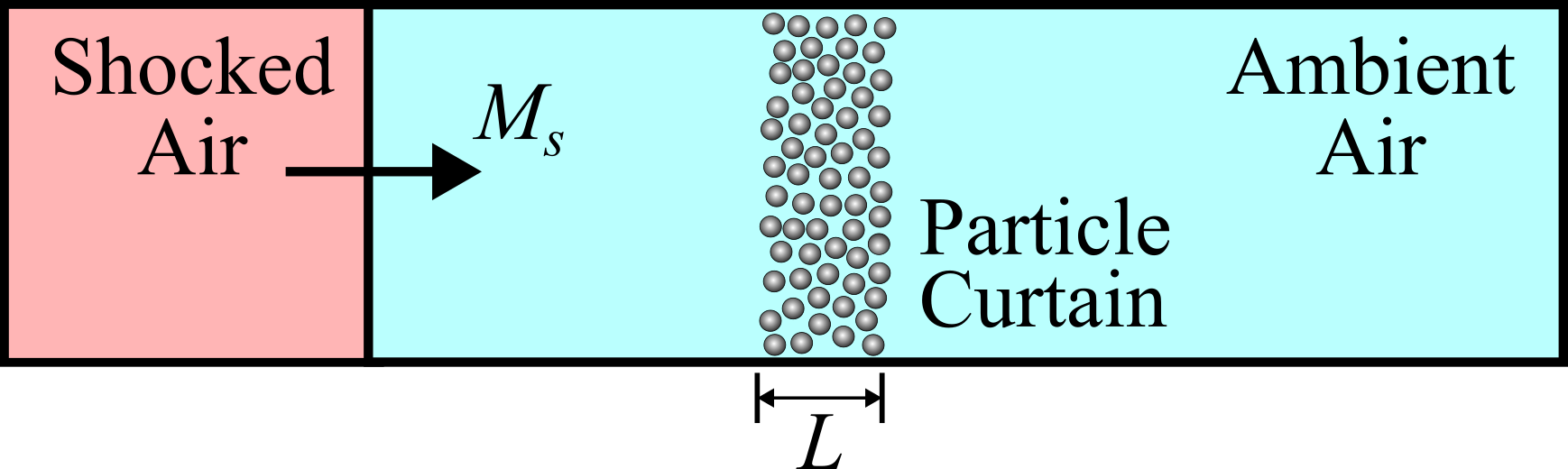}
    \caption{Initial conditions for the shock--particle curtain simulation.} \label{SPC}
\end{figure}
\begin{figure}[tbp]
    \centering
    \includegraphics[width=.4\textwidth]{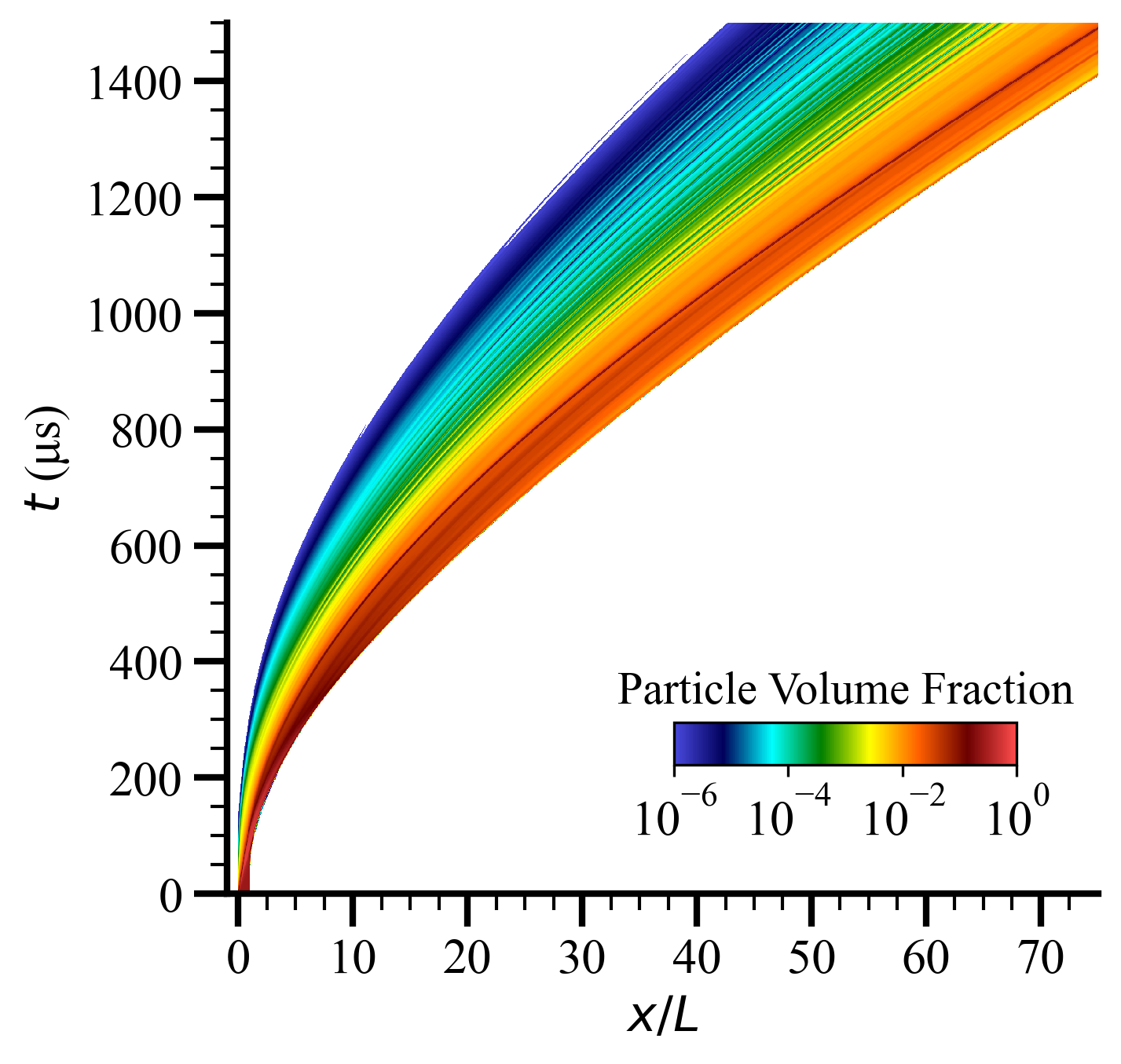}
    \includegraphics[width=.4\textwidth]{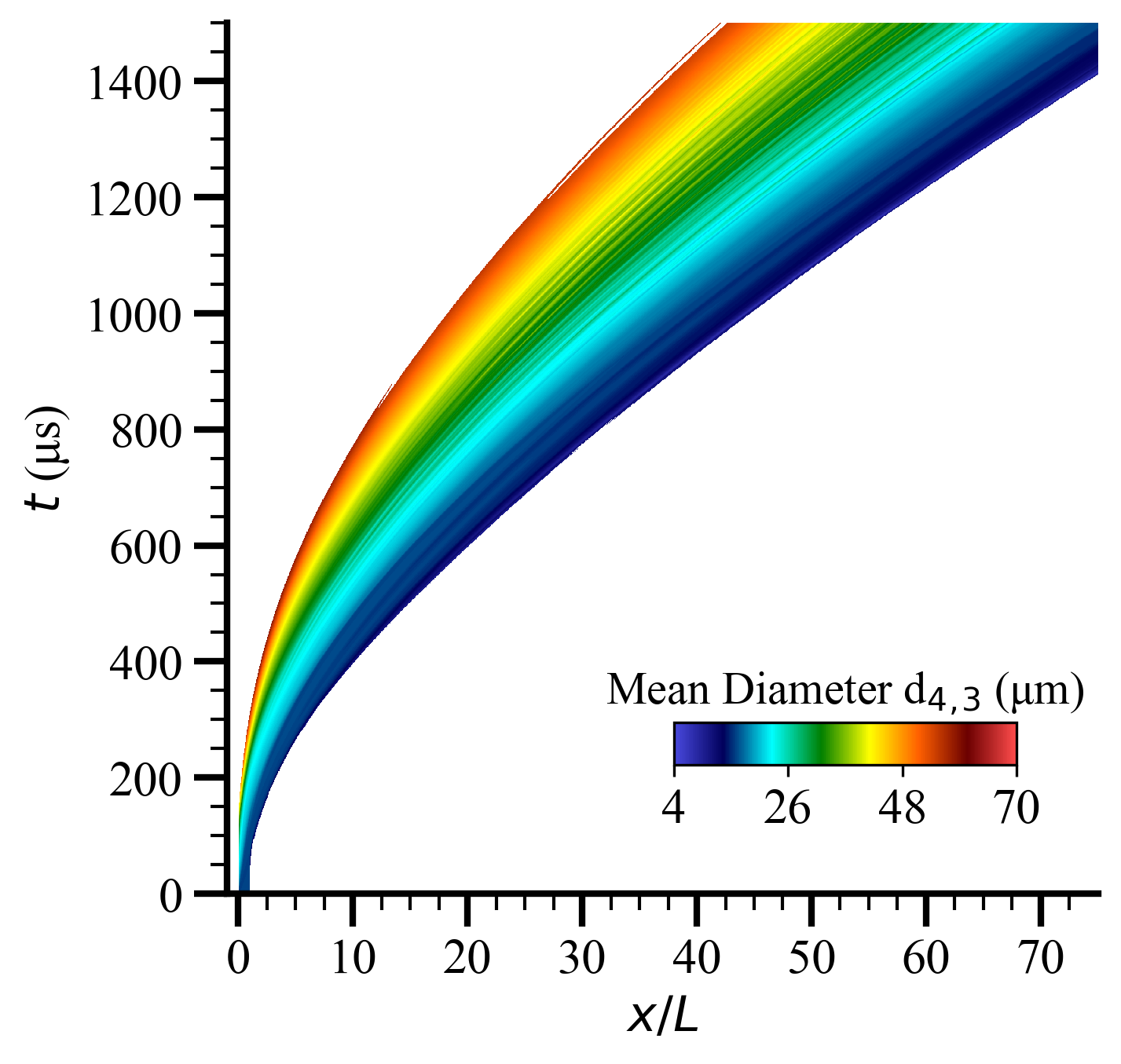}
    \caption{Position-time diagrams of (left) particle volume fraction and (right) volume-averaged mean diameter from the shock-particle-curtain interaction problem at $t=1.5$ ms after shock impact.} \label{SPC_xt}
\end{figure}

Figure~\ref{SPC_xt} shows position-time diagrams of particle volume fraction and $d_{4,3}$.  Here we see a widening of the particle curtain which is typical of shock-particle interactions \cite{Wagner2012,Daniel2022}.  Much like the polydisperse shock tube problem, the polydisperse particle curtain shows effects from size segregation.  The smaller particles accelerate more rapidly behind the shock than the larger particles.  Thus, the smaller particles are transported downstream first while the larger particles lag behind.  This results in a smooth gradient of particle size across the widening particle curtain. 

\subsection{Shock and Dilute Dust Layer Interaction}

\begin{figure}[tbp]
    \centering
    \includegraphics[width=.6\textwidth]{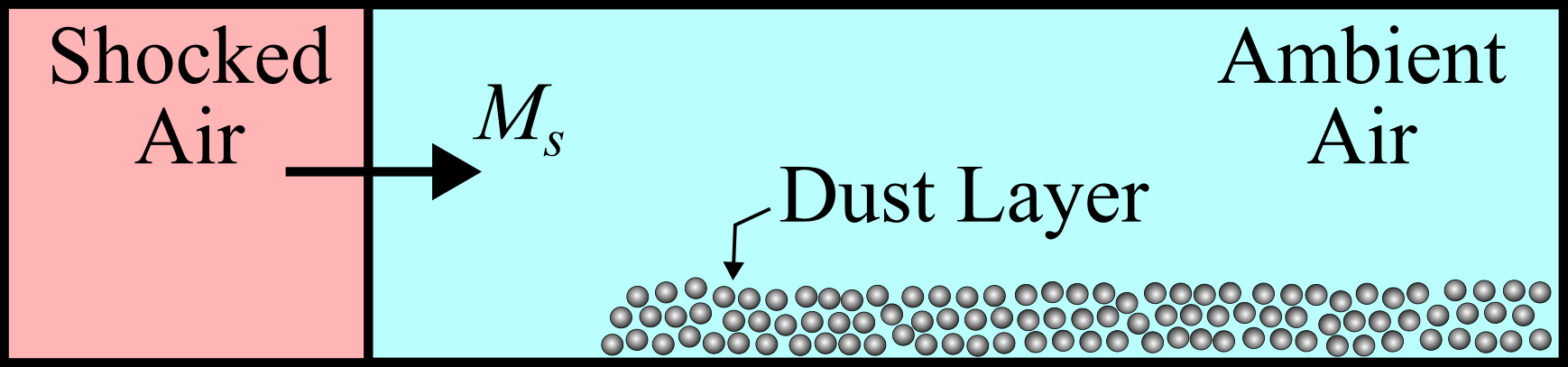}
    \caption{Illustration of the initial conditions for the shock--dust layer simulation. The shock is moving at Mach number $M_{s}$ towards the dust layer at the bottom of the domain.} \label{DustLayer_ICs}
\end{figure}

This test examines the interaction between a Mach 1.6 shock and a dilute dust layer particles \cite{Houim2016,Fedorov2010}. Figure~\ref{DustLayer_ICs} shows a sketch of the initial conditions for this problem. A Mach 1.6 shock propagates into air at 1 atm and 288 K and interacts with a dust layer.  The dust layer contains particles with the H-2 size distribution at a volume fraction of 0.04\%.  The height and length of the two-dimensional channel are 6.4 cm and 51.2 cm, respectively.  The left edge of dust layer and shock are located at 2 cm and 1 cm, respectively.  The height of the dust layer is 2 cm.  Adaptive mesh refinement is used with 2 levels of refinement with a corresponding grid spacing of 1 mm at the finest level.  
The left and right boundaries are non-reflecting and symmetry conditions are used on the top and bottom. 
\begin{figure}[tbp]
    \centering
    \includegraphics[width=.9\textwidth]{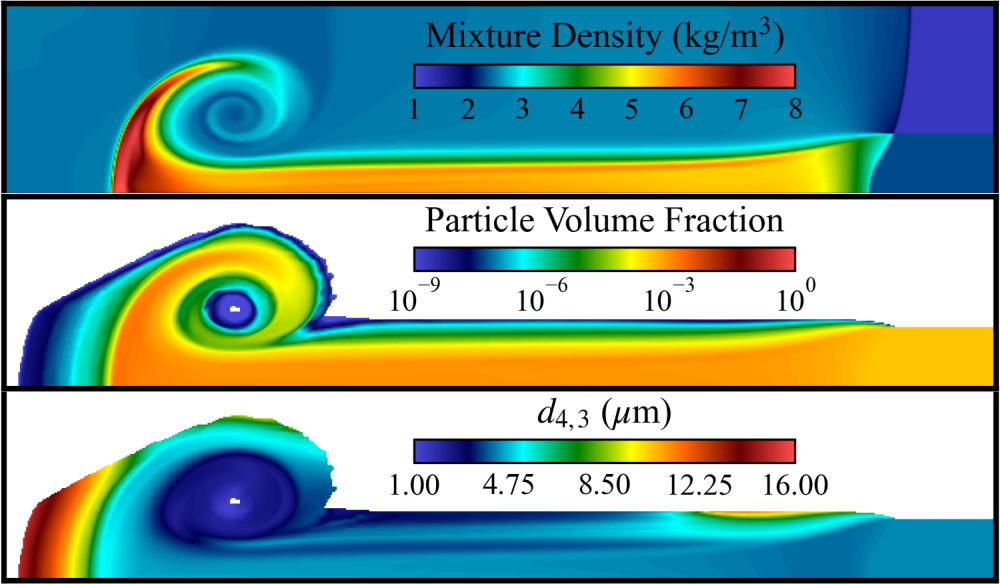}
    \caption{Simulation results of a shock-dilute dust layer interaction with H-2 aluminum particles at $t=900$ $\mu$s showing (top) the mixture density profile, (middle) aluminum particle volume fraction, and (bottom) volume-averaged particle diameter.} \label{H2_fig}
\end{figure}

Figure~\ref{H2_fig} shows the mixture density $\rho_{\text{mix}} = \alpha_g \rho_g + \alpha_p \rho_p$, particle volume fraction, and $d_{4,3}$ at $t=900$ $\mu$s. 
A large-scale vortex is produced at the leading edge of the dust layer due to the shear layer that forms between the fast-moving air above the dust layer and slower-moving air within the dust layer.  This vortex entrains causes the leading edge of the dust layer to roll up.  Earlier results using monodisperse models show two distinct vortices; one for the gas and the second particle \cite{Houim2016}.  The polydisperse model only shows one vortex that is more diffused.  This due to the size distribution where the particles have a range of velocity relaxation times where each particle size in the distribution follows its own trajectory, which blurs the gas and particle vortices. Size segregation is observed on the upstream edge of the particle layer due to the shock accelerating smaller particles first, much like the polydisperse shock tube and shock-particle curtain interaction problems.  Size segregation can also be observed near the shock on the top of the dust layer.  This is caused by high-pressure gases behind the shock infiltrating into the top of the dust layer.  This compacts the dust layer by preferentially dragging smaller particles from the top to the inside of the dust layer.

\subsection{Shock and Dense Dust Pile Interaction}
The interaction of a strong Mach 3 shock in air with a dense square-edge layer of dust is computed to demonstrate the capabilities of the model and solution algorithm.  The challenging problem contains all of the features of the previous problems, but also contains dense particle mixtures that approach the packing limit, strong shocks, and sharp granular interfaces.  Figure~\ref{DustLayer_ICs} shows the problem setup.  The initial volume fraction of the dust layer is 47\% and is comprised of H-2 particles.  The dust layer is 6.67 mm in height.  The Mach 3 shock propagates into air at 1 atm and 300 K.  The height and width of the domain are 10 cm and 50 cm, respectively.  The shock and left edge of the dust layer is initially located at $x$-locations of 19 cm and 20 cm, respectively.  Three levels of refinement are used corresponding to a grid spacing of 195 $\mu$m at the finest level.

\begin{figure}[tbp]
    \centering
    \includegraphics[width=.95\textwidth]{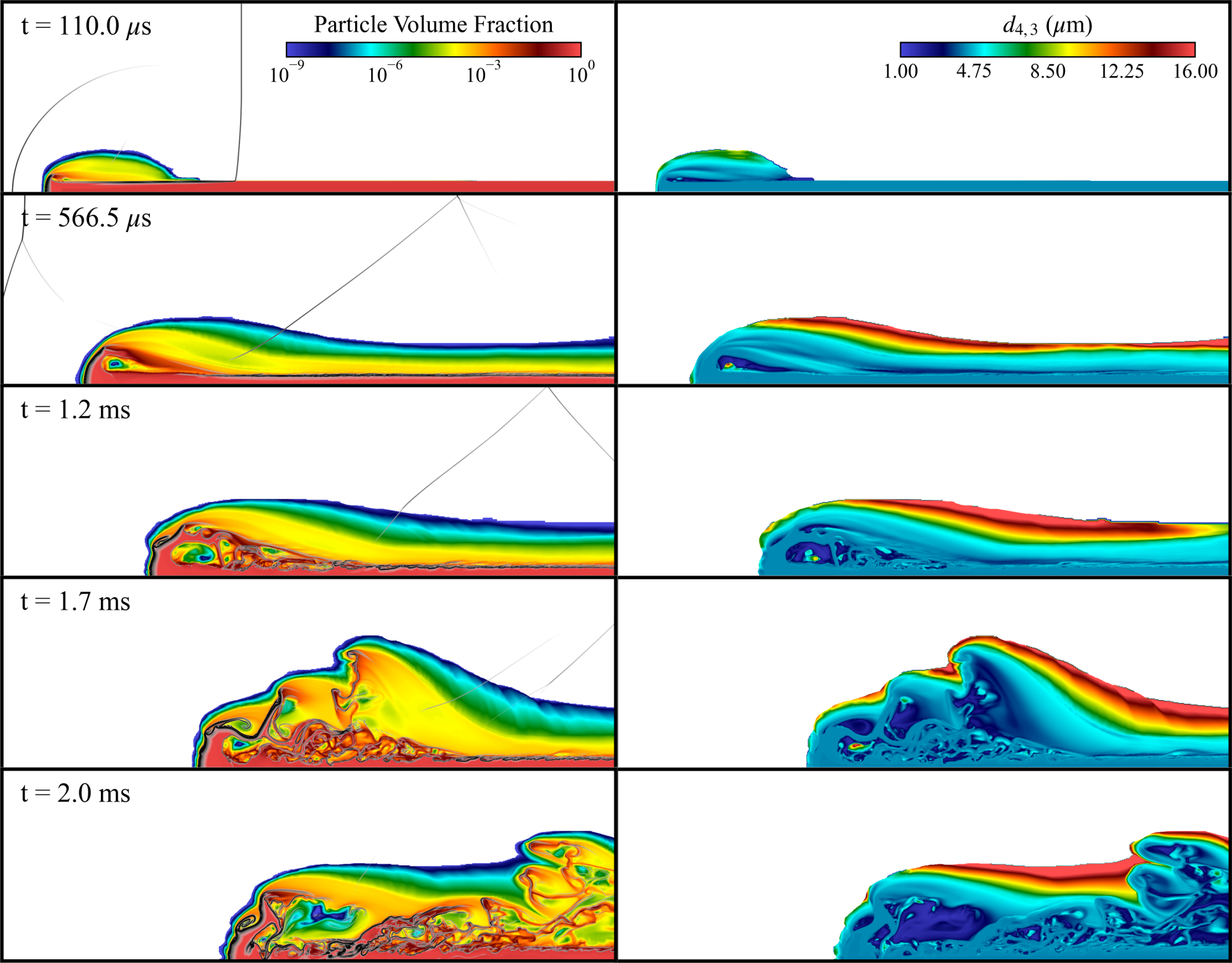}
    \caption{Time sequence of (left) particle volume fraction overlaid with a numerical schlieren and (right) number-averaged mean diameter for the shock-dust pile interaction simulation.} \label{DensePile_fig}
\end{figure}

Figure ~\ref{DensePile_fig} shows the evolution of the particle volume fraction and $d_{4,3}$ fields at selected times.  The initial interaction between the shock and dust layer produces intense forces acting on the particles due to drag and gas-phase pressure gradients  pushing on the leading edge of the dust layer.  This rapidly slows the gas in the dust layer and produces a strong reflected shock.  The post-shock gases flow around the leading edge of dust layer and entrain some of the particles.  The intense forces pushing on the leading edge dust layer pack the particles close to the packing limit.  As a result, the particle grains are ``locked'' together and the particles move together.  This can be seen by examining $d_{4,3}$ and $\alpha_p$ at 568 $\mu$s where the particle vortex is highly packed.  The mean size of the particles within the dust layer and in the dense portion of the particle vortex are nearly identical.  Aerodynamic lifting and drag forces strip some of the particles from the edge regions with dense particle mixtures.  These particles are entrained into a dilute cloud above the dust layer.  Similar to the other test problems aerodynamic drag forces segregate the particles by size.  At later times granular clusters (also called particle streamers)  are formed as a result of the nonlinearity in the drag model and inelastic particle collisions reducing the solids pressure where the particle volume fraction is higher \cite{Helland2000,Agrawal2001}.

\subsection{Shock-Induced Polydisperse Dust Layer Lifting}

\begin{figure}
    \centering
    \includegraphics[width=0.75\linewidth]{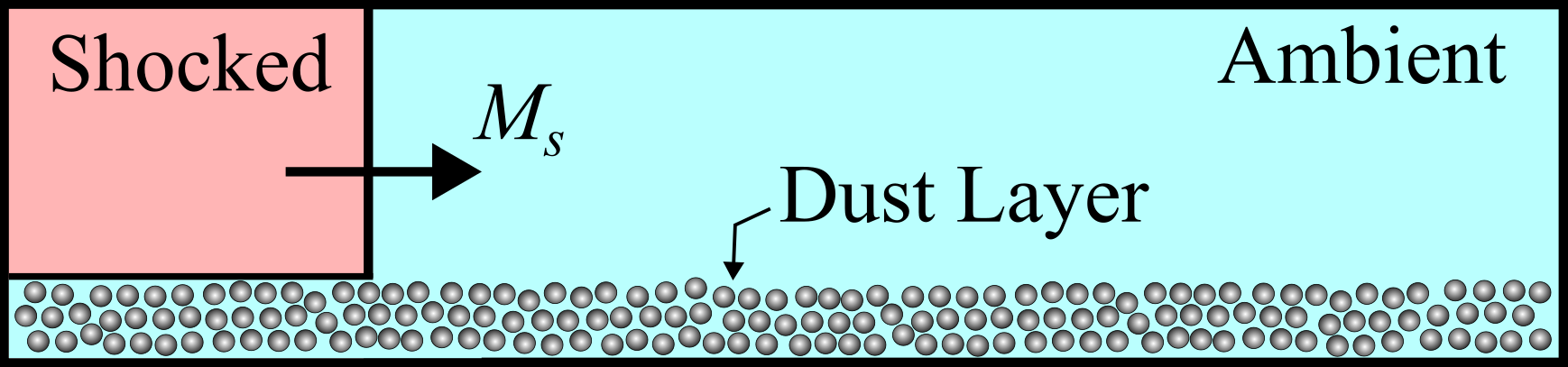} 
    \caption{Geometric setup for shock-induced lifting of dust layers.}
    \label{fig:DustLiftCartoon}
\end{figure}

\begin{figure}
    \centering
    \includegraphics[width=0.4\linewidth]{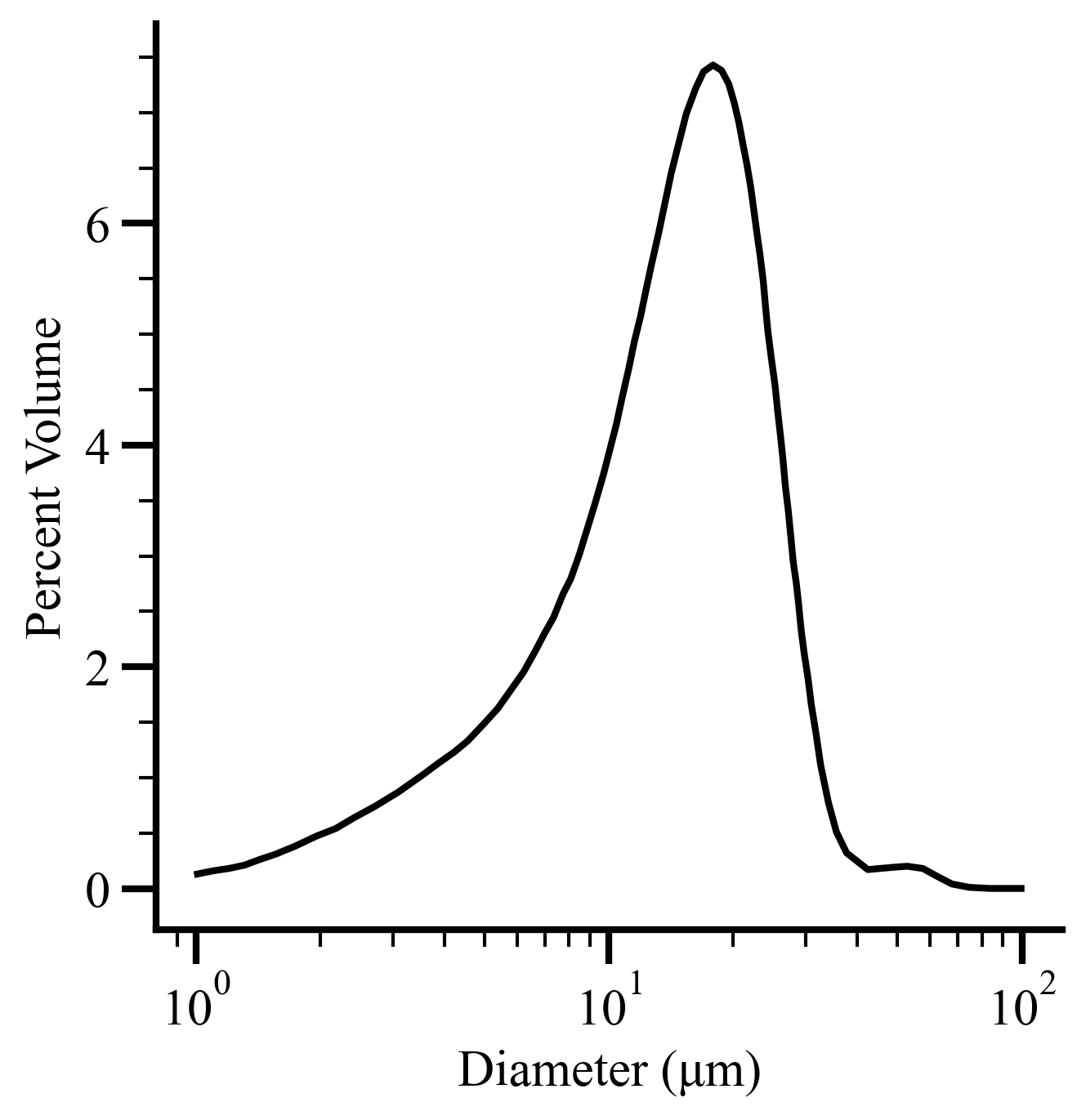} 
    \caption{Aluminum particle size distribution used in experiments of polydisperse dust-layer lifting \cite{Chowdhury2018}.}
    \label{fig:PSDChowd}
\end{figure}

\begin{figure}
    \centering
    \includegraphics[width=\linewidth]{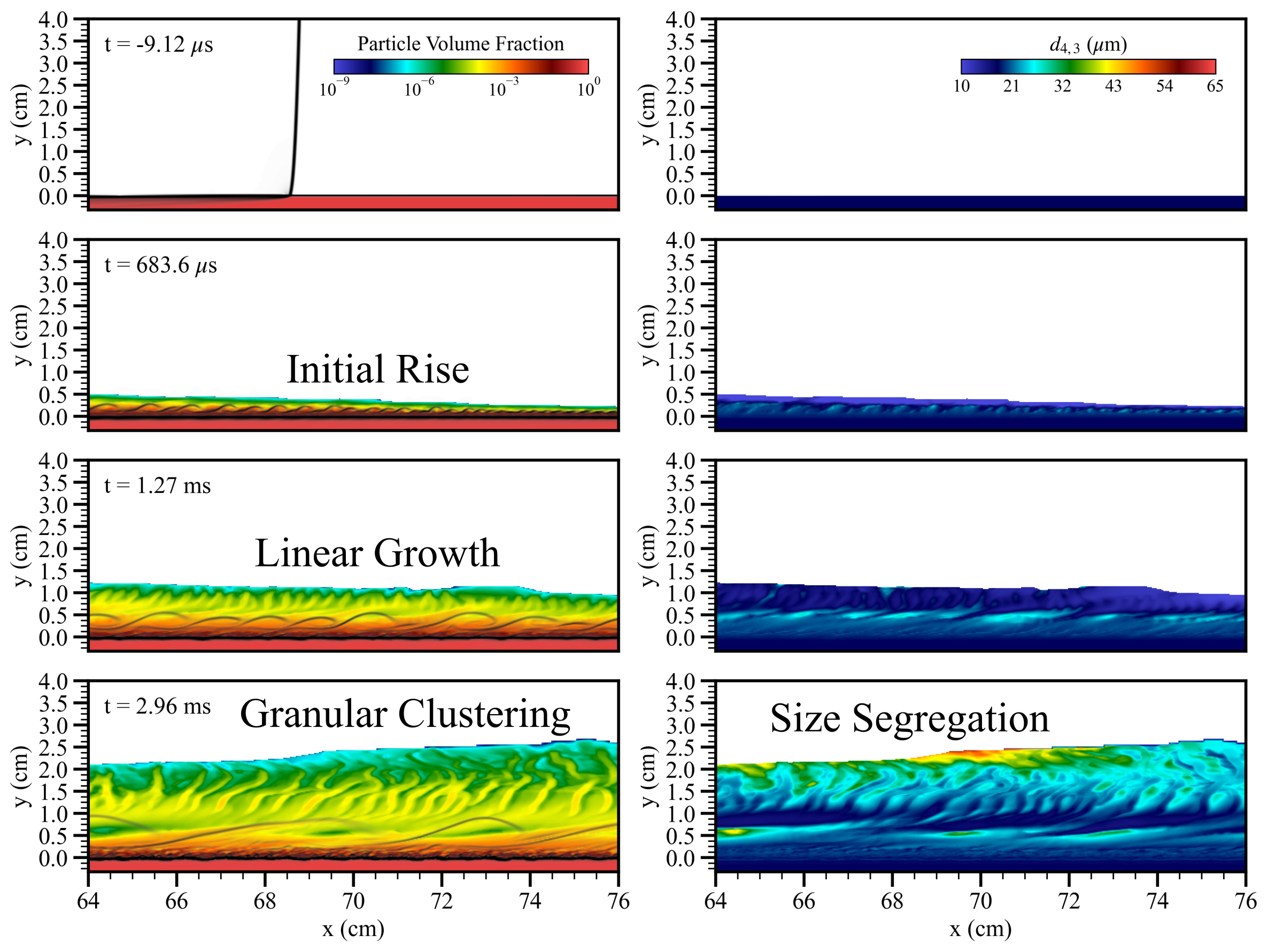} 
    \caption{Computed results of (left column) particle volume fraction overlaid with a numerical schlieren and (right column) $d_{4,3}$ of the particles as selected times.}
    \label{fig:TamuResults}
\end{figure}

\begin{figure}
    \centering
    \includegraphics[width=0.5\linewidth]{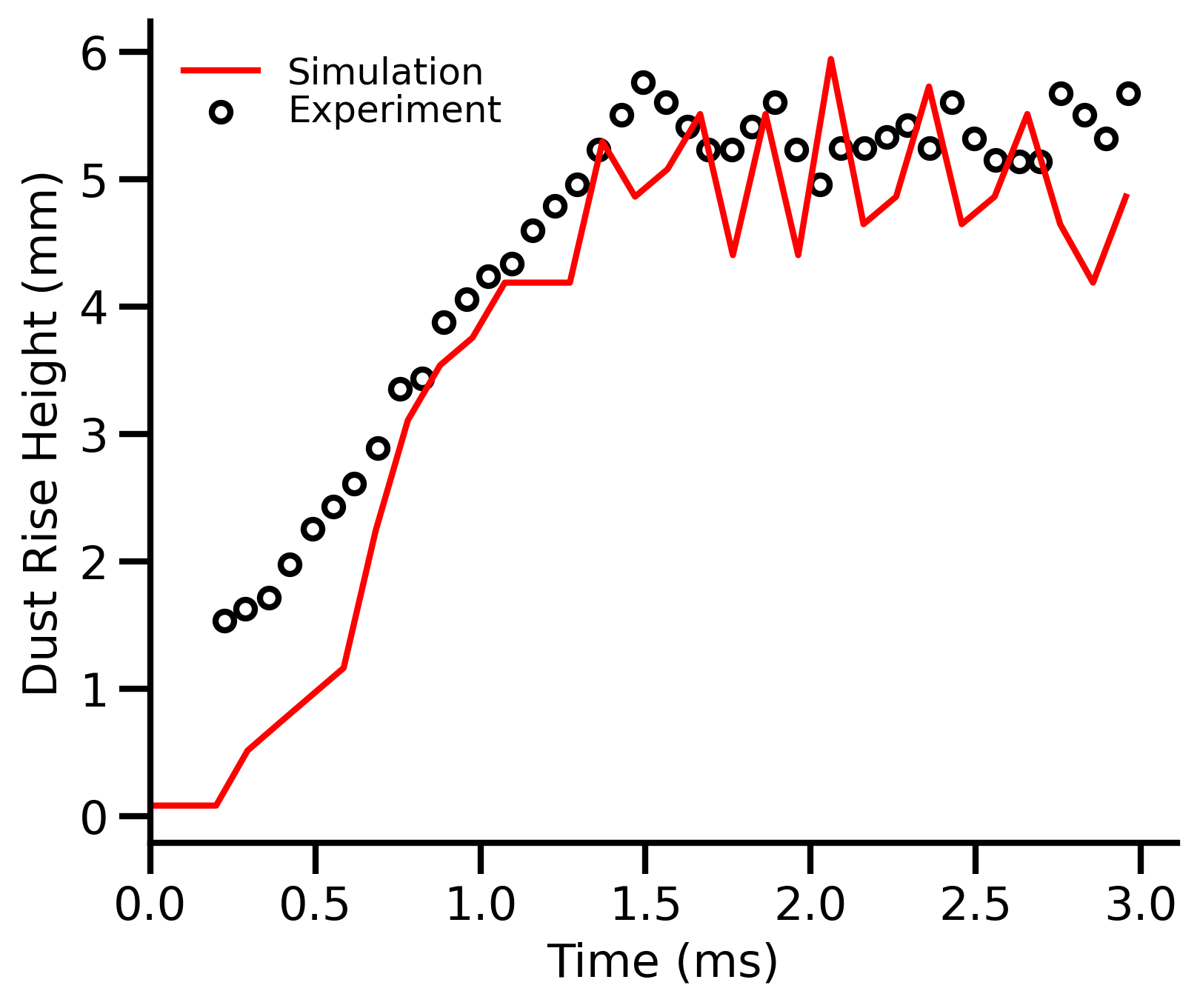} 
    \caption{Comparison between simulation results and experimental measurements \cite{Chowdhury2018} of the dust rise height as a function of time behind a Mach 1.53 shock traveling over a polydisperse dust layer.}
    \label{fig:TamuHeight}
\end{figure}

Here we simulate shock-induced dust-lifting experiments \cite{Chowdhury2018}. Figure~\ref{fig:DustLiftCartoon} shows the geometrical configuration for the simulation.  A Mach 1.53 shock in pure N$_2$ propagates over a layer of aluminum particles.   
The dust layer is 3.2~mm in height and has an initial volume fraction of 55.6\%.  Figure~\ref{fig:PSDChowd} shows the particle size distribution used in the experiments \cite{Chowdhury2018}. The nitrogen downstream of the shock is at 67 kPa and 295 K. The top and bottom boundaries are symmetry planes.  Non-reflecting conditions are used on the left and right boundaries.   The computational domain is 89 cm wide and 11.1 cm high. Adaptive mesh refinement is used with five levels of refinement that correspond to a grid spacing of 217 $\mu$m at the finest level. The value of $\alpha_{p,\text{crit}}$ used in the friction model cannot be below the initial packing of the dust layer.  The value of $\alpha_{p,\text{crit}}$ was increased to 56\% so $p_{p,f}$ is zero at the initial layer volume fraction.

Figure~\ref{fig:TamuResults}  shows a time sequence of particle volume fraction and mean diameter $d_{4,3}$. 
The dust initially rises behind the propagating shock due to a combination of aerodynamic lifting forces and particle pressure \cite{Ugarte2017,binning2}. The lofted dust continues to rise due to inertia.  Eventually, the lofted particles stop rising due to drag forces.  The results show size segregation where the smallest particles are trapped in the boundary-layer like flow immediately above the dust layer due to their low inertia. Larger particles can travel further into the core of the channel before drag forces stop them from rising further.   

Figure~\ref{fig:TamuHeight} shows the dust lifting height as a function of time behind the shock wave.  The dust lifting height is defined by where the lowest contour where $\alpha_p = 1.5\times 10^{-4}$.  Overall, the simulation results capture the experimental results relatively well.  The initial linear growth of the dust layer at $t=0$ is slightly underpredicted, but the later-time linear growth and mean plateau height of the lofted dust agree well with the data.

\subsection{Spherical Dispersal of a Dense Polydisperse Particle Bed}
\begin{figure}[tbp]
    \centering
    \includegraphics[width=.4\textwidth]{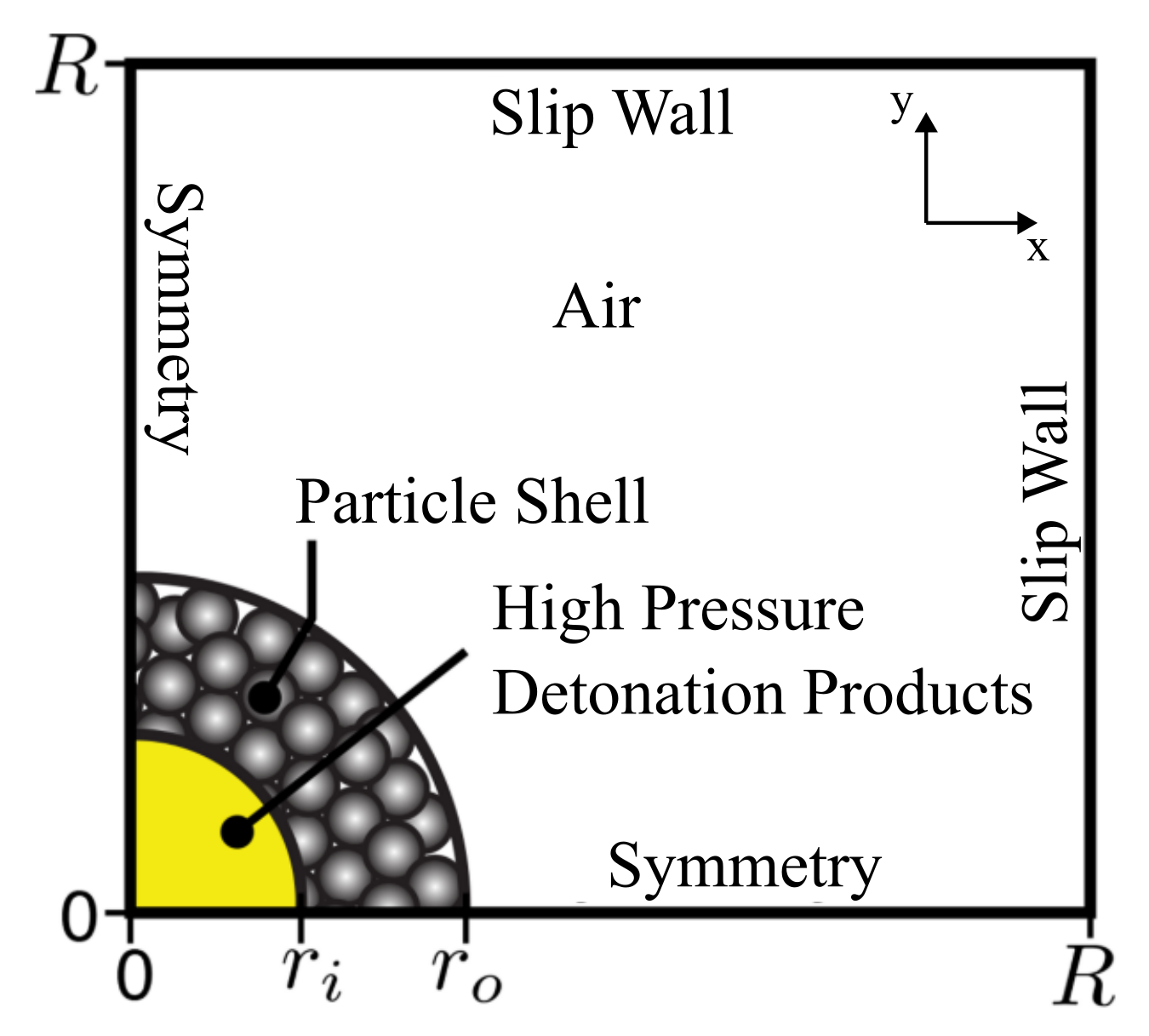}
    \caption{Illustration of the initial conditions for the inert dense particle shell dispersal problem.} \label{ICs_blast}
\end{figure}
\begin{figure}[tbp]
    \centering
    \includegraphics[width=\textwidth]{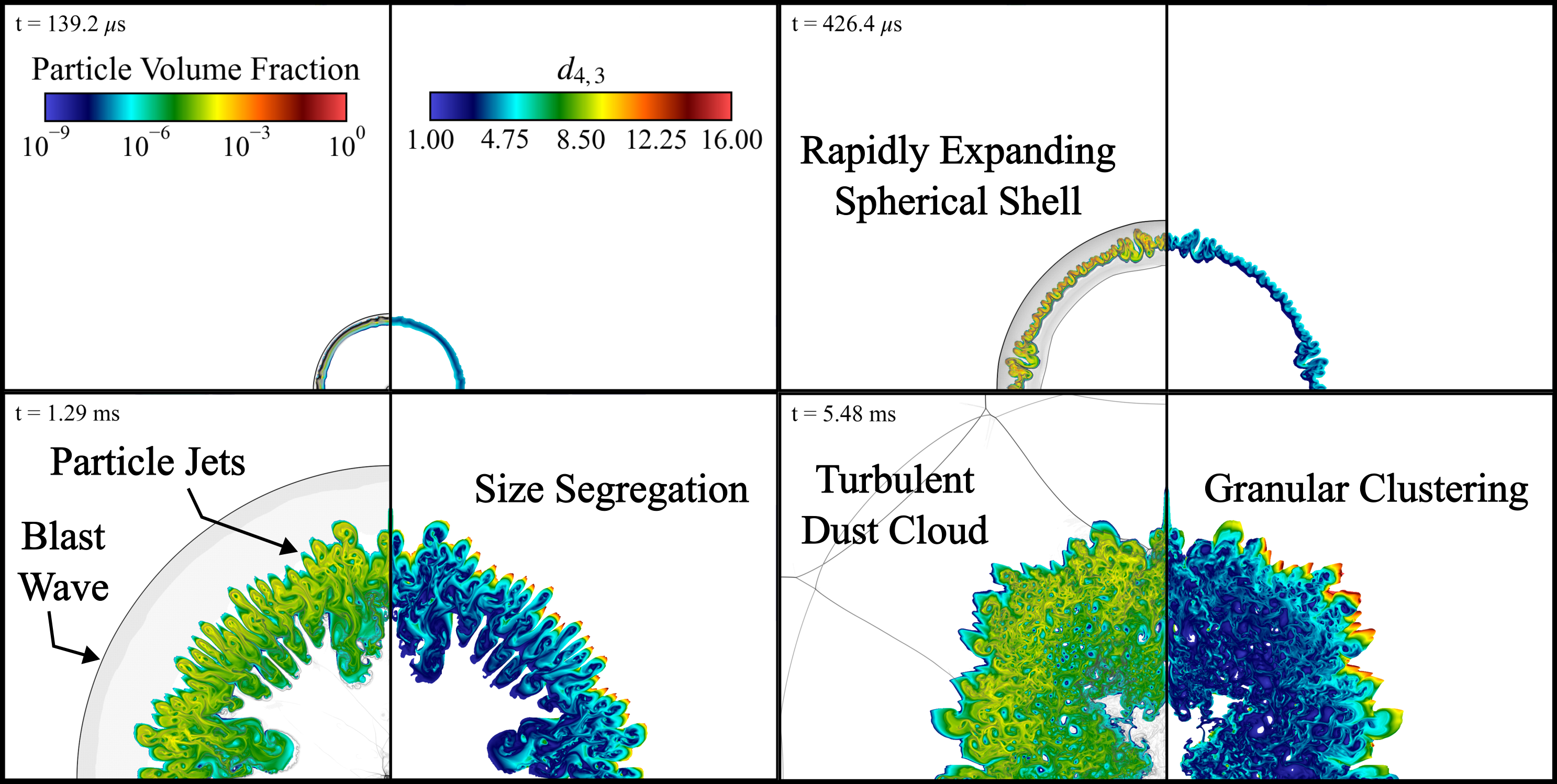}
    \caption{Time series of the particle volume fraction overlaid with a numerical schlieren and volume-averaged mean diameter for the inert dense particle shell dispersal problem. } \label{Blast_alpha_d43}
\end{figure}

Here we test robustness of the developed method for extreme problems involving very high pressures dispersal of particles by the detonation of an energetic material.  
Figure~\ref{ICs_blast} shows the geometrical setup for this extreme test. A spherical shell of particles with the H-2 size distribution are dispersed by a high-pressure, high-temperature spherical ball of gas. 
The volume fraction and thickness of the particle bed is 50\% and 1 cm, respectively.  The high pressure ball of gas has a radius of 5 cm and contains gas at 10.14 GPa and 4142 K, which corresponds to the constant-volume explosion limit of PETN \cite{Posey2021}. The gas outside of the high-pressure region and within the particle bed are air at 1 atm and 300 K.  We use a Jones-Wilkins-Lee (JWL) equation of state for gaseous mixtures \cite{Posey2021,Kuhl2007, Houim2021}.
The governing equations are solved in two-dimensional axisymmetric coordinates with rotational symmetry in the $x$-direction.  The domain measures 2 m and 2 m.  All boundaries use symmetry conditions, with the exception of the axis of rotation.  Four levels of refinement are used with a grid spacing of 976 $\mu$m at the finest level. 
The minimum particle volume fraction for particle removal was changed from $\alpha_{p,\min}=10^{-11}$ to $\alpha_{p,\min}=10^{-9}$ for this problem.  This was to alleviate difficulty in very low $\alpha_{p}$ impinging jets with significantly different abscissae, which is a similar to a fundamental issue of particle trajectory crossing problem that challenges QBMM approaches \cite{Desjardins2008}.

Figure ~\ref{Blast_alpha_d43} shows a time sequence of the particle volume fraction and volume-averaged particle diameter $d_{4,3}$. The expanding high-pressure gas initially pushes on the particle bed. The results in a granular shock \cite{Kamenetsky2000,Khmel2014a,Khmel2014b} that consolidates the particle bed close to the packing limit $\alpha_{p,\max}=65\%$. Next, the particle shell expands radially outwards, which acts as a porous piston that, in turn, pushes on the surrounding air.  The impulsive motion of the expanding particle bed results in an air shock that propagates away from the particles \cite{Posey2021}. 

Finger-like particle jets during the expansion of the particle bed due to instabilities originating from drag-induced vorticity generation \cite{Koneru2020,Frost2012}.  Significant size segregation starts occuring during the negative phase of the blast where the inner edge of the fireball reverses direction \cite{courtiaud2019analysis,zukas2013explosive}.  (See Fig.~\ref{Blast_alpha_d43} at 1.22 ms.) This inward velocity preferentially entrains smaller particles particles inwards while the larger particles continue to flow radially outwards. At even later times, we observe that the smaller particles are ``trapped'' in the fireball while the larger particles have enough inertia to escape it.

\section{Summary and Conclusions}
In this work we have extended our previous high-order numerical methods used for monodisperse Eulerian multiphase flows \cite{Houim2016} to our recently developed polydisperse Eulerian multiphase model \cite{Fox2024}.
The polydisperse model includes effects from drag, lift, convective heat transfer, particle collisions, friction, particle-fluid-particle pressure, and forces arising from finite-size particles \cite{Fox2019}.   Polydispersity is introduced to the Eulerian modeling approach with the quadrature method of moments \cite{Marchisio2013}. Continuous particle size distributions are included using the beta-generalized quadrature method of moments algorithm \cite{Fox2023}. 

One of the primary challenges was to develop a high-order solution algorithm necessary to capture sharp granular interfaces and effects such as particle size segregation while maintaining robustness for extreme flows.  The method computes the spatial derivatives in the governing equations by assembling the results from decoupled Riemann solvers for the gas and granular phases.  The gas phase used the HLLC Riemann solver with fifth-order MUSCL interpolation \cite{Kim2005}.  The granular phase Riemann problem was solved by transforming the moment equations into a form that resembles a set of decoupled monodisperse granular flow equations at each Gaussian quadrature node.  A modified AUSM$^{+}$-up solver was developed to solve the granular Riemann problems at each quadrature node.  The Gaussian quadrature abscissae were reconstructed using first-order upwinding, while all the other variables nominally used fifth-order WENO to maintain sharp granular interfaces.

The developed numerical solution method was tested against a variety of one-dimensional and two-dimensional test problems including polydisperse multiphase shock tubes, polydisperse shock and dust layer interactions, dispersal of dust layers by shock waves, and dispersal of spherical particle shells by very high pressure gas.  The method demonstrated robustness, resolved sharp interfaces, and had the ability to compute effects such as size segregation that are impossible to compute using monodisperse models.  

Future work will include developing methods to allow for high-order interpolation of the abscissae, including effects from phase change such as particle combustion and nucleation, and particle breakage.  In addition, models that include higher-order velocity moments to allow for particle trajectory crossing are will be explored.

\section*{Acknowledgments}
This work was supported by the United States Air Force Research Laboratory through contract FA8651-21-C-0015 and Sandia National Laboratories through contract DE-NA0003525.
Sandia National Laboratories is a multi-mission laboratory managed and operated by National Technology \& Engineering Solutions of Sandia, LLC (NTESS), a wholly owned subsidiary of Honeywell International Inc., for the U.S. Department of Energy’s National Nuclear Security Administration (DOE/NNSA) under contract DE-NA0003525. This paper describes objective technical results and analysis. Any subjective views or opinions that might be expressed in the paper do not necessarily represent the views of the U.S. Department of Energy or the United States Government.

\bibliographystyle{elsarticle-num} 
\bibliography{refs}

@article{Cimarelli2022,
author = {Cimarelli, C. and Genareau, K.},
journal = {Journal of Volcanology and Geothermal Research},
pages = {107449},
title = {{A review of volcanic electrification of the atmosphere and volcanic lightning}},
volume = {422},
year = {2022}
}

@article{Berry1974,
author = {Berry, E. X. and Reinhardt, R. L.},
journal = {Journal of the Atmospheric Sciences},
pages = {1814--1824},
title = {An Analysis of Cloud Drop Growth by Collection: Part {I.} Double Distributions},
volume = {31},
number = {7},
year = {1974}
}

@article{Youdin2005,
author = {Youdin, A. N. and Goodman, J.},
journal = {The Astrophysical Journal},
pages = {459--469},
title = {Streaming Instabilities in Protoplanetary Disks},
volume = {620},
number = {1},
year = {2005}
}

@article{Lambrechts2012,
author = {Lambrechts, M. and Johansen, A.},
journal = {Astronomy \& Astrophysics},
pages = {A32},
title = {{Rapid growth of gas-giant cores by pebble accretion}},
volume = {544},
year = {2012}
}

@article{Ingenito2004,
author = {Ingenito, A. and Bruno, C.},
journal = {Journal of Propulsion and Power},
title = {Using Aluminum for Space Propulsion},
pages = {1056--1063},
volume = {20},
number = {6},
year = {2004}
}

@inproceedings{Risha2009,
author = {Risha, G. A. and Connell Jr., T. L. and Yetter, R. A. and Yang, V. and Wood, T. D. and Pfeil, M. A. and Pourpoint, T. L. and Son, S. F.},
title = {Aluminum-Ice {(ALICE)} Propellants for Hydrogen Generation and Propulsion},
booktitle = {45th AIAA/ASME/SAE/ASEE Joint Propulsion Conference \& Exhibit},
year = {2009}
}

@article{Trzcinski2015,
author = {Trzci{\'{n}}ski, W. A. and Maiz, L.},
journal = {Propellants, Explosives, Pyrotechnics},
number = {5},
pages = {632--644},
title = {Thermobaric and Enhanced Blast Explosives - Properties and Testing Methods},
volume = {40},
year = {2015}
}

@article{Frost2018,
author = {Frost, D. L.},
journal = {Shock Waves},
number = {3},
pages = {439--449},
title = {{Heterogeneous/particle-laden blast waves}},
volume = {28},
year = {2018}
}

@article{Posey2021,
author = {Posey, J. W. and Roque, B. and Guhathakurta, S. and Houim, R. W.},
journal = {Physics of Fluids},
number = {11},
pages = {113308},
title = {{Mechanisms of prompt and delayed ignition and combustion of explosively dispersed aluminum powder}},
volume = {33},
year = {2021}
}

@article{Houim2022,
author = {Houim, R. W. and Posey, J. and Guhathakurta, S.},
journal = {International Journal of Energetic Materials and Chemical Propulsion},
number = {2},
pages = {1--14},
title = {IGNITION AND COMBUSTION OF {TNT}-DISPERSED ALUMINUM POWDER},
volume = {21},
year = {2022}
}

@article{Cimarelli2014,
author = {Cimarelli, C. and Alatorre-Ibarg{\"{u}}engoitia, M. A. and Kueppers, U. and Scheu, B. and Dingwell, D. B.},
journal = {Geology},
title = {{Experimental generation of volcanic lightning}},
pages = {79--82},
volume = {42},
number = {1},
year = {2014}
}

@article{MendezHarper2016,
author = {Méndez Harper, J. and Dufek, J.},
journal = {Journal of Geophysical Research: Atmospheres},
title = {{The effects of dynamics on the triboelectriﬁcation of volcanic ash}},
pages = {8209--8228},
volume = {121},
number = {14},
year = {2016}
}

@article{Neri2003,
author = {Neri, A. and Esposti Ongaro, T. and Macedonio, G. and Gidaspow, D.},
journal = {Journal of Geophysical Research: Solid Earth},
title = {{Multiparticle simulation of collapsing volcanic columns and pyroclastic flow}},
volume = {108},
number = {B4},
year = {2003}
}

@article{Saunders1993,
author = {Saunders, C. P. R.},
journal = {Journal of Applied Meteorology and Climatology},
title = {A Review of Thunderstorm Electrification Processes},
pages = {642--655},
volume = {32},
number = {4},
year = {1993}
}

@article{Squire2018,
author = {Squire, J. and Hopkins, P. F.},
journal = {Monthly Notices of the Royal Astronomical Society},
pages = {5011--5040},
title = {{Resonant drag instabilities in protoplanetary discs: the streaming instability and new, faster growing instabilities}},
volume = {477},
number = {4},
year = {2018}
}

@article{Squire2022,
author = {Squire, J. and Moroianu, S. and Hopkins, P. F.},
journal = {Monthly Notices of the Royal Astronomical Society},
pages = {110--130},
title = {{The acoustic resonant drag instability with a spectrum of grain sizes}},
volume = {510},
number = {1},
year = {2022}
}

@article{McNally2020,
author = {Paardekooper, S.-J. and McNally, C. P. and Lovascio, F.},
journal = {Monthly Notices of the Royal Astronomical Society},
pages = {4223--4238},
title = {{Polydisperse streaming instability – I. Tightly coupled particles and the terminal velocity approximation}},
volume = {499},
number = {3},
year = {2020}
}

@article{McNally2021,
author = {McNally, C. P. and Lovascio, F. and Paardekooper, S.-J.},
journal = {Monthly Notices of the Royal Astronomical Society},
pages = {1469--1486},
title = {{Polydisperse streaming instability – III. Dust evolution encourages fast instability}},
volume = {502},
number = {1},
year = {2021}
}

@book{Gidaspow1994,
author = {Gidaspow, D.},
publisher = {Academic},
title = {{Multiphase Flow and Fluidization}},
year = {1994}
}

@article{Lun1984,
author = {Lun, C. K.K. and Savage, S. B. and Jeffrey, D. J. and Chepurniy, N.},
journal = {Journal of Fluid Mechanics},
pages = {223--256},
title = {{Kinetic theories for granular flow: Inelastic particles in Couette flow and slightly inelastic particles in a general flowfield}},
volume = {140},
year = {1984}
}

@book{BrilliantovBook,
author = {Brilliantov, N. V. and P{\"{o}}schel, T.},
title = {{Kinetic Theory of Granular Gases}},
publisher = {Oxford University Press},
year = {2004}
}

@article{feroukas2023simplified,
title={Simplified interfacial area modeling in polydisperse two-phase flows under explosion situations},
author={Feroukas, Konstantinos and Chiapolino, Alexandre and Saurel, Richard},
journal={Fire},
volume={6},
number={1},
pages={21},
year={2023}
}

@article{binning,
title = {Effects of particle size and density on dust dispersion behind a moving shock},
journal = {Physical Review Fluids},
volume = {3},
pages = {064306},
year = {2018},
author = {Lai, S. and Houim, R. W. and Oran, E. S.}
}

@article{binning2,
title = {Dispersion of stratified dust layers by a moving shock wave},
journal = {International Journal of Multiphase Flow},
volume = {118},
pages = {87-96},
year = {2019},
author = {Lai, S. and Houim, R. W. and Oran, E. S.}
}

@article{McGraw1997,
author = {McGraw, R.},
journal = {Aerosol Science and Technology},
number = {2},
pages = {255--265},
title = {{Description of aerosol dynamics by the quadrature method of moments}},
volume = {27},
year = {1997}
}

@book{Marchisio2013,
address = {New York, USA},
author = {Marchisio, D. L. and Fox, R. O.},
publisher = {Cambridge University Press},
title = {{Computational Models for Polydisperse Particulate and Multiphase Systems}},
year = {2013}
}

@article{Fox2024,
author = {Fox, R. O. and Posey, J. W. and Houim, R. W. and Laurent, F.},
journal = {International Journal of Multiphase Flow},
pages = {104698},
title = {A kinetic-based model for polydisperse, high-speed, fluid–particle flows},
volume = {171},
year = {2024}
}

@article{Houim2016,
author = {Houim, R. W. and Oran, E. S.},
journal = {Journal of Fluid Mechanics},
pages = {166--220},
title = {A multiphase model for compressible granular–gaseous flows: formulation and initial tests},
volume = {789},
year = {2016}
}

@article{Fox2019,
author = {Fox, R. O.},
journal = {Journal of Fluid Mechanics},
pages = {282--329},
title = {A kinetic-based hyperbolic two-fluid model for binary hard-sphere mixtures},
volume = {877},
year = {2019}
}

@article{Fox2025,
author = {Fox, R. O.},
journal = {Journal of Fluid Mechanics},
volume = {1010},
title = {The particle–fluid–particle pressure tensor for ideal-fluid–particle flow},
pages = {A8},
year = {2025}
}

@phdthesis{Posey2025,
author = {Posey, J. W.},
school = {University of Florida},
title = {{High-Order Multiphase Modeling of Reactive Polydisperse Particles}},
type = {PhD},
year = {2025}
}

@techreport{McBride2002,
author = {McBride, B. J. and Zehe, M. J. and Gordon, S.},
institution = {National Aeronautics and Space Administration},
title = {{NASA Glenn coefficients for calculating thermodynamic properties of individual species}},
year = {2002},
number = {TP—2002-211556}
}

@book{Ern1994,
address = {Heidelberg},
author = {Ern, A. and Giovangigli, V.},
publisher = {Springer-Verlag},
title = {{Multicomponent Transport Algorithms}},
year = {1994}
}

@book{Kee2003,
address = {Hoboken, NJ},
author = {Kee, R. J. and Coltrin, M. E. and Glarborg, P.},
publisher = {John Wiley \& Sons},
title = {{Chemically Reacting Flow Theory and Practice}},
year = {2003}
}

@article{Houim2011a,
author = {Houim, R. W. and Kuo, K. K.},
journal = {Journal of Computational Physics},
pages = {8527--8553},
title = {{A low-dissipation and time-accurate method for compressible multi-component flow with variable specific heat ratios}},
volume = {230},
number = {23},
year = {2011}
}

@article{Johnson1987,
author = {Johnson, P C and Jackson, R},
journal = {Journal of Fluid Mechanics},
pages = {67--93},
title = {Frictional Collisional constitutive relations for granular materials, with application to plane shearing},
volume = {176},
year = {1987}
}

@article{Santos1999,
author = {Santos, A. and Yuste, S. B. and {L{\'{o}}pez De Haro}, Mariano},
journal = {Molecular Physics},
number = {1},
pages = {1--5},
title = {Equation of state of a multicomponent $d$-dimensional hard-sphere fluid},
volume = {96},
year = {1999}
}

@article{sinclair1989gas,
title={Gas-particle flow in a vertical pipe with particle-particle interactions},
author={Sinclair, J. L. and Jackson, R.},
journal={AIChE journal},
volume={35},
number={9},
pages={1473--1486},
year={1989}
}

@article{Ramirez1999,
author = {Ram{\'i}rez, R. and P{\"{o}}schel, T. and Brilliantov, N. V. and Schwager, T.},
journal = {Physical Review E},
pages = {4465--4472},
title = {{Coefficient of restitution of colliding viscoelastic spheres}},
volume = {60},
number = {4},
year = {1999}
}

@article{Hassani-Gangaraj2018,
author = {Hassani-Gangaraj, M. and Veysset, D. and Nelson, K. A. and Schuh, C. A.},
journal = {Scripta Materialia},
pages = {9--13},
title = {{In-situ observations of single micro-particle impact bonding}},
volume = {145},
year = {2018}
}

@article{Weir2005,
author = {Weir, G. and Tallon, S.},
journal = {Chemical Engineering Science},
pages = {3637--3647},
title = {{The coefficient of restitution for normal incident, low velocity particle impacts}},
volume = {60},
number = {13},
year = {2005}
}

@article{Huilin2003,
author = {Huilin, L. and Gidapsow, D.},
journal = {Chemical Engineering Science},
number = {16},
pages = {3777--3792},
title = {{Hydrodynamics of binary fluidization in a riser: CFD simulation using two granular temperatures}},
volume = {58},
year = {2003}
}

@article{Osnes2023,
author = {Osnes, A. N. and Vartdal, M. and Khalloufi, M. and Capecelatro, J. and Balachandar, S.},
journal = {International Journal of Multiphase Flow},
pages = {104485},
title = {Comprehensive quasi-steady force correlations for compressible flow through random particle suspensions},
volume = {165},
year = {2023}
}

@article{Ergun,
author = {Ergun, S.},
journal = {Chemical Engineering Progress},
pages = {89},
title = {Fluid flow through packed columns},
volume = {48},
number = {2},
year = {1952}
}

@article{Gunn1978,
author = {Gunn, D. J.},
journal = {International Journal of Heat and Mass Transfer},
number = {4},
pages = {467--476},
title = {{Transfer of heat or mass to particles in fixed and fluidised beds}},
volume = {21},
year = {1978}
}

@article{Andries2002,
author = {Andries, P. and Aoki, K. and Perthame, B.},
journal = {Journal of Statistical Physics},
number = {5-6},
pages = {993--1018},
title = {{A consistent BGK-type model for gas mixtures}},
volume = {106},
year = {2002}
}

@article{Zhang2021,
author = {Zhang, D. Z.},
journal = {Multiscale Modeling \& Simulation},
number = {2},
pages = {1066--1082},
title = {Stress from Long-Range Interactions in Particulate Systems},
volume = {19},
year = {2021}
}

@article{Abgrall1996,
author = {Abgrall, R{\'{e}}mi},
journal = {Journal of Computational Physics},
pages = {150--160},
title = {How to Prevent Pressure Oscillations in Multicomponent Flow Calculations: {A} Quasi Conservative Approach},
volume = {125},
year = {1996}
}

@article{Saurel1999,
author = {Saurel, R. and Abgrall, R.},
journal = {Journal of Computational Physics},
number = {2},
pages = {425--467},
title = {A Multiphase {G}odunov Method for Compressible Multifluid and Multiphase Flows},
volume = {150},
year = {1999}
}

@article{Liou2008,
author = {Liou, M.-S. and Chang, C.-H. and Nguyen, L. and Theofanous, T. G.},
journal = {AIAA Journal},
number = {9},
pages = {2345--2356},
title = {How to Solve Compressible Multifluid Equations: {A} Simple, Robust, and Accurate Method},
volume = {46},
year = {2008}
}

@book{Wilf1962,
author = {Wilf, H. S.},
title = {{Mathematics for the Physical Sciences}},
publisher = {John Wiley and Sons},
year = {1962}
}

@book{Golub1996,
author = {Golub, G. H. and Van Loan, C. F.},
publisher = {Johns Hopkins University Press},
title = {{Matrix Computations}},
year = {1996},
edition = {3}
}

@article{Wheeler1974,
author = {Wheeler, J. C.},
journal = {Rocky Mountain Journal of Mathematics},
pages = {287--296},
title = {{Modified moments and Gaussian quadratures}},
volume = {4},
year = {1974}
}

@book{gautschi2004orthogonal,
title={Orthogonal Polynomials: Computation and Approximation},
author={Gautschi, Walter},
year={2004},
publisher={OUP Oxford}
}

@article{Fox2023,
author = {Fox, R. O. and Laurent, F. and Passalacqua, A.},
journal = {Journal of Aerosol Science},
pages = {106096},
title = {{The generalized quadrature method of moments}},
volume = {167},
year = {2023}
}

@article{Yuan2011,
author = {Yuan, C. and Fox, R. O.},
journal = {Journal of Computational Physics},
number = {22},
pages = {8216--8246},
title = {Conditional quadrature method of moments for kinetic equations},
volume = {230},
year = {2011}
}

@article{Strang1968,
author = {Strang, G.},
journal = {SIAM Journal on Numerical Analysis},
number = {3},
pages = {506--517},
title = {On the Construction and Comparison of Difference Schemes},
volume = {5},
year = {1968}
}

@article{Courant1928,
author = {Courant, R. and Friedrichs, K. and Lewy, H.},
journal = {Mathematische Annalen},
pages = {32--74},
title = {{Über die partiellen Differenzengleichungen der mathematischen Physik}},
volume = {100},
year = {1928}
}

@article{Spiteri2002,
author = {Spiteri, R. J. and Ruuth, S. J.},
journal = {SIAM Journal on Numerical Analysis},
number = {2},
pages = {469--491},
title = {A new class of optimal high-order strong-stability-preserving time discretization methods},
volume = {40},
year = {2002}
}

@incollection{McGraw2012,
author = {McGraw, R.},
title = {Correcting Transport Errors During Advection of Aerosol and Cloud Moment Sequences in Eulerian Models},
booktitle = {Climate Models},
publisher = {InTech},
year = {2012}
}

@article{Toro1994,
author = {Toro, E. F. and Spruce, M. and Speares, W.},
journal = {Shock Waves},
pages = {25--34},
title = {{Restoration of the contact surface in the HLL-Riemann solver}},
volume = {4},
year = {1994}
}

@article{Davis1988,
title={Simplified second-order {G}odunov-type methods},
author={Davis, S. F.},
journal={SIAM Journal on Scientific and Statistical Computing},
volume={9},
number={3},
pages={445--473},
year={1988}
}

@article{Kim2005,
author = {Kim, K. H. and Kim, C.},
journal = {Journal of Computational Physics},
number = {2},
pages = {570--615},
title = {{Accurate, efficient and monotonic numerical methods for multi-dimensional compressible flows. Part II: Multi-dimensional limiting process}},
volume = {208},
year = {2005}
}

@article{Thornber2008,
author = {Thornber, B. and Mosedale, A. and Drikakis, D. and Youngs, D. and Williams, R. J. R.},
journal = {Journal of Computational Physics},
number = {10},
pages = {4873--4894},
title = {An improved reconstruction method for compressible flows with low {M}ach number features},
volume = {227},
year = {2008}
}

@article{Liou1993,
author = {Liou, Meng-Sing and Steffen, Christopher J.},
journal = {Journal of Computational Physics},
number = {1},
pages = {23--39},
title = {A New Flux Splitting Scheme},
volume = {107},
year = {1993}
}

@article{Liou1996,
author = {Liou, M.-S.},
journal = {Journal of Computational Physics},
number = {2},
pages = {364--382},
title = {{A Sequel to AUSM: AUSM+}},
volume = {129},
year = {1996}
}

@article{Liou2006,
author = {Liou, M.-S.},
journal = {Journal of Computational Physics},
number = {1},
pages = {137--170},
title = {{A sequel to AUSM, Part II: AUSM+-up for all speeds}},
volume = {214},
year = {2006}
}

@article{Desjardins2008,
author = {Desjardins, O. and Fox, R. O. and Villedieu, P.},
journal = {Journal of Computational Physics},
pages = {2514--2539},
title = {{A quadrature-based moment method for dilute fluid-particle flows}},
volume = {227},
number = {4},
year = {2008}
}

@article{Vikas2011,
author = {Vikas, V. and Wang, Z. J. and Passalacqua, A. and Fox, R. O.},
journal = {Journal of Computational Physics},
pages = {5328--5352},
title = {{Realizable high-order finite-volume schemes for quadrature-based moment methods}},
volume = {230},
number = {13},
year = {2011}
}

@article{Liu1994,
author = {Liu, X. D. and Osher, S. and Chan, T.},
journal = {Journal of Computational Physics},
pages = {200--212},
title = {Weighted Essentially Non-oscillatory Schemes},
volume = {115},
year = {1994}
}

@article{Jiang1996,
author = {Jiang, G. S. and Shu, C. W.},
journal = {Journal of Computational Physics},
number = {126},
pages = {202--228},
title = {{Efficient implementation of weighted ENO schemes}},
volume = {126},
year = {1996}
}

@article{Rusanov1962,
author = {Rusanov, V. V.},
journal = {USSR Computational Mathematics and Mathematical Physics},
number = {2},
pages = {304--320},
title = {{The calculation of the interaction of non-stationary shock waves and obstacles}},
volume = {1},
year = {1962}
}

@article{Shen2016,
author = {Shen, Z. and Yan, W. and Yuan, G.},
journal = {Journal of Computational Physics},
pages = {185--206},
title = {{A robust HLLC-type Riemann solver for strong shock}},
volume = {309},
year = {2016}
}

@book{Zwillinger1989,
author = {Zwillinger, D.},
title = {{Handbook of Differential Equations}},
publisher = {Academic Press},
year = {1989}
}

@article{Moler2003,
author = {Moler, C. and Van Loan, C.},
journal = {SIAM Review},
pages = {3--49},
title = {Nineteen Dubious Ways to Compute the Exponential of a Matrix, Twenty-Five Years Later},
volume = {45},
number = {1},
year = {2003}
}

@article{Haff1983,
author = {Haff, P. K.},
journal = {Journal of Fluid Mechanics},
pages = {401--430},
title = {{Grain flow as a fluid mechanical phenomenon}},
volume = {134},
year = {1983}
}

@article{Brown1989,
author = {Brown, P. N. and Byrne, G. D. and Hindmarsh, A. C.},
journal = {SIAM Journal on Scientific and Statistical Computing},
pages = {40--91},
title = {{VODE}, A VARIABLE-COEFFICIENT ODE SOLVER},
volume = {31},
year = {1989}
}

@article{Blais2020,
author = {Blais, F. and Julien, P. and Palecka, J. and Goroshin, S. and Bergthorson, J. M.},
journal = {Combustion Science and Technology},
number = {8},
pages = {1513--1526},
title = {Effect of Initial Reactant Temperature on Flame Speeds in Aluminum Dust Suspensions},
volume = {194},
year = {2020}
}

@article{Julien2015,
author = {Julien, P. and Vickery, J. and Goroshin, S. and Frost, D. L. and Bergthorson, J. M.},
journal = {Combustion and Flame},
pages = {4241--4253},
title = {{Freely-propagating flames in aluminum dust clouds}},
volume = {162},
number = {11},
year = {2015}
}

@article{Zhang2019,
author = {Zhang, W. and Almgren, A. and Beckner, V. and Bell, J. and Blaschke, J. and Chan, C. and Day, M. and Friesen, B. and Gott, K. and Graves, D. and Katz, M. and Myers, A. and Nguyen, T. and Nonaka, A. and Rosso, M. and Williams, S. and Zingale, M.},
journal = {Journal of Open Source Software},
number = {37},
pages = {1370},
title = {{AMReX:} a framework for block-structured adaptive mesh refinement},
volume = {4},
year = {2019}
}

@article{Miura1982,
author = {Miura, H. and Glass, I. I.},
journal = {Proceedings of the Royal Society A},
pages = {373--388},
title = {On a dusty-gas shock tube},
volume = {382},
number = {1783},
year = {1982}
}

@article{Saito2003,
author = {Saito, T. and Marumoto, M. and Takayama, K.},
journal = {Shock Waves},
number = {4},
pages = {299--322},
title = {Numerical investigations of shock waves in gas–particle mixtures},
volume = {13},
year = {2003}
}

@article{Fedorov2010,
author = {Fedorov, A. V. and Fedorchenko, I. A.},
journal = {Combustion, Explosion and Shock Waves},
number = {5},
pages = {578--588},
title = {Numerical simulation of shock wave propagation in a mixture of a gas and solid particles},
volume = {46},
year = {2010}
}

@article{Wagner2012,
author = {Wagner, J. L. and Beresh, S. J. and Kearney, S. P. and Trott, W. M. and Castaneda, J. N. and Pruett, B. O. and Baer, M. R.},
journal = {Experiments in Fluids},
number = {6},
pages = {1507--1517},
title = {{A multiphase shock tube for shock wave interactions with dense particle fields}},
volume = {52},
year = {2012}
}

@article{Daniel2022,
author = {Daniel, K. A. and Wagner, J. L.},
journal = {International Journal of Multiphase Flow},
pages = {104082},
title = {{The shock-induced dispersal of particle curtains with varying material density}},
volume = {152},
year = {2022}
}

@article{Helland2000,
author = {Helland, E. and Occelli, R. and Tadrist, L.},
journal = {Powder Technology},
number = {3},
pages = {210--221},
title = {{Numerical study of cluster formation in a gas–particle circulating fluidized bed}},
volume = {110},
year = {2000}
}

@article{Agrawal2001,
author = {Agrawal, K. and Loezos, P. N. and Syamlal, M. and Sundaresan, S.},
journal = {Journal of Fluid Mechanics},
pages = {151--185},
title = {{The role of meso-scale structures in rapid gas–solid flows}},
volume = {445},
year = {2001}
}

@article{Chowdhury2018,
author = {Chowdhury, A. and Johnston, H. G. and Mashuga, C. V. and Mannan, M. S. and Petersen, E. L.},
journal = {Experimental Thermal and Fluid Science},
pages = {1--10},
title = {{Effect of particle size and polydispersity on dust entrainment behind a moving shock wave}},
volume = {93},
year = {2018}
}

@article{Ugarte2017,
author = {Ugarte, O. J. and Houim, R. W. and Oran, E. S.},
journal = {Physical Review Fluids},
issue = {7},
title = {Examination of the forces controlling dust dispersion by shock waves},
volume = {2},
year = {2017},
pages = {074304}
}

@inproceedings{Kuhl2007,
address = {Karlsruhe},
author = {Kuhl, A. L. and Khasainov, B.},
booktitle = {37th International Annual Conference Energetic Materials Characterisation and Performance of Advanced Systems},
title = {{Quadratic model of thermodynamic states in SDF explosions}},
year = {2007}
}

@article{Houim2021,
author = {Houim, R. W.},
journal = {Shock Waves},
number = {8},
title = {{A simplified burn model for simulating explosive effects and afterburning}},
volume = {31},
year = {2021},
pages = {851--875}
}

@article{Kamenetsky2000,
author = {Kamenetsky, V. and Goldshtein, A. and Shapiro, M. and Degani, D.},
journal = {Physics of Fluids},
number = {11},
pages = {3036--3049},
title = {{Evolution of a shock wave in a granular gas}},
volume = {12},
year = {2000}
}

@article{Khmel2014a,
author = {Khmel', T. A. and Fedorov, A. V.},
journal = {Combustion, Explosion, and Shock Waves},
pages = {196--207},
title = {Description of dynamic processes in two-phase colliding media with the use of molecular-kinetic approaches},
volume = {50},
year = {2014}
}

@article{Khmel2014b,
author = {Khmel', T. A. and Fedorov, A. V.},
journal = {Combustion, Explosion, and Shock Waves},
pages = {547--555},
title = {Modeling of propagation of shock and detonation waves in dusty media with allowance for particle collisions},
volume = {50},
year = {2014}
}

@article{Koneru2020,
author = {Koneru, R. B. and Rollin, B. and Durant, B. and Ouellet, F. and Balachandar, S.},
journal = {Physics of Fluids},
number = {9},
pages = {093301},
title = {{A numerical study of particle jetting in a dense particle bed driven by an air-blast}},
volume = {32},
year = {2020}
}

@article{Frost2012,
author = {Frost, D. L. and Gr{\'{e}}goire, Y. and Petel, O. and Goroshin, S. and Zhang, F.},
journal = {Physics of Fluids},
pages = {091109},
title = {{Particle jet formation during explosive dispersal of solid particles}},
volume = {24},
year = {2012}
}

@article{courtiaud2019analysis,
  title={Analysis of mixing in high-explosive fireballs using small-scale pressurised spheres},
  author={Courtiaud, S{\'e}bastien and Lecysyn, Nicolas and Damamme, Gilles and Poinsot, Thierry and Selle, Laurent},
  journal={Shock Waves},
  volume={29},
  number={2},
  pages={339--353},
  year={2019}
}

@book{zukas2013explosive,
title={Explosive effects and applications},
author={Zukas, Jonas A and Walters, William},
year={2013},
publisher={Springer Science \& Business Media}
}




% %% For numbered reference style

%   \textit{\LaTeX: a document preparation system},
%   Addison Wesley, Massachusetts,

% \end{thebibliography}
\end{document}